\documentclass[pra,longbibliography,twocolumn,showpacs,nofootinbib,superscriptaddress,notitlepage]{revtex4-2}

\pdfoutput=1 

\usepackage[ruled,vlined]{algorithm2e}

\usepackage{dsfont}
\usepackage{amsmath}
\usepackage{amssymb,bm}
\usepackage{amsthm}
\usepackage{mathtools}
\usepackage{url}

\usepackage{array, makecell}
\usepackage{boldline}
\usepackage{enumitem}

\usepackage{color,dsfont}
\usepackage{graphicx}
\usepackage{ragged2e}
\usepackage[colorlinks=true, hyperindex, breaklinks, linkcolor=blue, urlcolor=blue, citecolor=blue]{hyperref} 
\usepackage[normalem]{ulem}
\usepackage[capitalise]{cleveref}
\usepackage{mathrsfs}
\usepackage[caption=false]{subfig}
\usepackage{mathtools}
\usepackage{float}
\usepackage{verbatim}
\usepackage{latexsym}
\usepackage{amsmath}
\usepackage{amssymb}
\usepackage{setspace}
\usepackage{amsfonts}
\usepackage{stmaryrd}
\usepackage{xcolor}
\usepackage{enumitem}
\usepackage{lipsum}

\DeclareMathOperator*{\argmin}{arg\,min}

\usepackage{multirow}
\usepackage{booktabs}

\begin{document}
\title{Low overhead fault-tolerant quantum error correction with the surface-GKP code} 
\author{Kyungjoo Noh}\email{nkyungjo@amazon.com}
\affiliation{AWS Center for Quantum Computing, Pasadena, CA 91125, USA}
\affiliation{IQIM, California Institute of Technology, Pasadena, CA 91125, USA}
\author{Christopher Chamberland}\email{cchmber@amazon.com}
\affiliation{AWS Center for Quantum Computing, Pasadena, CA 91125, USA}
\affiliation{IQIM, California Institute of Technology, Pasadena, CA 91125, USA}
\author{Fernando G.S.L. Brand\~ao}\email{fbrandao@amazon.com}
\affiliation{AWS Center for Quantum Computing, Pasadena, CA 91125, USA}
\affiliation{IQIM, California Institute of Technology, Pasadena, CA 91125, USA}
\begin{abstract}
Fault-tolerant quantum error correction is essential for implementing quantum algorithms of significant practical importance. In this work, we propose a highly effective use of the surface-GKP code, i.e., the surface code consisting of bosonic GKP qubits instead of bare two-level qubits. In our proposal, we use error-corrected two-qubit gates between GKP qubits and introduce a maximum likelihood decoding strategy for correcting shift errors in the two-GKP-qubit gates. Our proposed decoding reduces the total CNOT failure rate of the GKP qubits, e.g., from $0.87\%$ to $0.36\%$ at a GKP squeezing of $12$dB, compared to the case where the simple closest-integer decoding is used. Then, by concatenating the GKP code with the surface code, we find that the threshold GKP squeezing is given by $9.9$dB under the the assumption that finite-squeezing of the GKP states is the dominant noise source. More importantly, we show that a low logical failure rate $p_{L} < 10^{-7}$ can be achieved with moderate hardware requirements, e.g., $291$ modes and $97$ qubits at a GKP squeezing of $12$dB as opposed to $1457$ bare qubits for the standard rotated surface code at an equivalent noise level (i.e., $p=0.36\%$). Such a low failure rate of our surface-GKP code is possible through the use of space-time correlated edges in the matching graphs of the surface code decoder. Further, all edge weights in the matching graphs are computed dynamically based on analog information from the GKP error correction using the full history of all syndrome measurement rounds. We also show that a highly-squeezed GKP state of GKP squeezing $\gtrsim 12$dB can be experimentally realized by using a dissipative stabilization method, namely, the Big-small-Big method, with fairly conservative experimental parameters. Lastly, we introduce a three-level ancilla scheme to mitigate ancilla decay errors during a GKP state preparation.                  
\end{abstract}
\maketitle

\section{Introduction}
\label{section:Introduction}

Despite various opportunities offered by noisy intermediate-scale quantum (NISQ) devices \cite{Preskill2018_quantum}, fault-tolerant quantum error correction techniques \cite{Gottesman2009_introduction} are essential for executing quantum algorithms intractable by classical computers such as integer factorization \cite{Shor1994_algorithms} and the simulation of real-time dynamics of large quantum systems \cite{Lloyd1996_universal}. One of the most promising approaches towards fault-tolerant quantum computing is to implement the surface code \cite{FMMC12} (or its variants) using bare two-level qubits such as transmons \cite{Koch2007_charge_insensitive,Schreier2008_suppressing} or internal states of trapped ions \cite{Egan2020_fault_tolerant}. 

Recently, however, it has become increasingly clear that bosonic qubits (error corrected via bosonic quantum error correction \cite{Albert2018_performance,Joshi2020_quantum,Cai2020_bosonic}) provide unique advantages that are not available to bare two-level qubits. For instance, two-component cat codes (consisting of two coherent states $|\pm \alpha\rangle$) \cite{Cochrane1999_macroscopically,Jeong2002_efficient,Mirrahimi2014_dynamically,Leghtas2015_confining,Lescanne2020_exponential} naturally realize noise-biased qubits whose bit-flip error rate is exponentially suppressed in the size of the code $|\alpha|^{2}$, whereas the phase-flip error rate increases only linearly in $|\alpha|^{2}$. Most importantly, a CNOT gate between these two noise-biased cat qubits can be performed in a bias-preserving way. That is, high noise bias (towards phase-flip errors) can be maintained during the entire execution of the CNOT gate through a suitably designed control scheme \cite{Guillaud2019_repetition,Puri2020_bias_preserving,Guillaud2020_error,Chamberland2020_building}. On the other hand, as shown in Ref.\ \cite{Aliferis2008_fault_tolerant}, a bias-preserving CNOT gate is not possible with strictly two-dimensional bare qubits. In recent proposals \cite{Guillaud2019_repetition,Guillaud2020_error,Chamberland2020_building}, the unique noise-bias feature of bosonic cat qubits has been utilized to significantly reduce the required resource overheads for implementing fault-tolerant quantum computation.           

GKP qubits \cite{Gottesman2001_encoding,Fluhmann2019_encoding,Fluhmann2020_direct,CampagneIbarcq_2020_quantum,deNeeve2020_error} are another example of bosonic qubits which enjoy unique advantages unavailable to bare two-level qubits: if GKP qubits are used to implement a next-level error-correcting code (e.g., the surface code), extra analog information gathered from GKP error correction \cite{Fukui2017_analog} can inform us which GKP qubits are more likely to have had an error. Thus, by incorporating the extra analog information to the decoder of the next-level code, one can significantly boost the performance of the next-level error-correcting code \cite{Fukui2017_analog,Fukui2018_high_threshold,Vuillot2019_quantum,Noh2020_fault_tolerant,Yamasaki2020_polylog_overhead,Fukui2020_all,Larsen2021_fault_tolerant,Bourassa2021_blueprint,Rozpedek2021_quantum}. While it has been shown that bare two-level qubits can also benefit from analog information in a limited context of qubit readout \cite{Xue2020_repetitive,DAnjou2021_generalized}, bare two-level qubits do not have access to analog information in a more general setting such as during gates. 

In this paper, we propose a highly optimized version of the surface-GKP code (i.e., concatenation of the GKP code with the surface code), building on remarkable recent progress in the study of GKP codes. In particular, assuming that the finite squeezing of the GKP states is the only noise source (see \cref{subsection:Maximum likelihood decoding for GKP} for a justification), we show that the threshold GKP squeezing of our surface-GKP code is given by $9.9$dB, which is lower than all reported threshold values of the GKP squeezing in the literature under the same noise model and without the use of post-selection. Our threshold $9.9$dB is also close to the GKP squeezing of $9.5$dB which was experimentally achieved in a circuit QED system \cite{CampagneIbarcq_2020_quantum}. 

While the fault-tolerance threshold is an important metric, it is even more important to study the behavior of logical error rates below the threshold. That is, it is crucial to understand how many hardware elements are needed to achieve a target logical error rate to estimate an overall cost of a fault-tolerance method. For instance, while there are measurement-based schemes that achieve a very low GKP squeezing threshold (e.g., $7.8$dB) with the help of post-selection \cite{Fukui2018_high_threshold,Yamasaki2020_polylog_overhead}, it is not yet clear if the overall resource overhead is manageable in practice despite the use of post-selection. 

Hence, going beyond the discussion of fault-tolerant thresholds, we further demonstrate that our surface-GKP code can achieve low logical error rates with moderate resource costs at a reasonable value of GKP squeezing. For instance, we show that a logical error rate per code cycle $p_{L}<10^{-7}$ can be achieved with only $291$ oscillator modes and $97$ qubits at a moderately high GKP squeezing of $12$dB. On the other hand, to achieve a similar logical failure rate $p_{L}<10^{-7}$, the standard surface code approach is estimated to require at least $1457$ bare qubits at an equivalent noise level using a toy circuit-level depolarizing noise model (see \cref{fig:GKPl1AlldAnalog,tab:overhead comparison}). Moreover, we show in \cref{section:Experimental realization of a highly squeezed GKP qunaught state} that a highly-squeezed GKP state with GKP squeezing $\gtrsim 12$dB can be experimentally realized by using dissipative stabilization methods \cite{Royer2020_stabilization} (and our modification of them to mitigate ancilla qubit errors) assuming fairly conservative experimental parameters. We also consider the use of noise-biased ancilla qubits (in a similar spirit as in Ref.\ \cite{Puri2019_stabilized}) by using a rectangular-lattice GKP code for the surface code ancilla qubits. We conclude that using such biased qubits provide no improvements when the GKP squeezing is below $11$dB and a very minor improvement when the GKP squeezing is $\ge 11$dB (see \cref{fig:CompareLambdas}).         

To achieve a low logical error rate with only moderate resource requirements, we adopt the teleportation-based GKP error correction \cite{Walshe2020_continuous_variable} in our surface-GKP code proposal, instead of the more widely used Steane-type GKP error correction \cite{Gottesman2001_encoding}. In the literature, discussions on the use of teleportation-based method have been centered around the fact that it does not require online squeezing operations which is particularly convenient for optical systems \cite{Walshe2020_continuous_variable,Fukui2020_all,Larsen2021_fault_tolerant}. In \cref{subsection:Teleportation-based GKP error correction}, however, we emphasize the performance aspect of the teleportation-based method and thus motivate the use of it even in systems where online squeezing operations are not much harder to implement than beam-splitter interactions. We also adopt the recent idea of performing GKP error correction four times for each surface code stabilizer measurement \cite{Larsen2021_fault_tolerant}, as opposed to applying them only once per every syndrome extraction \cite{Vuillot2019_quantum,Noh2020_fault_tolerant}. By doing so, every two-qubit gate between GKP qubits is error corrected.   

The key contributions of our work are as follows: first, we introduce a maximum likelihood decoding method for decoding shift errors in the error-corrected two-qubit gates (i.e., CNOT and CZ gates) between GKP qubits. We then demonstrate that our maximum likelihood decoding method significantly outperforms the simple, more widely used closest-integer decoding method. For instance, while the closest-integer decoding yields a total CNOT failure rate of $0.87\%$ at a GKP squeezing of $12$dB, our maximum likelihood decoding achieves a CNOT failure rate of $0.36\%$ at the same GKP squeezing (the gap gets wider as we increase the GKP squeezing; see \cref{tab:CNOT logical failure rates}). Secondly, we add space-time correlated edges in the matching graphs of the surface code decoder. In addition, we provide a detailed method for dynamically computing all edge weights after incorporating the extra analog information from all GKP error corrections over the full syndrome measurement history. Adding space-time correlated edges is especially crucial in the case where every two-qubit gate is error corrected because doing so introduces errors that are correlated both in space and time in the surface code error correction. Lastly, we systematically analyze dissipative methods \cite{Royer2020_stabilization} for stabilizing a highly-squeezed GKP state and show that the Big-small-Big method is particularly well suited for preparing a highly-squeezed GKP state. Moreover, we propose to use a three-level ancilla to mitigate ancilla decay errors and demonstrate that a high decay rate in the ancilla can be tolerated. Note that the ancilla here refers to an ancilla for stabilizing a GKP state and should be distinguished from the ancilla GKP qubits in the surface-GKP code.   

Our paper is organized as follows. In \cref{section:Rectangular-lattice GKP code}, we review basic properties of the GKP qubits and compare two GKP error correction methods, namely, Steane-type and teleportation-based GKP error correction methods. Most importantly, in \cref{subsection:Maximum likelihood decoding for GKP}, we introduce a maximum likelihood decoding method for error-corrected two qubit gates between GKP qubits and show that it significantly outperforms the closest-integer decoding. In \cref{section:Surface-GKP code}, we present the main results regarding the logical failure rates of our highly optimized surface-GKP code (see \cref{fig:GKPl1AlldAnalog}) and compare the resource overheads of our scheme with the standard surface code approach based on bare qubits (see \cref{tab:overhead comparison}). In \cref{section:Experimental realization of a highly squeezed GKP qunaught state}, we discuss the experimental feasibility of a highly-squeezed GKP state and introduce a three-level ancilla technique for mitigating ancilla decay errors. Lastly, we conclude the paper with discussion and outlook in \cref{section:Discussion and outlook}.

\section{Rectangular-lattice GKP code}
\label{section:Rectangular-lattice GKP code}

Here, we review the rectangular-lattice GKP code, Steane-type, and teleportation-based GKP error correction schemes. We also propose a maximum likelihood decoding method for correcting shift errors during an error-corrected two-qubit gate between GKP qubits. In particular, for two-GKP qubit gates (e.g., CNOT gate), we consider a hybrid scheme where the control qubit (to be used as an ancilla qubit in the surface code) is encoded in a rectangular-lattice GKP code and the target qubit (to be used as a data qubit) is encoded in the square-lattice GKP code. 

We choose to focus on the hybrid scheme instead of the homogeneous scheme where both qubits are encoded in the same rectangular-lattice GKP code. The homogeneous scheme was considered in Ref.\ \cite{Hanggli2020_enhanced} to bias the noise of the GKP qubits and was shown to be advantageous in the high noise regime for a code capacity noise model (i.e., without considering finite squeezing of GKP states). In our case, however, we are mainly concerned with the finite GKP squeezing as it is the most dominant noise source in practical scenarios (see \cref{subsection:Maximum likelihood decoding for GKP} for a more quantitative discussion). In this case, in the practically relevant regime of GKP squeezing between $9$dB and $13$dB, we found that the enhanced phase-flip rate in the homogeneous scheme is too large to be compensated by any advantages from a reduced bit-flip rate.

On the other hand, a key motivation behind the hybrid scheme is to bias the noise only in the ancilla qubits of the surface code such that the noise back-propagation from an ancilla qubit to the data block is reduced at the expense of increased syndrome measurement error rate (in the same spirit as in Ref.\ \cite{Puri2019_stabilized}; see also \cref{tab:GKP CNOT CZ detailed error rates}). In this case, we do observe a very minor improvement over the case where we use the noise-unbiased square-lattice GKP qubits everywhere. Although we end up focusing on the square-lattice GKP code, we review the more general rectangular-lattice GKP codes to explicitly demonstrate the optimality (or near-optimality) of the square-lattice GKP code.              

\subsection{Logical states, operations. and measurements}
\label{subsection:Logical states and operations}

Rectangular-lattice GKP code states are stabilized by two commuting displacement operators 
\begin{align}
    \hat{S}_{q,\lambda} &= \exp\Big{[} i \frac{ 2\sqrt{\pi} }{ \lambda }  \hat{q} \Big{]} ,  
    \nonumber\\
    \hat{S}_{p,\lambda} &= \exp[ -i 2\sqrt{\pi}  \lambda  \hat{p} ] . 
\end{align}
Note that the $\lambda = 1$ case corresponds to the square-lattice GKP code and as will be made clear shortly, the parameter $\lambda$ determines the aspect ratio of the underlying rectangular lattice. The logical Pauli operators are given by 
\begin{align}
    \hat{Z}_{\lambda} &\equiv \hat{S}_{q,\lambda}^{\frac{1}{2}} = \exp\Big{[} i \frac{ \sqrt{\pi} }{ \lambda }  \hat{q} \Big{]} ,  
    \nonumber\\
    \hat{X}_{\lambda} &\equiv \hat{S}_{p,\lambda}^{\frac{1}{2}} = \exp[ -i \sqrt{\pi}  \lambda  \hat{p} ] . 
\end{align}
Thus, the logical states in the computational basis (eigenstates of $\hat{Z}_{\lambda}$) are given by 
\begin{align}
    |0_{\lambda} \rangle &\propto \sum_{n\in\mathbb{Z}} |\hat{q} = 2n\sqrt{\pi}\lambda\rangle, 
    \nonumber\\
    |1_{\lambda} \rangle &\propto \sum_{n\in\mathbb{Z}} |\hat{q} = (2n+1)\sqrt{\pi}\lambda\rangle, \label{eq:GKP computational basis states}
\end{align}
and the logical states in the complementary basis (eigenstates of $\hat{X}_{\lambda}$) are given by 
\begin{align}
    |+_{\lambda} \rangle &\propto \sum_{n\in\mathbb{Z}} |\hat{p} = 2n\sqrt{\pi}/ \lambda\rangle, 
    \nonumber\\
    |-_{\lambda} \rangle &\propto \sum_{n\in\mathbb{Z}} |\hat{p} = (2n+1)\sqrt{\pi} / \lambda\rangle . \label{eq:GKP complementary basis states}
\end{align}
Note that the position and momentum quadratures of the logical states sit on a rectangular lattice $\lbrace (n_{q} \sqrt{\pi}\lambda  , n_{p} \sqrt{\pi}/\lambda  ) | n_{q},n_{p} \in \mathbb{Z} \rbrace$. Thus, compared to the square-lattice case (i.e., $\lambda = 1$), the spacing in the position quadrature is elongated by a factor of $\lambda$ and the spacing in the momentum quadrature is contracted by the same factor. In terms of the error-correcting capability, this means that the rectangular-lattice GKP code can correct any small shift error $\exp[ i (\xi_{p}\hat{q} - \xi_{q} \hat{p} )]$ with 
\begin{align}
    |\xi_{q}| < \frac{\sqrt{\pi}\lambda}{2} , \quad |\xi_{p}| < \frac{\sqrt{\pi}}{2\lambda} . 
\end{align}
As a result, choosing $\lambda > 1$ and assuming that noise is symmetric in both the position and the momentum quadratures (which is typically the case; see, e.g., \cref{subsection:Teleportation-based GKP error correction}), the rectangular-lattice GKP code has a higher chance of having a logical $Z$ error than a logical $X$ error. The opposite is true for $\lambda < 1$.    

Regardless of the underlying lattice structure of the GKP code, logical Clifford operations of the GKP code can be performed by using a Gaussian operation (or a quadratic Hamiltonian in the quadrature operators). The most relevant Clifford operations to the implementation of the surface code are the CNOT and the CZ gates. In our surface code architecture, we consider a hybrid code where the ancilla qubits of the surface code are encoded in a rectangular-lattice GKP code but the data qubits are always encoded in the square-lattice GKP code. Thus, we assume that the control qubit of the CNOT gate is a rectangular-lattice GKP qubit with $0.8\le \lambda \le 1.2$ and the target qubit is a square-lattice GKP qubit with $\lambda = 1$. Such a CNOT gate can be implemented by using a rescaled SUM gate (or a $qp$ coupling)   
\begin{align}
    \mathrm{CNOT}_{j\rightarrow k} &= \exp[-\frac{ i}{\lambda }\hat{q}_{j}\hat{p}_{k}] ,  \label{eq:GKP logical CNOT}
\end{align}
where $j$ is the control qubit and $k$ is the target qubit. Similarly, the CZ gate between a rectangular-lattice GKP qubit and a square-lattice GKP qubit can be realized via a $qq$ coupling: 
\begin{align}
    \mathrm{CZ}_{j,k} &= \exp\Big{[} \frac{i}{\lambda}\hat{q}_{j}\hat{q}_{k} \Big{]} , \label{eq:GKP logical CZ}
\end{align}
where the qubit $j$ is encoded in a rectangular-lattice GKP code and the qubit $k$ is encoded in the square-lattice GKP qubit. 

Lastly, we remark that the $Z$-basis measurement (distinguishing $|0_{\lambda}\rangle$ from $|1_{\lambda}\rangle$) can be performed by a position homodyne measurement. More specifically, if the position homodyne measurement outcome is in the range $ \lbrace q_{m} | (n_{q} - 1/2) \sqrt{\pi}\lambda < q_{m} < (n_{q} + 1/2) \sqrt{\pi}\lambda  \rbrace $ for an even (odd) $n_{q}$, we infer that the state is in $|0_{\lambda}\rangle$ ($|1_{\lambda}\rangle$). Similarly, the $X$-basis measurement can be done by a momentum homodyne measurement: if the measurement outcome $p_{m}$ is in the range $ \lbrace p_{m} | (n_{p} - 1/2) \sqrt{\pi} / \lambda < p_{m} < (n_{p} + 1/2) \sqrt{\pi} / \lambda  \rbrace $ for an even (odd) $n_{p}$, we infer that the state is in $|+_{\lambda}\rangle$ ($|-_{\lambda}\rangle$). 

\subsection{Teleportation-based GKP error correction}
\label{subsection:Teleportation-based GKP error correction}

\begin{figure*}
    \centering
    \includegraphics[width=0.8\textwidth]{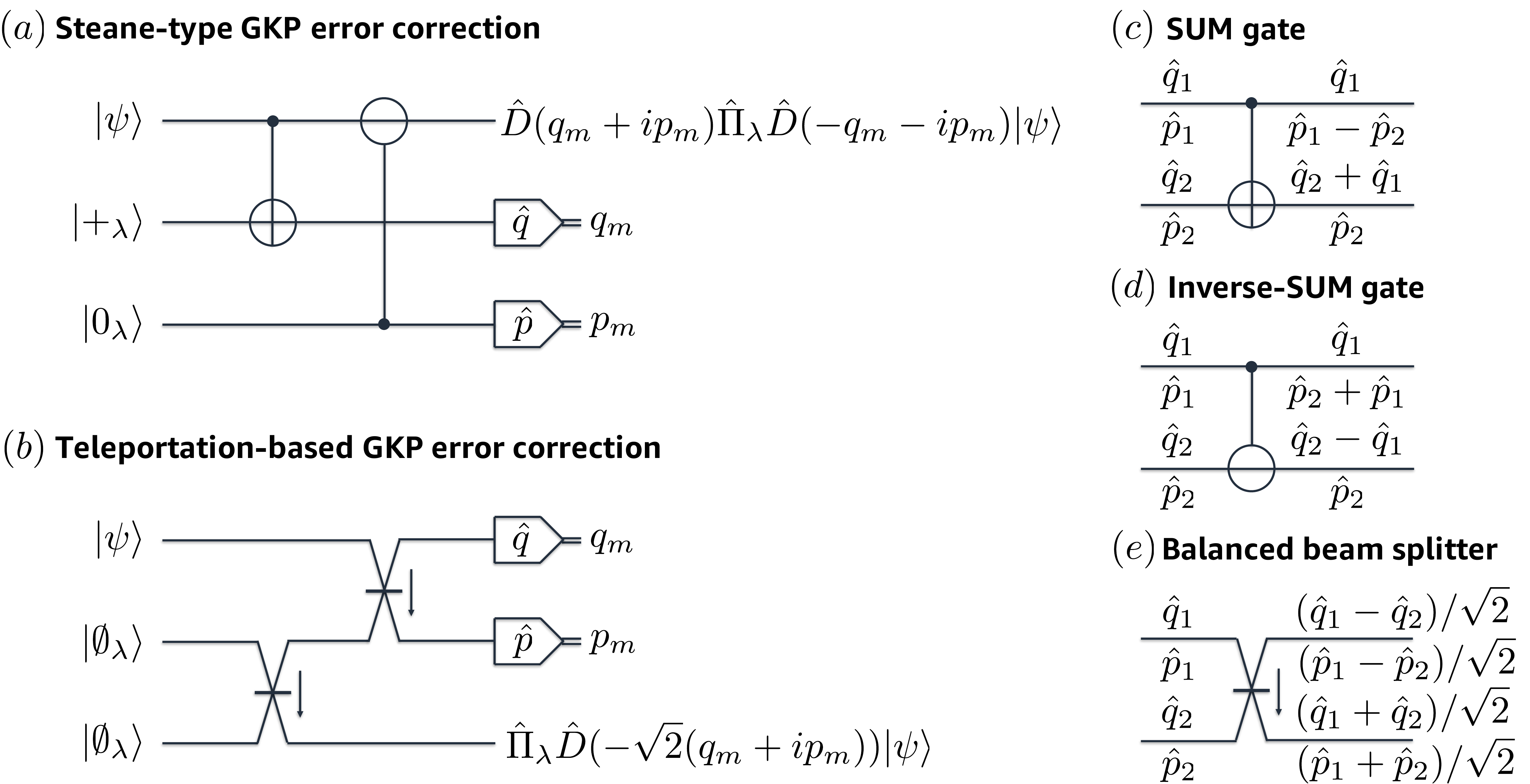}
    \caption{Quantum circuits of (a) the Steane-type and (b) the teleportation-based GKP error correction protocols. The evolution of the quadrature operators due to the circuit elements (SUM gate, inverse-SUM gate, and balanced beam-splitter interaction) in the Heisenberg picture are described in (c)--(e). The measurements at the end of the circuits are a homodyne measurement of the position ($\hat{q}$) or the momentum ($\hat{p}$) operator. The ancilla input states $|+_{\lambda}\rangle$, $|0_{\lambda}\rangle$, $|\emptyset_{\lambda}\rangle$ are defined in \cref{eq:GKP computational basis states,eq:GKP complementary basis states,eq:GKP qunaught state}. Conditioned on the measurement outcomes $q_{m}$ and $p_{m}$, the output state is given by $\hat{D}(q_{m}+ip_{m})\hat{\Pi}_{\lambda}\hat{D}(-q_{m}-ip_{m})|\psi\rangle $ in the Steane-type scheme and by $\hat{\Pi}_{\lambda}\hat{D}(-\sqrt{2}(q_{m}+ip_{m}))|\psi\rangle$ in the teleportation-based scheme (see \cref{appendix:GKP error correction} for the derivation). Here, $\hat{\Pi}_{\lambda} \equiv |0_{\lambda}\rangle\langle 0_{\lambda}| + |1_{\lambda}\rangle\langle 1_{\lambda}|$ is the projection operator to the GKP code space and $\hat{D}(\xi_{q} + i\xi_{p}) \equiv \exp[ i(\xi_{p}\hat{q} - \xi_{q}\hat{p} ) ]$ is the displacement operator. Note that our convention for the displacement operator is different from the usual convention by a factor of $\sqrt{2}$ in the shift argument, i.e., $\hat{D}(\xi) = \exp[  ( \xi \hat{a}^{\dagger} - \xi^{*}\hat{a} ) / \sqrt{2} ]$. We use this different convention because it is more convenient for describing the GKP code.   }
    \label{fig:GKP error correction}
\end{figure*}

To correct for the shift errors with the GKP code, it is essential to non-destructively measure quadrature operators modulo an appropriate spacing (e.g., $\hat{q}$ modulo $\sqrt{\pi}\lambda$ and $\hat{p}$ modulo $\sqrt{\pi}\lambda$ in the case of the rectangular-lattice GKP code). In the original GKP paper \cite{Gottesman2001_encoding}, it was proposed to achieve this via a Steane-type error correction by using two ancilla GKP states $|0_{\lambda}\rangle$, $|+_{\lambda}\rangle$, the SUM gate and its inverse, and homodyne measurements (see \cref{fig:GKP error correction} (a)). Recently, Ref.\ \cite{Walshe2020_continuous_variable} proposed an alternative method, namely a teleportation-based GKP error correction method, which requires two identical ancilla GKP states $|\emptyset_{\lambda}\rangle$ (to be defined below), beam-splitter interactions, and homodyne measurements (see \cref{fig:GKP error correction} (b)).  

One of the advantages of the teleportation-based GKP error correction, as advocated in Refs.\ \cite{Walshe2020_continuous_variable,Fukui2020_all,Larsen2021_fault_tolerant}, is that it works with beam-splitter interactions and does not require any online squeezing operations. On the other hand, the Steane-type GKP error correction is based on the SUM gate and its inverse which do require squeezing operations in addition to beam-splitter interactions. The absence of squeezing operations in the teleportation-based method is especially convenient for optical systems since beam-splitter interactions are much easier to realize than squeezing operations in the optics setting. In superconducting systems, however, engineering the SUM gate is not quite harder than engineering a beam-splitter interaction because one can simply add more drive tones to realize the extra squeezing operations on top of the beam-splitter interaction \cite{Gao2018_programmable,Zhang2019_engineering,Gao2019_entanglement}. 

In this section, we review both the Steane-type and the teleportation-based GKP error correction schemes. In particular, we pay particular attention to the performance aspect of these two schemes and emphasize that the teleportation-based method performs better than the Steane-type method under the same condition on the quality of the ancilla GKP states. That is, the teleportation-based method is a better choice than the Steane-type method for any systems regardless of whether online squeezing operations are hard to implement or not.   

Since the teleportation-based GKP error correction was only recently introduced but is crucial for boosting the performance of the GKP code, we review it as well as the Steane-type GKP error correction in great detail. In \cref{subsubsection:Noiseless ancilla GKP states}, we first focus on the case where ancilla GKP states are noiseless and present the key ideas behind both GKP error correction schemes. The adverse effects of noisy ancilla GKP states will be discussed in \cref{subsubsection:Noisy ancilla GKP states}.  

\subsubsection{Noiseless ancilla GKP states}
\label{subsubsection:Noiseless ancilla GKP states}

In the Steane-type GKP error correction (shown in \cref{fig:GKP error correction} (a)), the ancilla which is initially prepared in $|+_{\lambda}\rangle$ non-destructively measures the position of the state $|\psi\rangle$ modulo $\sqrt{\pi}\lambda$. Similarly the other ancilla which is initially in $|0_{\lambda}\rangle$ measures the momentum of the state $|\psi\rangle$ modulo $\sqrt{\pi}/\lambda$ in a non-destructive manner. As a result, conditioned on the measurement outcomes $q_{m}$ and $p_{m}$, the output state is given by 
\begin{align}
    |\psi(q_{m},p_{m})\rangle_{\mathrm{Steane}} &\propto \hat{D}(q_{m}+ip_{m})\hat{\Pi}_{\lambda}\hat{D}(-q_{m}-ip_{m})|\psi\rangle,  \label{eq:ideal output state of the GKP error correction Steane}
\end{align}
where $\hat{\Pi}_{\lambda} \equiv |0_{\lambda}\rangle\langle 0_{\lambda}| + |1_{\lambda}\rangle\langle 1_{\lambda}|$ is the projection operator to the GKP code space and $\hat{D}(\xi_{q} + i\xi_{p}) \equiv \exp[ i(\xi_{p}\hat{q} - \xi_{q}\hat{p} ) ]$ is the displacement operator (see \cref{appendix:GKP error correction} for the derivation). Note that our convention for the displacement operator, which is suitable for describing the GKP code, is different from the usual convention by a factor of $\sqrt{2}$ in the shift argument, i.e., $\hat{D}(\xi) = \exp[ ( \xi \hat{a}^{\dagger} - \xi^{*}\hat{a} ) / \sqrt{2} ]$. Note also that the non-destructiveness of the measurement is evident once we view $\hat{D}(q_{m}+ip_{m})\hat{\Pi}_{\lambda}\hat{D}(-q_{m}-ip_{m})$ as the projection operator to a displaced GKP code space by $\hat{D}(q_{m}+ip_{m})$. 

Since the output state is in a displaced GKP code space, such shift needs to be corrected. This is done by applying a correction shift $\hat{D}(-\xi_{q} - i\xi_{p})$, where the size of the correction shift is given by   
\begin{align}
    \xi_{q} &= R_{\sqrt{\pi}\lambda}(q_{m}), \quad \xi_{p} = R_{\sqrt{\pi}/\lambda}(p_{m}) . 
\end{align}
Here, the remainder function $R_{s}(z)$ is defined as 
\begin{align}
    R_{s}(z) &\equiv \min_{n\in\mathbb{Z}} | z-ns | =  z - s \Big{\lfloor} \frac{z}{s} + \frac{1}{2} \Big{\rfloor}.  
\end{align}
By using the remainder function in the correction shifts, we are employing a maximum likelihood decoding strategy under the assumption that smaller shifts are more likely to happen than larger shifts (this assumption is not necessarily true in the case of the CNOT and the CZ gates; see \cref{subsection:Maximum likelihood decoding for GKP} for more details). Through this decoding strategy, we can correct any small shifts $ \hat{D}(\xi_{q} + i\xi_{p}) =  \exp[ i (\xi_{p}\hat{q} - \xi_{q} \hat{p} )]$ such that $|\xi_{q}| < \sqrt{\pi}\lambda / 2$ and $|\xi_{p}| < \sqrt{\pi}/(2\lambda)$.    

In the teleportation-based GKP error correction scheme (shown in \cref{fig:GKP error correction} (b)), both ancilla modes are initialized in the GKP qunaught state
\begin{align}
    |\emptyset_{\lambda}\rangle &\propto \sum_{n\in\mathbb{Z}} | \hat{q} = n\sqrt{2\pi}\lambda \rangle \propto \sum_{n\in\mathbb{Z}} | \hat{p} = n\sqrt{2\pi} / \lambda \rangle .  \label{eq:GKP qunaught state}
\end{align}
Ref.\ \cite{Walshe2020_continuous_variable} called $|\emptyset_{\lambda}\rangle$ a ``qunaught'' state because it is the unique state of the rectangular-lattice GKP code that encodes only one logical state, hence carrying no quantum information (note the additional factor of $\sqrt{2}$ in the lattice spacing in \cref{eq:GKP qunaught state} compared to the logical states of a GKP qubit in \cref{eq:GKP computational basis states,eq:GKP complementary basis states}). The GKP qunaught state has also been called the grid state \cite{Duivenvoorden2017_single_mode} or the canonical GKP state \cite{Noh2020_encoding,Hannggli2021_oscillator_to_oscillator} in different contexts. 

Define the beam-splitter unitary operator $\hat{B}_{j\rightarrow k}(\theta)$ between the modes $j$ and $k$ as 
\begin{align}
    \hat{B}_{j\rightarrow k}(\theta) &\equiv \exp[ -i\theta(\hat{q}_{j}\hat{p}_{k} - \hat{p}_{j}\hat{q}_{k}) ] \nonumber\\
    &= \exp[ \theta( \hat{a}_{j}^{\dagger} \hat{a}_{k} - \hat{a}_{j}\hat{a}_{k}^{\dagger} ) ] . 
\end{align}
In the Heisenberg picture, the quadrature operators are transformed via the beam-splitter unitary $\hat{B}_{j\rightarrow k}(\theta)$ as follows: 
\begin{align}
    \begin{bmatrix}
    \hat{q}_{j} \\
    \hat{q}_{k} \\
    \hat{p}_{j} \\
    \hat{p}_{k} 
    \end{bmatrix} \rightarrow \begin{bmatrix}
    \cos\theta & -\sin\theta & 0 & 0\\
    \sin\theta & \cos\theta & 0 & 0\\
    0 & 0 & \cos\theta & -\sin\theta\\
    0 & 0 & \sin\theta & \cos\theta
    \end{bmatrix} \begin{bmatrix}
    \hat{q}_{j} \\
    \hat{q}_{k} \\
    \hat{p}_{j} \\
    \hat{p}_{k} 
    \end{bmatrix} \label{eq:quadrature transformation balanced beam-splitter interaction}
\end{align}
Thus, we have a balanced beam-splitter interaction at $\theta = \pi/4$. Applying the balanced beam-splitter interaction $\hat{B}_{2 \rightarrow 3}(\pi/4)$ between the two GKP qunaught states $|\emptyset_{\lambda}\rangle_{2}$ and $|\emptyset_{\lambda}\rangle_{3}$, we get an encoded GKP-Bell state, i.e., 
\begin{align}
    &\hat{B}_{2 \rightarrow 3}\Big{(}\frac{\pi}{4}\Big{)}|\emptyset_{\lambda}\rangle_{2}|\emptyset_{\lambda}\rangle_{3} 
    \nonumber\\
    &\propto |\Phi^{+}_{\lambda}\rangle_{2,3} \equiv \frac{1}{\sqrt{2}} ( |0_{\lambda}\rangle_{2}|0_{\lambda}\rangle_{3} + |1_{\lambda}\rangle_{2}|1_{\lambda}\rangle_{3} ) . 
\end{align}
Once the GKP-Bell state is prepared, we apply the beam-splitter interaction between the modes $1$ and $2$ which is then followed by homodyne measurements of the quadrature oepratores $\hat{q}_{1}$, $\hat{p}_{2}$. Such a measurement is the continuous-variable analog of the Bell measurement. Thus, the circuit in \cref{fig:GKP error correction} (b) implements quantum teleportation from the mode $1$ to mode $3$. This is why it is called a teleportation-based GKP error correction scheme. As a result, conditioned on the measurement outcomes $q_{m}$ and $p_{m}$, the output state is given by (see \cref{appendix:GKP error correction} for the derivation) 
\begin{align}
    |\psi(q_{m},p_{m})\rangle_{\mathrm{Teleport}} &\propto \hat{\Pi}_{\lambda}\hat{D}(-\sqrt{2}(q_{m}+ip_{m}))|\psi\rangle .  \label{eq:ideal output state of the GKP error correction teleportation}
\end{align}

In contrast to the Steane error correction case, the output state of the teleportation-based GKP error correction always is in the GKP code space, not a displaced one, regardless of the measurement outcomes (assuming noiseless ancilla GKP states). However, just like in the case of the discrete-variable qubit teleporation, even if the input state $|\psi\rangle$ is an ideal GKP encoded state, the teleported state may differ from the input state by a logical Pauli operator depending on the measurement outcome $q_{m}$ and $p_{m}$. More specifically, if the input state is an ideal GKP state, $\sqrt{2}q_{m}$ and $\sqrt{2}p_{m}$ can only be an integer multiple of $\sqrt{\pi}\lambda$ and $\sqrt{\pi}/\lambda$, respectively, i.e., $\sqrt{2}q_{m} = n_{q}\sqrt{\pi}\lambda$ and $\sqrt{2}p_{m} = n_{p}\sqrt{\pi}/\lambda$. Then, the teleported state may differ from the input state by a logical Pauli operator $\hat{Z}_{\lambda}^{n_{p}}\hat{X}_{\lambda}^{n_{q}}$. Such an extra Pauli operator may be physically corrected by applying a suitable correction shift to the teleported state or be kept track of in software via a Pauli frame \cite{Knill2005_quantum,DiVincenzo2007_effective,Terhal2015_quantum,Chamberland2018_fault_tolerant} (until we have to apply a non-Clifford gate).   

If the input state $|\psi\rangle$ is not an ideal GKP state and is instead shifted from the GKP code space, the measurement outcomes multiplied by $\sqrt{2}$ (i.e., $\sqrt{2}q_{m}$ and $\sqrt{2}p_{m}$) will no longer be an integer multiple of $\sqrt{\pi}\lambda$ and $\sqrt{\pi}/\lambda$. In this case, we find the closest integers to $\sqrt{2}q_{m} / (\sqrt{\pi}\lambda)$ and $\sqrt{2}p_{m} / (\sqrt{\pi}/\lambda)$ to determine the extra Pauli operator. That is, we set $n_{q}$ and $n_{p}$ as 
\begin{align}
    n_{q} &= \Big{\lfloor} \frac{\sqrt{2}q_{m}}{\sqrt{\pi}\lambda} + \frac{1}{2} \Big{\rfloor} , \quad n_{p} = \Big{\lfloor} \frac{\sqrt{2}p_{m}}{\sqrt{\pi}/\lambda} + \frac{1}{2} \Big{\rfloor},  
\end{align}
and either physically apply a Pauli correction $\hat{Z}_{\lambda}^{n_{p}}\hat{X}_{\lambda}^{n_{q}}$ through a correction shift or keep track of it via a Pauli frame. Similarly as in the case of the Steane-type GKP error correction, by searching for the closest integer, we are employing a maximum likelihood decoding strategy under the same assumption that smaller shifts are more likely than larger shifts (which, again, needs to be revisited in the case of the two-qubit gates between two GKP qubits). Through this decoding strategy, we can correct any small shifts $\hat{D}(\xi_{q} + i\xi_{p}) = \exp[ i (\xi_{p}\hat{q} - \xi_{q} \hat{p} )]$ such that $|\xi_{q}| < \sqrt{\pi}\lambda / 2$ and $|\xi_{p}| < \sqrt{\pi}/(2\lambda)$, similarly as in the case of the Steane-type GKP error correction.     

\subsubsection{Noisy ancilla GKP states}
\label{subsubsection:Noisy ancilla GKP states}

\begin{figure*}
    \centering
    \includegraphics[width=0.8\textwidth]{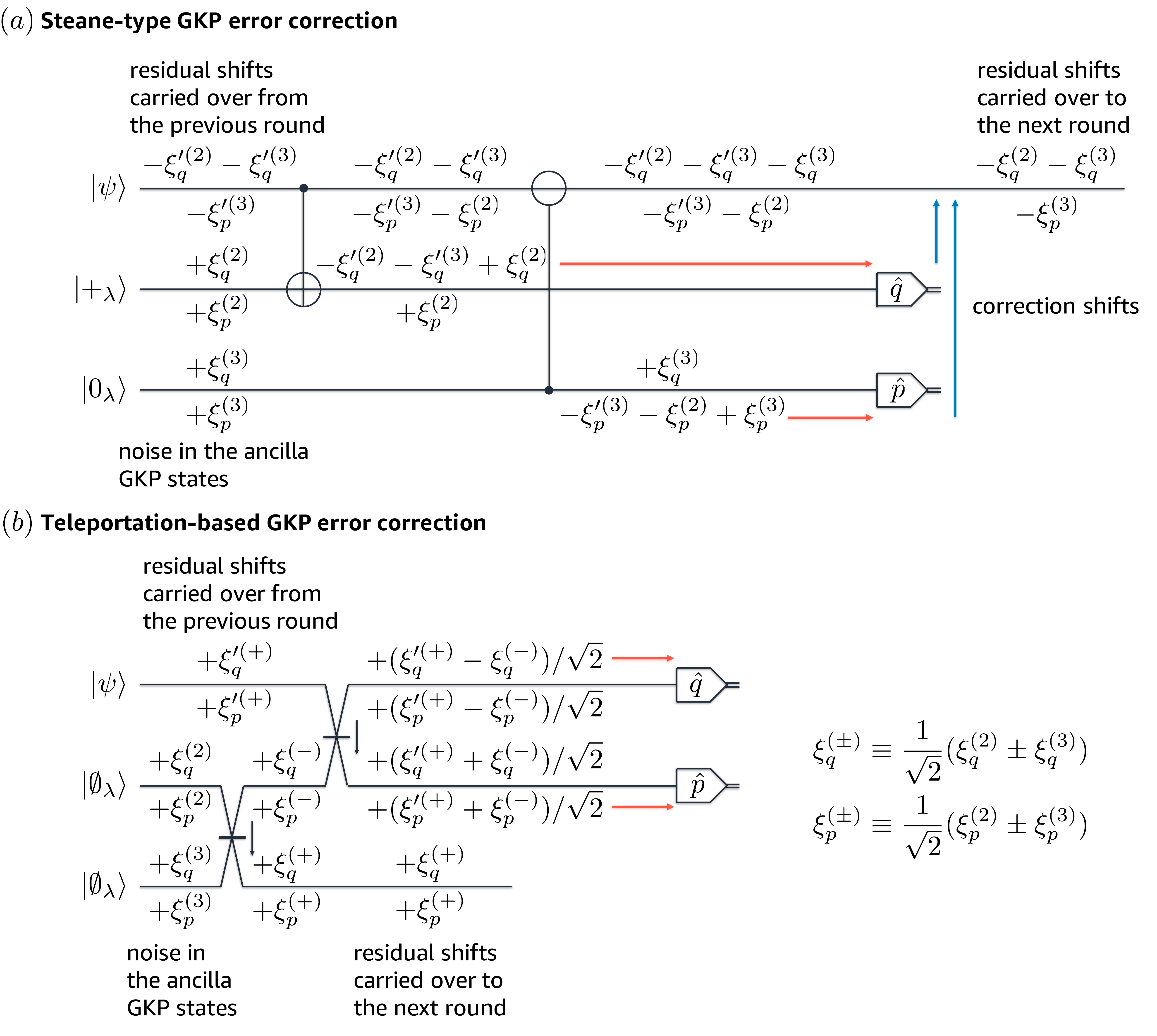}
    \caption{Propagation of the shift errors (due to noisy ancilla GKP states) in (a) the Steane-type GKP error correction and (b) the teleportation-based GKP error correction scheme. We assume that each ancilla GKP state is corrupted by a random shift error $\hat{D}(\xi_{q} + i\xi_{p}) = \exp[ i(\xi_{p}\hat{q} -\xi_{q} \hat{p} ) ]$ where $\xi_{q}$ and $\xi_{p}$ are drawn from an independent and identically distributed Gaussian distribution with zero mean and variance $\sigma_{\mathrm{gkp}}^{2}$, i.e., $\xi_{q},\xi_{p} \sim_{\mathrm{iid}} \mathcal{N}(0, \sigma_{\mathrm{gkp}}^{2})$. In both schemes, $(\xi_{q}^{(2)}, \xi_{p}^{(2)})$ and $(\xi_{q}^{(3)}, \xi_{p}^{(3)})$ are the shifts due to the noisy ancilla GKP states in the second and the third modes, respectively. Also, $\xi'^{(2)}_{q},\xi'^{(2)}_{p},\xi'^{(2)}_{p},\xi'^{(3)}_{p}$ denote the ancilla GKP shifts in the previous round of the GKP error correction, which are partially carried over to the current round. In the Steane-type method, the position and the momentum homodyne measurements are corrupted by net shift $-\xi'^{(2)}_{q} - \xi'^{(3)}_{q} + \xi_{q}^{(2)}$ and $-\xi'^{(3)}_{p} - \xi^{(2)}_{p} + \xi_{p}^{(3)}$, respectively, whose variances are given by $3\sigma_{\mathrm{gkp}}^{2}$. In the teleportation-based method, the homodyne measurements are corrupted by net shifts $(\xi'^{(+)}_{q} - \xi^{(-)}_{q})/\sqrt{2}$ and $(\xi'^{(+)}_{p} + \xi^{(-)}_{p})/\sqrt{2}$. Note that the actual induced shifts on the GKP qubit are effectively given by $\xi'^{(+)}_{q} - \xi^{(-)}_{q}$ and $\xi'^{(+)}_{p} + \xi^{(-)}_{p}$ due to the extra $\sqrt{2}$ factor in \cref{eq:ideal output state of the GKP error correction teleportation}. Thus, the variances of the induced shifts are given by $2\sigma_{\mathrm{gkp}}^{2}$, which is smaller than $3\sigma_{\mathrm{gkp}}^{2}$ in the case of the Steane-type protocol.       }
    \label{fig:GKP error correction noisy ancilla}
\end{figure*}

We now discuss the adverse effects of the noise in ancilla GKP states. Since realistic GKP states are only finitely squeezed, they introduce shift errors when they are injected to a GKP error correction circuit. More specifically, a finitely-squeezed GKP state can be understood as a state resulting from applying a coherent Gaussian random shift error (with a noise variance $\sigma_{\mathrm{gkp}}^{2}$ in both the position and the momentum quadratures) to an ideal GKP state. For the purpose of efficient classical simulation of the shift errors, such coherent shift errors are often approximated as incoherent shift errors (i.e., twirling approximation; see, e.g., Appendix A of Ref.\ \cite{Noh2020_fault_tolerant}). 

Here, we also make the twirling approximation and assume that the shifts due to finite squeezing are incoherent. 
More specifically, we assume that each finitely-squeezed GKP state introduces a random shift error $\hat{D}(\xi_{q} + i\xi_{p}) = \exp[ i( \xi_{p}\hat{q} - \xi_{q}\hat{p} )]$ where $\xi_{q}$ and $\xi_{p}$ are randomly drawn from an independent and identically distributed Gaussian distribution $\mathcal{N}( 0, \sigma_{\mathrm{gkp}}^{2})$ with zero mean and noise variance $\sigma_{\mathrm{gkp}}^{2}$. That is, in both GKP error correction schemes, we assume that the ancilla modes (supporting a GKP state) are corrupted by shift errors $\xi_{q}^{(2)},\xi_{p}^{(2)},\xi_{q}^{(3)},\xi_{p}^{(3)} \sim_{\mathrm{iid}}\mathcal{N}(0,\sigma_{\mathrm{gkp}}^{2})$ (see \cref{fig:GKP error correction noisy ancilla}). Following the same convention in the literature, we define the GKP squeezing $\sigma_{\mathrm{gkp}}^{(\mathrm{dB})}$ (in the unit of decibel) as 
\begin{align}
    \sigma_{\mathrm{gkp}}^{(\mathrm{dB})} = 10 \log_{10}\Big{(} \frac{ 1/2 }{ \sigma_{\mathrm{gkp}}^{2} } \Big{)}, \label{eq:GKP squeezing}
\end{align}
where $1/2$ is the variance of the quadrature noise of a vacuum state $|\hat{n} =0\rangle$. 

In the case of the Steane-type GKP error correction (shown in \cref{fig:GKP error correction noisy ancilla} (a)), the data mode inherits residual shifts $-\xi'^{(2)}_{q} - \xi'^{(3)}_{q}$ and $-\xi'^{(3)}_{p}$ from the previous round of the GKP error correction, which will become clear shortly. The position shift $-\xi'^{(2)}_{q} - \xi'^{(3)}_{q}$ is then tranferred to the second ancilla mode via the SUM gate and is added to the ancilla position shift $\xi_{q}^{(2)}$. Thus, the net position shift in the second mode is given by $-\xi'^{(2)}_{q} - \xi'^{(3)}_{q} + \xi_{q}^{(2)}$ and the measurement outcome of the second mode is given by $q_{m} = -\xi'^{(2)}_{q} - \xi'^{(3)}_{q} + \xi_{q}^{(2)}$ modulo $\sqrt{\pi}\lambda$. In the meantime, an extra position shift $-\xi_{q}^{(3)}$ is added to the data mode due to transfer of the position shift in the third mode via the inverse-SUM gate. As a result, after applying a correction shift for the position quadrature $-R_{\sqrt{\pi}\lambda}(q_{m})$, the data mode has a residual position shift (modulo $\sqrt{\pi}\lambda$)
\begin{align}
    &-\xi'^{(2)}_{q} - \xi'^{(3)}_{q} - \xi_{q}^{(3)} -  ( -\xi'^{(2)}_{q} - \xi'^{(3)}_{q} + \xi_{q}^{(2)} ) 
    \nonumber\\
    &= -\xi_{q}^{(2)}- \xi_{q}^{(3)} , 
\end{align}
which is carried over to the next round of the GKP error correction (hence the initial shift $-\xi'^{(2)}_{q} - \xi'^{(3)}_{q}$ in the current round). 

In the momentum quadrature, the initial shift $-\xi'^{(3)}_{p}$ is added by an extra shift error $-\xi_{p}^{(2)}$ transferred from the second mode via the SUM gate. Then, this shift is tranferred to the third mode and is added to the momentum shift $\xi_{p}^{(3)}$ in the third mode. Hence, the third mode is corrupted by a net momentum shift $-\xi'^{(3)}_{p} - \xi_{p}^{(2)} + \xi_{p}^{(3)}$ and the measurement outcome is given by $p_{m} = -\xi'^{(3)}_{p} - \xi_{p}^{(2)} + \xi_{p}^{(3)}$ modulo $\sqrt{\pi}/\lambda$. After applying a correction momentum shift $-R_{\sqrt{\pi}/\lambda}(p_{m})$ to the date mode, the data mode has a residual momentum shift (modulo $\sqrt{\pi}/\lambda$)
\begin{align}
    &-\xi'^{(3)}_{p} - \xi_{p}^{(2)} - ( -\xi'^{(3)}_{p} - \xi_{p}^{(2)} + \xi_{p}^{(3)} )
    \nonumber\\
    &= -\xi_{p}^{(3)} , 
\end{align}
which is carried over to the next round (hence $-\xi'^{(3)}_{p}$ initially in the current round). 

Note that both the net position shift in the second mode $-\xi'^{(2)}_{q} - \xi'^{(3)}_{q} + \xi_{q}^{(2)}$ and the net momentum shift in the third mode $ -\xi'^{(3)}_{p} - \xi_{p}^{(2)} + \xi_{p}^{(3)}$ have a noise variance $3\sigma_{\mathrm{gkp}}^{2}$. If these net shifts are smaller than the correctable shift $\sqrt{\pi}\lambda/2$ and $\sqrt{\pi}/(2\lambda)$, the correction shifts do not cause any logical Pauli errors. However, if they are not, $R_{\sqrt{\pi}\lambda}(q_{m})$ and $R_{\sqrt{\pi}/\lambda}(p_{m})$ are given by  
\begin{align}
    R_{\sqrt{\pi}\lambda}(q_{m}) &= -\xi'^{(2)}_{q} - \xi'^{(3)}_{q} + \xi_{q}^{(2)} + n_{q}\sqrt{\pi}\lambda, 
    \nonumber\\
    R_{\sqrt{\pi}/\lambda}(p_{m}) &= -\xi'^{(3)}_{p} - \xi_{p}^{(2)} + \xi_{p}^{(3)} + n_{p}\sqrt{\pi} / \lambda , 
\end{align}
with some $(n_{q},n_{p})\neq (0,0)$. Thus, the data GKP qubit is corrupted by a logical Pauli error $\hat{X}_{\lambda}^{n_{q}}\hat{Z}_{\lambda}^{n_{p}}$. The probability of having such a Pauli error is closely related to the weight of the tail of the Gaussian distribution $\mathcal{N}(0, 3\sigma_{\mathrm{gkp}}^{2})$. Note that the factor of $3$ in the noise variance is also directly related to the fact that the Glancy-Knill threshold is given by $\sqrt{\pi}/6 = (\sqrt{\pi}/2)/3$ for the square-lattice GKP code \cite{Glancy2006_error}. In what follows, we show that the relevant noise variance is given by $2\sigma_{\mathrm{gkp}}^{2}$ (instead of $3\sigma_{\mathrm{gkp}}^{2}$) in the case of the teleportation-based GKP error correction, hence it performs better than the Steane-type method under the same GKP squeezing $\sigma_{\mathrm{gkp}}^{(\mathrm{dB})}$.  

In the case of the teleportation-based GKP error correction (shown in \cref{fig:GKP error correction noisy ancilla} (b)), the ancilla shift errors are mixed via the balanced beam-splitter interaction. Thus, the third mode (where the state is teleported to) has residual shift errors 
\begin{align}
    \xi^{(+)}_{q} &= (\xi^{(2)}_{q} + \xi^{(3)}_{q})/\sqrt{2}, 
    \nonumber\\
    \xi^{(+)}_{p} &= (\xi^{(2)}_{p} + \xi^{(3)}_{p})/\sqrt{2}
\end{align} 
which is then carried over to the next round of the GKP error correction. This also means that, in the current round, the data mode inherits residual shifts $\xi'^{(+)}_{q}$ and $\xi'^{(+)}_{p}$ from the previous round ($\xi'^{(+)}_{q}$ and $\xi'^{(+)}_{p}$ are defined similarly as $\xi^{(+)}_{q}$ and $\xi^{(+)}_{p}$). After the balanced beam-splitter interaction between the two ancilla modes, the second mode has shift errors
\begin{align}
    \xi^{(-)}_{q} &= (\xi^{(2)}_{q} - \xi^{(3)}_{q})/\sqrt{2} , 
    \nonumber\\
    \xi^{(-)}_{p} &= (\xi^{(2)}_{p} - \xi^{(3)}_{p})/\sqrt{2} . 
\end{align}
Note that since the quadrature transformation matrix of a beam-splitter interaction is an orthogonal matrix (see \cref{eq:quadrature transformation balanced beam-splitter interaction}), the four shifts $\xi^{(+)}_{q},\xi^{(+)}_{p},\xi^{(-)}_{q},\xi^{(-)}_{p}$ are mutually independent just like the untranformed ancilla shifts $\xi^{(2)}_{q},\xi^{(2)}_{p},\xi^{(3)}_{q},\xi^{(3)}_{p}$. The shifts in the second mode are then mixed with the shifts in the data mode and result in a net shift error $(\xi'^{(+)}_{q} - \xi^{(-)}_{q} ) /\sqrt{2}$ in the position quadrature and $(\xi'^{(+)}_{p} + \xi^{(-)}_{p} ) /\sqrt{2}$ in the momentum quadrature. Thus, the measurement outcomes are given by 
\begin{align}
    q_{m} &= (\xi'^{(+)}_{q} - \xi^{(-)}_{q} ) /\sqrt{2}\textrm{ modulo } \sqrt{\pi}\lambda/\sqrt{2} ,
    \nonumber\\
    p_{m} &= (\xi'^{(+)}_{p} + \xi^{(-)}_{p} ) /\sqrt{2} \textrm{ modulo } \sqrt{\pi}/(\sqrt{2}\lambda) . 
\end{align}
As can be seen from \cref{eq:ideal output state of the GKP error correction teleportation}, the actual shifts on the data GKP qubit are given by $\sqrt{2}q_{m}$ and $\sqrt{2}p_{m}$ with an extra $\sqrt{2}$ factor. Thus, the relevant shifts (to be compared with the lattice spacings $\sqrt{\pi}\lambda$ and $\sqrt{\pi}/\lambda$) are $\xi'^{(+)}_{q} - \xi^{(-)}_{q} $ and $\xi'^{(+)}_{p} + \xi^{(-)}_{p}$, whose noise variances are given by $2\sigma_{\mathrm{gkp}}^{2}$. Since the relevant noise variance is smaller in the teleportation-based GKP error correction scheme by a factor of $3/2$, logical Pauli errors are less likely to occur in the teleportation-based method than in the Steane-type method, given the same GKP squeezing $\sigma_{\mathrm{gkp}}^{(\mathrm{dB})}$.

Note that since the weight of the Gaussian tail decreases exponentially in the inverse noise variance, the constant factor improvement in the noise variance can bring about a significant decrease in the logical Pauli error probabilities. Apart from the absence of the online squeezing operations and the enhanced resilience against the ancilla GKP noise, we also remark that the teleportation-based scheme is more advantageous for keeping the energy of the encoded state bounded than the Steane-type method. This is thanks to the symmetry between the position and the momentum quadratures in the teleportation-based scheme (which is related to the absence of online squeezing operations). Such a symmetry prevents the Gaussian envelope of the finitely-squeezed GKP states from being distorted during the GKP error correction, and hence always minimizes the energy of the encoded GKP states (see \cref{subappendix:Finitely squeezed ancilla GKP states} for more details).             

\subsection{Maximum likelihood decoding for the CNOT and CZ gates between two GKP qubits}
\label{subsection:Maximum likelihood decoding for GKP}

\begin{figure*}
    \centering
    \includegraphics[width=0.9\textwidth]{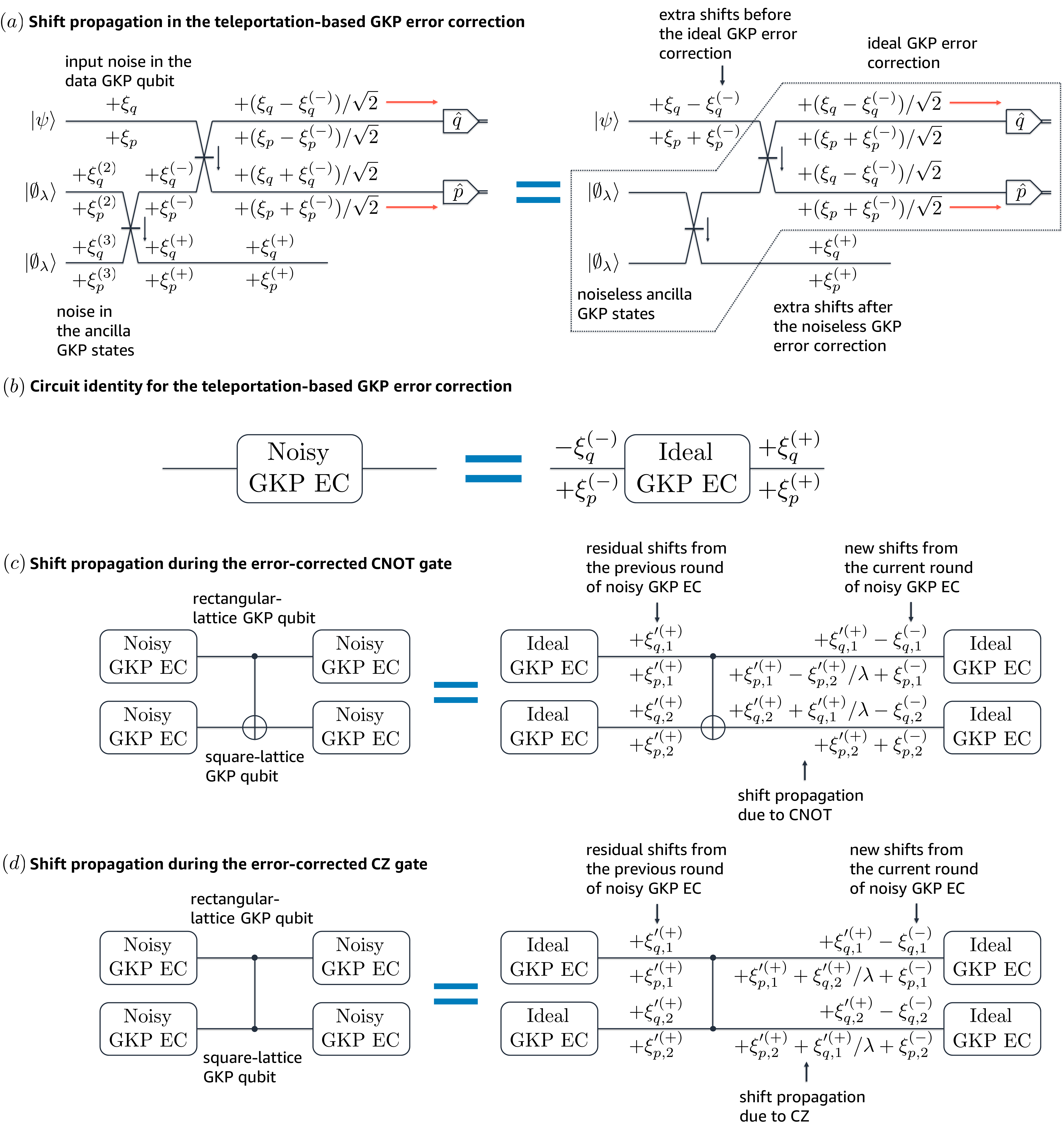}
    \caption{(a) Propagation of the shift errors (due to noisy ancilla GKP states) in the teleportation-based GKP error correction scheme. As shown in the right hand side of (a), one can alternatively assume that the ancilla GKP states are noiseless but instead there are extra shift errors $(-\xi^{(-)}_{q} , \xi^{(-)}_{p} )$ and $(\xi^{(+)}_{q} , \xi^{(+)}_{p} )$ before and after the ideal GKP error correction, respectively. Hence, as summarized in (b), a noisy teleportation-based GKP error correction understood as an ideal GKP error correction preceded and followed by extra random shift errors $(-\xi^{(-)}_{q} , \xi^{(-)}_{p} )$ and $(\xi^{(+)}_{q} , \xi^{(+)}_{p} )$ (see also Eq.\ (57) in Ref.\ \cite{Walshe2020_continuous_variable}). Here, $\xi^{(-)}_{q} , \xi^{(-)}_{p} , \xi^{(+)}_{q} , \xi^{(+)}_{p} \sim_{\mathrm{iid}} \mathcal{N}( 0, \sigma_{\mathrm{gkp}}^{2}) $. Shift propagation in the error-corrected (c) CNOT and (d) CZ gates between two GKP qubits, where the first qubit is a rectangular-lattice GKP qubit with $\lambda \ge 1$ and the second qubit is the square-lattice GKP qubit. In both cases, the two GKP qubits inherit shift errors $\xi'^{(+)}_{q,1} , \xi'^{(+)}_{p,1} , \xi'^{(+)}_{q,2} , \xi'^{(+)}_{p,2}$ from the previous round of noisy GKP error correction. These shift errors are propagated through the CNOT and the CZ gates. Lastly, extra shift errors $-\xi^{(-)}_{q,1} , \xi^{(-)}_{p,1} , -\xi^{(-)}_{q,2} , \xi^{(-)}_{p,2}$ are added due to the noisy GKP error correction in the current round. As a result, the net accumulated shift errors have noise variances $2\sigma_{\mathrm{gkp}}^{2}$ and $(2 + 1/\lambda^{2})\sigma_{\mathrm{gkp}}^{2}$.         }
    \label{fig:GKP error corrected two qubit gates}
\end{figure*}

Here we consider error-corrected logical two-qubit gates between GKP qubits. That is, we apply GKP error correction right after each logical two-qubit gate. Thus, when analyzing the performance of an error-corrected gate, we assume that each gate is preceded by a GKP error correction from the previous time step and is followed by another GKP error correction which corrects for the shifts accumulated by the end of the gate. In particular, we propose a maximum likelihood decoding for the logical CNOT and CZ gate for the GKP qubits and show that it significantly outperforms the simple closest-integer decoding scheme used, e.g., in Ref.\ \cite{Larsen2021_fault_tolerant}.    

In the rest of the paper (except for the appendices), we exclusively consider the teleportation-based GKP error correction scheme due to its superior performance over the Steane-based method under the same GKP squeezing. More specifically, we will assume that each GKP error correction is noisy because the ancilla GKP states are only finitely squeezed with a noise variance $\sigma_{\mathrm{gkp}}^{2}$. Note that in practice, there may be extra noise sources other than the finite GKP squeezing, with photon loss being the most notable example. However, we neglect such extra noise sources in part for simplicity but also because these other noise sources are not appreciable compared to the noise due to the finite GKP squeezing in realistic scenarios. 

For instance, the highest GKP squeezing $\sigma_{\mathrm{gkp}}^{(\mathrm{dB})}$ that has been experimentally achieved so far is slightly below $10\mathrm{dB}$. Converting this to the noise variance $\sigma_{\mathrm{gkp}}^{2}$, we find 
\begin{align}
    \sigma_{\mathrm{gkp}}^{2} = \frac{1}{2}10^{- \sigma_{\mathrm{gkp}}^{(\mathrm{dB})} / 10} = 0.05
\end{align}
at $\sigma_{\mathrm{gkp}}^{(\mathrm{dB})} =  10 \mathrm{dB}$. As explained in the previous section, this noise variance is doubled because there is one carried over from the previous round of the GKP error correction and another one that is added from the current round. Thus, the relevant noise variance due to the finite GKP squeezing is given by $2\sigma_{\mathrm{gkp}}^{2} = 0.1$ at $\sigma_{\mathrm{gkp}}^{(\mathrm{dB})} =  10 \mathrm{dB}$. On the other hand, as shown in Ref.\ \cite{Noh2020_fault_tolerant}, the additional noise variance due to the photon loss is given by $\kappa t_{G}$ (assuming amplification with the same rate), where $\kappa$ is the photon loss rate and $t_{G}$ is the gate time. Thus, if the gate time $t_{G}$ is $100$ times shorter than the single photon relaxation time $1/\kappa$, the extra noise variance due to the photon loss is only about $\kappa t_{G} = 0.01$. Note that even at $\sigma_{\mathrm{gkp}}^{(\mathrm{dB})} =  13 \mathrm{dB}$, we have $2\sigma_{\mathrm{gkp}}^{2} \simeq 0.05$ and thus the noise variance due to finite GKP squeezing dominates. Hence, as a zeroth order approximation, we focus on the case where the finite squeezing of the ancilla GKP states is the only noise source. However, the maximum likelihood decoding method we present here applies to more general cases, e.g., with extra noise due to photon losses.     

Before analyzing the error-corrected logical CNOT and CZ gate for the GKP qubits, we remark that a noisy (teleportation-based) GKP error correction with finitely-squeezed ancilla GKP states can be understood as an ideal GKP error correction preceded and followeded by extra shift errors $(-\xi^{(-)}_{q} , \xi^{(-)}_{p} )$ and $(\xi^{(+)}_{q} , \xi^{(+)}_{p} )$, respectively, where $\xi^{(-)}_{q} , \xi^{(-)}_{p} , \xi^{(+)}_{q} , \xi^{(+)}_{p}  \sim_{\mathrm{iid}}\mathcal{N}( 0, \sigma_{\mathrm{gkp}}^{2} ) $ (see \cref{fig:GKP error corrected two qubit gates} (a) and (b)). This circuit identity will be used throughout the analysis of the error-corrected two qubit gates between GKP qubits.

\subsubsection{Error-corrected CNOT gate}
\label{subsubsection:Error-corrected CNOT gate}

Recall that the logical CNOT gate between two GKP qubits can be realized by a $qp$ coupling (see \cref{eq:GKP logical CNOT}). Here, we assume that the first qubit is a rectangular-lattice GKP qubit with $\lambda \ge 1$ and the second qubit is the suquare-lattice GKP qubit. As shown in \cref{fig:GKP error corrected two qubit gates} (c), before the CNOT gate is applied, the GKP qubits inherit shift errors $( \xi'^{(+)}_{q,1} , \xi'^{(+)}_{p,1} )$ and $( \xi'^{(+)}_{q,2} , \xi'^{(+)}_{p,2} )$ from the previous round of the noisy GKP error correction. These shifts are propagated through the CNOT gate as $( \xi'^{(+)}_{q,1} , \xi'^{(+)}_{p,1} - \xi'^{(+)}_{p,2}/\lambda )$ and $( \xi'^{(+)}_{q,2} +  \xi'^{(+)}_{q,1} /\lambda , \xi'^{(+)}_{p,2} )$. Then, due to the additional noise from the GKP error correction after the CNOT gate, extra shift errors $( -\xi^{(-)}_{q,1} , \xi^{(-)}_{p,1} )$ and $( -\xi^{(-)}_{q,2} , \xi^{(-)}_{p,2} )$ are added to the quadratures. Consequently, the two GKP qubits have net shifts 
\begin{align}
    \xi_{q}^{(1)} &= \xi'^{(+)}_{q,1} -\xi^{(-)}_{q,1}, 
    \nonumber\\
    \xi_{p}^{(1)} &= \xi'^{(+)}_{p,1} - \frac{1}{\lambda}\xi'^{(+)}_{p,2} + \xi^{(-)}_{p,1}, 
    \nonumber\\
    \xi_{q}^{(2)} &= \xi'^{(+)}_{q,2} + \frac{1}{\lambda} \xi'^{(+)}_{q,1} -\xi^{(-)}_{q,2}, 
    \nonumber\\
    \xi_{p}^{(2)} &= \xi'^{(+)}_{p,2} + \xi^{(-)}_{p,2}. \label{eq:CNOT shift propagation qq main text}
\end{align}
Since all eight underlying shifts $\xi'^{(+)}_{q,1} , \cdots, \xi'^{(+)}_{p,2}$ and $\xi^{(-)}_{q,1} , \cdots, \xi^{(-)}_{p,2}$ are randomly drawn from an independent and identically distributed Gaussian distribution $\mathcal{N}( 0, \sigma_{\mathrm{gkp}}^{2})$, we have noise variances $\mathrm{Var}(\xi_{q}^{(1)}) = \mathrm{Var}(\xi_{p}^{(2)}) = 2\sigma_{\mathrm{gkp}}^{2}$ and $\mathrm{Var}(\xi_{q}^{(2)}) = \mathrm{Var}(\xi_{p}^{(1)}) = (2 + 1/\lambda^{2})\sigma_{\mathrm{gkp}}^{2}$. That is, the position quadrature of the target mode and the momentum quadrature of the control mode have higher noise variances than the other quadratures due to the noise propagation. A motivation for using a rectangular-lattice GKP qubit with $\lambda \ge 1$ in the first qubit is to mitigate the enhanced noise variance via the $1/\lambda$ suppression. However, since the momentum spacing of the first qubit is decreased by a factor of $\lambda$, the $Z$ error rate on the first qubit is enhanced as a result.

\begin{figure*}
    \centering
    \includegraphics[width=0.95\textwidth]{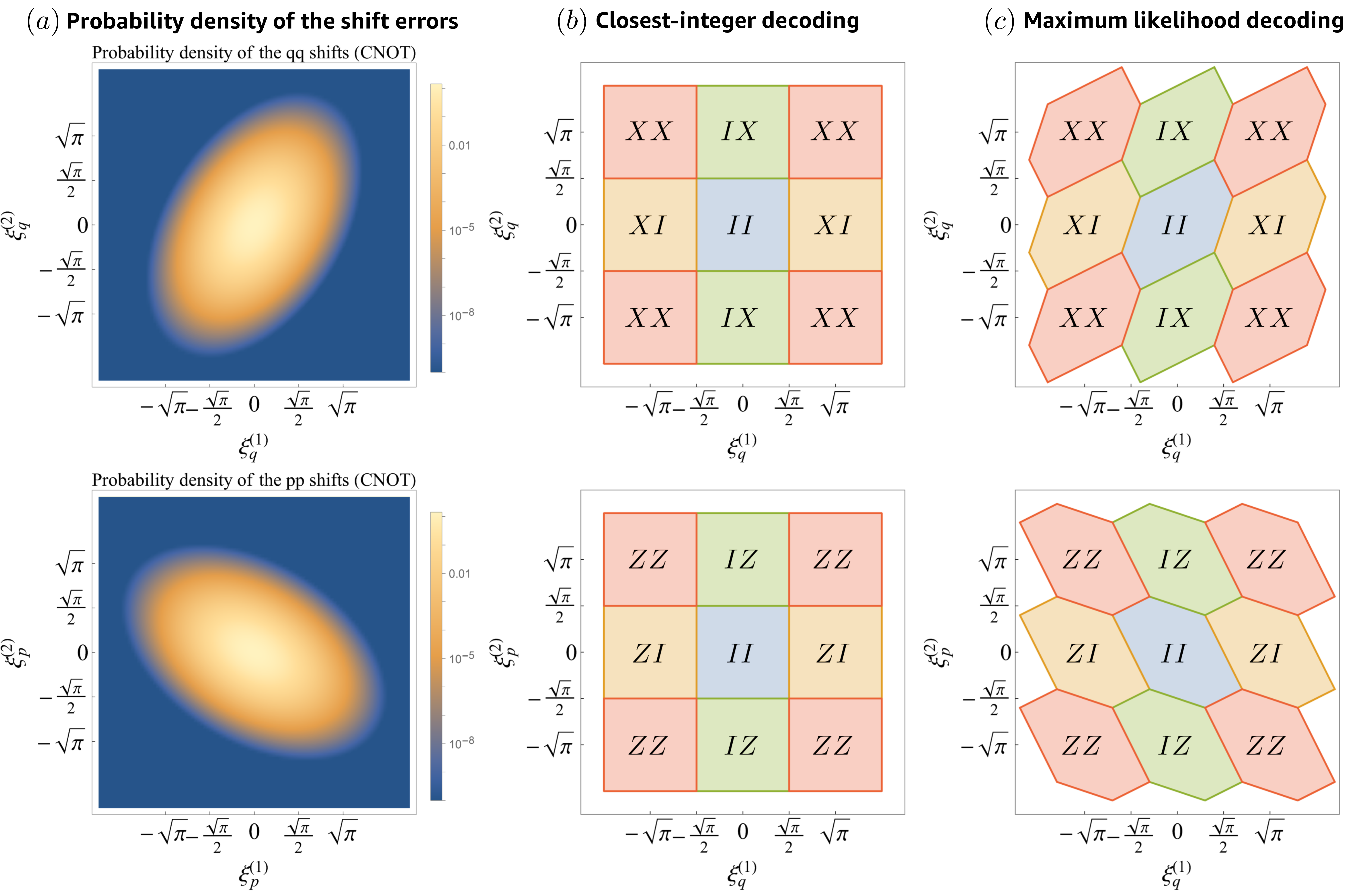}
    \caption{(a) Probability density functions of the position shifts (top panel) and momentum shifts (bottom panel) in the error-corrected CNOT gate between two square-lattice GKP qubits (i.e., $\lambda = 1$) of GKP squeezing $10$dB. The first and second qubits are control and target qubits, respectively. The decision boundaries of the closest-integer decoding and maximum likelihood decoding are shown in (b) and (c). In the top (bottom) panel, the blue, orange, green, and red regions correspond to shifts that result in logical $II$, $XI$, $IX$, $XX$ ($II$, $ZI$, $IZ$, $ZZ$) errors, respectively. Note that the decision boundaries in the maximum likelihood decoding method are better aligned with the corresponding probability density functions in (a) than the ones in the simple closest-integer decoding scheme. }
    \label{fig:GKP CNOT hexagon decoding}
\end{figure*}

Given these shift errors, the measurement outcomes in the GKP error correction (at the end of the CNOT gate) are given by 
\begin{align}
    \sqrt{2}q_{m}^{(1)} &= \xi_{q}^{(1)} + n_{q}^{(1)} \sqrt{\pi}\lambda, 
    \nonumber\\
    \sqrt{2}p_{m}^{(1)} &= \xi_{p}^{(1)} + n_{p}^{(1)} \sqrt{\pi} / \lambda, 
    \nonumber\\
    \sqrt{2}q_{m}^{(2)} &= \xi_{q}^{(2)} + n_{q}^{(2)} \sqrt{\pi}, 
    \nonumber\\
    \sqrt{2}p_{m}^{(2)} &= \xi_{p}^{(2)} + n_{p}^{(2)} \sqrt{\pi} .  
\end{align}
Given these measurement outcomes, the goal of the decoding is to correctly infer the four integers $n_{q}^{(1)} , n_{p}^{(1)} , n_{q}^{(2)} , n_{p}^{(2)}$. A simple and widely used decoding strategy is to find the closest integers to $\sqrt{2}q_{m}^{(1)} / (\sqrt{\pi}\lambda)$, $\sqrt{2}p_{m}^{(1)} / (\sqrt{\pi} / \lambda)$, $\sqrt{2}q_{m}^{(2)} / (\sqrt{\pi})$, $\sqrt{2}p_{m}^{(2)} / (\sqrt{\pi} )$, i.e., 
\begin{align}
    \bar{n}_{q}^{(1)} &= \Big{\lfloor} \frac{\sqrt{2}q_{m}^{(1)}}{\sqrt{\pi}\lambda} + \frac{1}{2} \Big{\rfloor} , \quad \bar{n}_{p}^{(1)} = \Big{\lfloor} \frac{\sqrt{2}p_{m}^{(1)}}{\sqrt{\pi} / \lambda} + \frac{1}{2} \Big{\rfloor} , 
    \nonumber\\
    \bar{n}_{q}^{(2)} &= \Big{\lfloor} \frac{\sqrt{2}q_{m}^{(2)}}{\sqrt{\pi}} + \frac{1}{2} \Big{\rfloor} , \quad \bar{n}_{p}^{(2)} = \Big{\lfloor} \frac{\sqrt{2}p_{m}^{(2)}}{\sqrt{\pi} } + \frac{1}{2} \Big{\rfloor} . 
\end{align}
For instance, Ref.\ \cite{Larsen2021_fault_tolerant} used the closest-integer decoding to correct for the shift errors in several logical two-qubit gates between GKP qubits (which are equivalent to the CNOT gate up to single-qubit Clifford gates). This way, any net shifts such that $|\xi_{q}^{(1)}| < \sqrt{\pi}\lambda /2$, $|\xi_{p}^{(1)}| < \sqrt{\pi} /(2\lambda)$, and $|\xi_{q}^{(2)}|, |\xi_{p}^{(2)}| < \sqrt{\pi}/2$ can be corrected. Note that the rationale behind the closest-integer decoding is that smaller shifts are more likely to happen than larger shifts. However, we will show below that this is not necessarily the case for the logical two-qubit gates.

Here, we propose a different decoding strategy, a maximum likelihood decoding, which outperforms the simple closet-integer decoding. Our maximum likelihood decoding is based on an observation that the net shifts errors are mutually correlated. For instance, the net position shifts $\xi_{q}^{(1)}$ and $\xi_{q}^{(2)}$ are correlated via the propagated shift $\xi'^{(+)}_{q,1}$. More specifically, the position shift vector $\boldsymbol{\xi_{q}} \equiv ( \xi_{q}^{(1)} , \xi_{q}^{(2)} )^{T}$ follows a two-variable Gaussian distribution $\mathcal{N}(0, V_{\boldsymbol{ \xi_{q} }} )$ with zero mean and a covariance matrix 
\begin{align}
    V_{\boldsymbol{ \xi_{q} }} &= \begin{bmatrix}
    \mathrm{Var}( \xi_{q}^{(1)} ) & \mathrm{Cov}( \xi_{q}^{(1)} , \xi_{q}^{(2)} ) \\
    \mathrm{Cov}( \xi_{q}^{(1)} , \xi_{q}^{(2)} ) & \mathrm{Var}( \xi_{q}^{(2)} )
    \end{bmatrix} 
    \nonumber\\
    &= \sigma_{\mathrm{gkp}}^{2} \begin{bmatrix}
    2 & 1/\lambda \\
    1/\lambda& 2+ 1/\lambda^{2}
    \end{bmatrix} . 
\end{align}
Due to the correlation, shifts in a certain direction are more likely to happen than others. In other words, a larger shift in a preferred direction can occur more often than a smaller shift in a less preferred direction (see the top panel of \cref{fig:GKP CNOT hexagon decoding} (a)). Such a possibility is not taken into account in the simple closest-integer decoding.

Note that the probability density function of the position shifts $\xi_{q}^{(1)} ,\xi_{q}^{(2)}$ is given by 
\begin{align}
    &P_{\mathrm{CNOT}}^{[qq]}( \xi_{q}^{(1)} ,\xi_{q}^{(2)} )  = \frac{1}{2\pi \sqrt{ | V_{\boldsymbol{ \xi_{q} }} | } } \exp\Big{[} -\frac{1}{2} \boldsymbol{ \xi_{q} }^{T} V_{\boldsymbol{ \xi_{q} }}^{-1} \boldsymbol{ \xi_{q} } \Big{]} 
    \nonumber\\
    &= \frac{1}{2\pi\sqrt{ (4 + \frac{1}{\lambda^{2}} ) \sigma_{\mathrm{gkp}}^{4} }} 
    \nonumber\\
    & \times \exp\Big{[} -\frac{ (2 + \frac{1}{\lambda^{2}}) (\xi_{q}^{(1)})^{2} + 2(\xi_{q}^{(2)})^{2} -\frac{2}{\lambda}\xi_{q}^{(1)}\xi_{q}^{(2)} }{2( 4 + \frac{1}{\lambda^{2}}  )\sigma_{\mathrm{gkp}}^{2} } \Big{]} , 
\end{align}
where $| V_{\boldsymbol{ \xi_{q} }} |$ is the determinant of $V_{\boldsymbol{ \xi_{q} }}$. Then, the goal of our maximum likelihood decoding is as follows. Given the measurement outcomes $\sqrt{2}q_{m}^{(1)}$ and $\sqrt{2}q_{m}^{(2)}$, we want to find $n_{q}^{(1)}$ and $n_{q}^{(2)}$ such that $P_{\mathrm{CNOT}}^{[qq]}( \xi_{q}^{(1)} ,\xi_{q}^{(2)} )$ is maximized, where $\xi_{q}^{(1)} = \sqrt{2}q_{m}^{(1)} - n_{q}^{(1)}\sqrt{\pi}\lambda$ and $\xi_{q}^{(2)} = \sqrt{2}q_{m}^{(2)} - n_{q}^{(2)}\sqrt{\pi}$. Since $P_{\mathrm{CNOT}}^{[qq]}( \xi_{q}^{(1)} ,\xi_{q}^{(2)} )$ is maximized when $ ( 2 + 1/\lambda^{2} ) (\xi_{q}^{(1)})^{2} + 2(\xi_{q}^{(2)})^{2} -(2/\lambda)\xi_{q}^{(1)}\xi_{q}^{(2)}$ is minimized, this is equivalent to solving the following optimization problem: 
\begin{align}
    &(n^{\star(1)}_{q} , n^{\star(2)}_{q} ) = \argmin_{ n_{1} , n_{2}} \Big{(} (2 + \frac{1}{\lambda^{2}})x_{1}^{2} + 2x_{2}^{2} -\frac{ 2}{ \lambda } x_{1}x_{2}  \Big{)} , 
    \nonumber\\
    &\mathrm{where}\,\,\, x_{1} \equiv  \sqrt{2}q_{m}^{(1)} - n_{1}\sqrt{\pi}\lambda , \,\,\, x_{2} \equiv  \sqrt{2}q_{m}^{(2)} - n_{2}\sqrt{\pi}, \label{eq:CNOT qq optimization}
\end{align}
given $\sqrt{2}q_{m}^{(1)}$ and $\sqrt{2}q_{m}^{(2)}$. Our algorithm for solving this optimization problem is given in \cref{alg:CNOTnq1nq2} in \cref{appendix:Maximum likelihood decoding of the logical two-qubit gates between two GKP qubits}.

The decision boundaries of the simple closest-integer decoding and our maximum likelihood decoding are shown in \cref{fig:GKP CNOT hexagon decoding} (b) and (c), respectively. Note that the decision boundaries in the maximum likelihood method are better aligned with the the corresponding probability density functions than the ones in the closest-integer decoding. Thus, there are shifts that can be correctly countered by the maximum likelihood method but are incorrectly mapped to a different lattice point, hence causing a logical Pauli error, in the closest-integer decoding. 

Similarly as in the case of the position shifts, the momentum shifts $\xi_{p}^{(1)}$ and $\xi_{p}^{(2)}$ are also mutually correlated via the propagated shift $\xi'^{(+)}_{p,2}$. Hence, the momentum shift vector $\boldsymbol{\xi_{p}} \equiv ( \xi_{p}^{(1)} , \xi_{p}^{(2)} )^{T} $ follows a two-variable Gaussian distribution $\mathcal{N}( 0, V_{\boldsymbol{ \xi_{p} }} )$ with zero mean and a covariance matrix 
\begin{align}
    V_{\boldsymbol{ \xi_{p} }} &= \sigma_{\mathrm{gkp}}^{2} \begin{bmatrix}
    2 + 1/\lambda^{2} & -1/\lambda \\
    -1/\lambda & 2
    \end{bmatrix} . 
\end{align}
The probability density function of the momentum shifts $ \xi_{p}^{(1)}$ and $ \xi_{p}^{(2)}$ is thus given by (see the bottom panel of \cref{fig:GKP CNOT hexagon decoding} (a))
\begin{align}
    &P_{\mathrm{CNOT}}^{[pp]}( \xi_{p}^{(1)} ,\xi_{p}^{(2)} ) =\frac{1}{2\pi\sqrt{ ( 4 + \frac{1}{\lambda^{2}} )\sigma_{\mathrm{gkp}}^{4} }}
    \nonumber\\
    &\times \exp\Big{[} -\frac{ 2(\xi_{p}^{(1)})^{2} + ( 2 + \frac{1}{\lambda^{2}}) (\xi_{p}^{(2)})^{2} + \frac{ 2}{ \lambda} \xi_{p}^{(1)}\xi_{p}^{(2)} }{2(4+\frac{1}{\lambda^{2}})\sigma_{\mathrm{gkp}}^{2} } \Big{]} . 
\end{align}
The goal is the decoding is then to find two integers $n^{(1)}_{p}$ and $n^{(2)}_{p}$, given measurement outcomes $\sqrt{2}p_{m}^{(1)}$ and  $\sqrt{2}p_{m}^{(2)}$, such that $P_{\mathrm{CNOT}}^{[pp]}( \xi_{p}^{(1)} ,\xi_{p}^{(2)} ) $ is maximized where $\xi_{p}^{(1)} = \sqrt{2}p_{m}^{(1)} - n_{p}^{(1)} \sqrt{\pi}/\lambda$ and $\xi_{p}^{(2)} = \sqrt{2}p_{m}^{(2)} - n_{p}^{(2)} \sqrt{\pi}$. This can be done by solving the following optimization problem
\begin{align}
    &(n^{\star(1)}_{p} , n^{\star(2)}_{p} ) = \argmin_{ n_{1} , n_{2}} \Big{(} 2x_{1}^{2} + ( 2+\frac{1}{\lambda^{2}} )x_{2}^{2} + \frac{2}{\lambda} x_{1}x_{2}  \Big{)} , 
    \nonumber\\
    &\mathrm{where}\,\,\, x_{1} \equiv  \sqrt{2}p_{m}^{(1)} - n_{1}\sqrt{\pi}/\lambda , \,\,\, x_{2} \equiv  \sqrt{2}p_{m}^{(2)} - n_{2}\sqrt{\pi}, \label{eq:CNOT pp optimization}
\end{align}
given $\sqrt{2}p_{m}^{(1)}$ and $\sqrt{2}p_{m}^{(2)}$, and our algorithm for solving this optimization problem is given in \cref{alg:CNOTnp1np2} in \cref{appendix:Maximum likelihood decoding of the logical two-qubit gates between two GKP qubits}.


\subsubsection{Error-corrected CZ gate}
\label{subsubsection:Error-corrected CZ gate}

The logical CZ gate between a rectangular-lattice GKP qubit and a square-lattice GKP qubit can be realized by a $qq$ coupling (see \cref{eq:GKP logical CZ}). We assume that the first qubit is encoded in a rectangular-GKP code with $\lambda \ge 1$ and the second qubit is encoded in the square-lattice GKP code. Similarly as in the case of the CNOT gate, the GKP qubits inherit shift errors $( \xi'^{(+)}_{q,1} , \xi'^{(+)}_{p,1} )$ and $( \xi'^{(+)}_{q,2} , \xi'^{(+)}_{p,2} )$ from the previous round of the noisy GKP error correction before the CZ gate is applied. These shifts are propagated via the CZ gate as $( \xi'^{(+)}_{q,1} , \xi'^{(+)}_{p,1} +  \xi'^{(+)}_{q,2}/\lambda )$ and $( \xi'^{(+)}_{q,2} , \xi'^{(+)}_{p,2} +  \xi'^{(+)}_{q,1}/\lambda )$. Then, due to the additional noise from the GKP error correction after the CZ gate, extra shift errors $( -\xi^{(-)}_{q,1} , \xi^{(-)}_{p,1} )$ and $( -\xi^{(-)}_{q,2} , \xi^{(-)}_{p,2} )$ are added to the quadratures. Consequently, the two GKP qubits have net shifts 
\begin{align}
    \xi_{q}^{(1)} &= \xi'^{(+)}_{q,1} -\xi^{(-)}_{q,1}, 
    \nonumber\\
    \xi_{p}^{(1)} &= \xi'^{(+)}_{p,1} + \frac{1}{\lambda}\xi'^{(+)}_{q,2} + \xi^{(-)}_{p,1}, 
    \nonumber\\
    \xi_{q}^{(2)} &= \xi'^{(+)}_{q,2} -\xi^{(-)}_{q,2}, 
    \nonumber\\
    \xi_{p}^{(2)} &= \xi'^{(+)}_{p,2} + \frac{1}{\lambda}\xi'^{(+)}_{q,1} + \xi^{(-)}_{p,2}. 
\end{align}
See \cref{fig:GKP error corrected two qubit gates} (d). Then, since the noise correlation structure is equivalent to that of the CNOT gate (up to permutation and sign change), the rest of the analysis, including the maximum likelihood decoding, can be performed in an analogous way.

\subsection{Performance of the maximum-likelihood decoding for error-corrected two-GKP-qubit gates}
\label{subsection:Performance of the maximum-likelihood decoding for error-corrected two-GKP-qubit gates}

\begin{table*}[t!]
    \begin{center}
        \begin{tabular}{c | c | c | c | c | c}
            \toprule
            CNOT logical failure rate & $9$dB & $10$dB & $11$dB & $12$dB & $13$dB \\
            \midrule
            Closest-integer decoding &  $1.01\times 10^{-1}$ & $5.23\times 10^{-2}$ &  $2.33\times 10^{-2}$ &  $8.69\times 10^{-3}$ & $2.60\times 10^{-3}$ \vspace{0.05cm} \\
            Maximum likelihood decoding & $6.89\times 10^{-2}$  &  $3.12\times 10^{-2}$ & $1.18\times 10^{-2}$  & $3.61\times 10^{-3}$  & $8.53\times 10^{-4}$  \\
            \bottomrule 
        \end{tabular}
        \newline
        \vspace{0.1 cm}
        \newline
        \begin{tabular}{c | c | c | c | c }
            \toprule
            CNOT logical failure rate & $9.5$dB & $10.5$dB & $11.5$dB & $12.5$dB  \\
            \midrule
            Closest-integer decoding & $7.39\times 10^{-2}$ & $3.57\times 10^{-2}$ & $1.46\times 10^{-2}$  &  $4.90\times 10^{-3}$ \vspace{0.05cm} \\
            Maximum likelihood decoding & $4.73\times 10^{-2}$ & $1.96\times 10^{-2}$ & $6.71\times 10^{-3}$  & $1.82\times 10^{-3}$ \\
            \bottomrule
        \end{tabular}
    \end{center}
    \caption{Failure rate of the logical CNOT gate between two square-lattice GKP qubits (i.e., $\lambda = 1$) for various values of the GKP squeezing of the ancilla GKP states used in noisy (teleportation-based) GKP error corrections. Note that the gap between the closest-integer decoding and our maximum likelihood decoding gets wider as we increase the GKP squeezing $\sigma_{\mathrm{gkp}}^{(\mathrm{dB})}$. }
    \label{tab:CNOT logical failure rates}
\end{table*}

\begin{table}[t!]
    \begin{center}
    \begin{tabular}{c | c | c | c }
            \toprule
             & CX ($\lambda = 0.8$) & CX ($\lambda = 1$) & CX ($\lambda = 1.2$)  \\
            \midrule
            $II$ & $1- 9.98\times 10^{-3}$ & $1- 6.71\times 10^{-3}$ & $1- 1.31\times 10^{-2}$   \\
            $ZI$ & $4.06\times 10^{-4}$ & $2.89\times 10^{-3}$ & $1.03\times 10^{-2}$   \\
            $IZ$ & $1.56\times 10^{-4}$ & $2.43\times 10^{-4}$ & $3.13\times 10^{-4}$ \\
            $ZZ$ & $1.56\times 10^{-4}$ & $2.43\times 10^{-4}$ & $3.13\times 10^{-4}$ \\
            $XI$ & $2.36\times 10^{-3}$ & $2.43\times 10^{-4}$ & $1.69\times 10^{-5}$ \\
            $IX$ & $4.54\times 10^{-3}$ & $2.87\times 10^{-3}$ & $2.08\times 10^{-3}$ \\
            $ZX$ & $1.86\times 10^{-6}$ & $8.26\times 10^{-6}$ & $2.19\times 10^{-5}$  \\
            $XX$ & $2.36\times 10^{-3}$ & $2.43\times 10^{-4}$ & $1.69\times 10^{-5}$  \\
            \bottomrule
        \end{tabular}
        \begin{tabular}{c | c | c | c }
            \toprule
             & CZ ($\lambda = 0.8$) & CZ ($\lambda = 1$) & CZ ($\lambda = 1.2$)  \\
            \midrule
            $II$ & $1- 9.98\times 10^{-3}$ & $1- 6.71\times 10^{-3}$ & $1- 1.31\times 10^{-2}$   \\
            $ZI$ & $4.06\times 10^{-4}$ & $2.87\times 10^{-3}$ & $1.03\times 10^{-2}$    \\
            $IZ$ & $4.54\times 10^{-3}$ & $2.87\times 10^{-3}$ & $2.08\times 10^{-3}$ \\
            $ZZ$ & $1.86\times 10^{-6}$ & $8.26\times 10^{-6}$ & $2.19\times 10^{-5}$ \\
            $XI$ & $2.36\times 10^{-3}$ & $2.43\times 10^{-4}$ & $1.69\times 10^{-5}$  \\
            $XZ$ & $2.36\times 10^{-3}$ & $2.43\times 10^{-4}$ & $1.69\times 10^{-5}$  \\
            $IX$ & $1.56\times 10^{-4}$ & $2.43\times 10^{-4}$ & $3.13\times 10^{-4}$ \\
            $ZX$ & $1.56\times 10^{-4}$ & $2.43\times 10^{-4}$ & $3.13\times 10^{-4}$ \\
            \bottomrule
        \end{tabular}
    \end{center}
    \caption{Logical Pauli error probabilities of the CNOT (or CX) and CZ gates between two GKP qubits with $\sigma_{\mathrm{gkp}}^{(\mathrm{dB})} = 11.5$dB. The first qubit is a rectangular-lattice GKP qubit with $\lambda \in \lbrace 0.8, 1, 1.2 \rbrace$ and the second qubit is always a square-lattice GKP qubit. These Pauli error probabilities are computed as described in \cref{subappedix:Noise model}. All the Pauli error probabilities that are not shown in the table are smaller than $10^{-6}$.     }
    \label{tab:GKP CNOT CZ detailed error rates}
\end{table}

Since the decision boundaries in our maximum-likelihood decoding is better aligned with the underlying probability distribution of the shift errors, the maximum likelihood would outperform the closest-integer decoding. To explicitly show this, we compare in \cref{tab:CNOT logical failure rates} the performance of the closest-integer decoding scheme and the maximum likelihood decoding method for the error-corrected CNOT gate between two square-lattice GKP qubits (i.e., $\lambda = 1$). For all values of the GKP squeezing (from $9$dB to $13$dB), the maximum likelihood decoding method outperforms the simple closest-integer decoding method. In particular, as the GKP squeezing becomes larger, the advantage offered by our maximum likelihood decoding scheme becomes more significant. For instance, while the failure rate of the logical CNOT gate is reduced only by a factor of $1.46$ by using the maximum likelihood decoding at $\sigma_{\mathrm{gkp}}^{(\mathrm{dB})} = 9\mathrm{dB}$, the reduction factor becomes $3.05$ at $\sigma_{\mathrm{gkp}}^{(\mathrm{dB})} = 13\mathrm{dB}$. Note that this constant factor reduction significantly reduces the logical error rate when the GKP code is concatenated with the surface code.    

Going beyond the square-lattice GKP qubits, we also show how the different error Pauli error rates are changed if the control GKP qubits are replaced by rectangular-GKP qubits. In \cref{tab:GKP CNOT CZ detailed error rates}, we provide $8$ dominant Pauli error probabilities of the error-corrected CNOT and CZ gates between two GKP qubits of GKP squeezing $\sigma_{\mathrm{gkp}}^{(\mathrm{dB})} = 11.5$dB. The first qubit is a control qubit and is encoded in the rectangular-lattice GKP code with $\lambda = \in \lbrace 0.8, 1, 1.2 \rbrace$. Regardless, the second qubit (i.e., the target qubit) is always encoded in the square-lattice GKP code. For both CNOT and CZ gates, we observe that the $Z$ error on the control qubit is enhanced as we choose $\lambda > 1$. For instance, $p_{ZI}$ is increased from $0.287\%$ to $1.03\%$ as we increase $\lambda = 1$ to $\lambda = 1.2$. In the surface code architecture, this results in more frequent syndrome measurement errors. On the other hand, the $X$ ($Z$) error rate on the target qubit in the CNOT (CZ) gate is reduced when we use $\lambda > 1$. In particular, $p_{IX}$ ($p_{IZ}$) of the CNOT (CZ) gate is decreased from $0.287\%$ to $0.208\%$ as we increase $\lambda = 1$ to $\lambda =1.2$, thanks to the $1/\lambda$ suppression factor in the noise propagation (see, e.g., \cref{eq:CNOT shift propagation qq main text}). In the surface code architecture, this translates to a reduced noise back-propagation to the data block during a surface code stabilizer measurement. Note that the opposite is true for $\lambda < 1$, i.e., increased noise back-propagation to the surface code data block ($0.287\%$ to $0.453\%$ as we go from $\lambda = 1$ to $\lambda = 0.8$) and reduced ancilla $Z$ error, hence reducing the syndrome measurement error rate ($0.287\%$ to $0.041\%$ via $\lambda =1$ to $\lambda = 0.8$). It is also important to observe that the total CNOT (or CZ) failure rate, i.e., $1-p_{II}$ is minimum at $\lambda = 1$.  



\section{Surface-GKP code}
\label{section:Surface-GKP code}

In this section, we briefly review the surface-GKP code and present results for logical failure rates of states encoded in the surface-GKP code. In particular, we show that a very low logical error rate (e.g., $<10^{-7}$) can be achieved by using the square-lattice GKP qubits with a moderately high GKP squeezing of only $12$dB and the surface code distance $d = 7$. We also investigate if the use of noise-biased GKP qubits (with $\lambda \neq 1$) in the surface code ancilla can further reduce the logical error rate. We conclude that, under the noise model considered (i.e. when finite squeezing of the GKP states is the only noise source), it is optimal (or very close to optimal) to use the square-lattice GKP qubits everywhere. 

\begin{figure*}
    \centering
    \includegraphics[width = 0.8\textwidth]{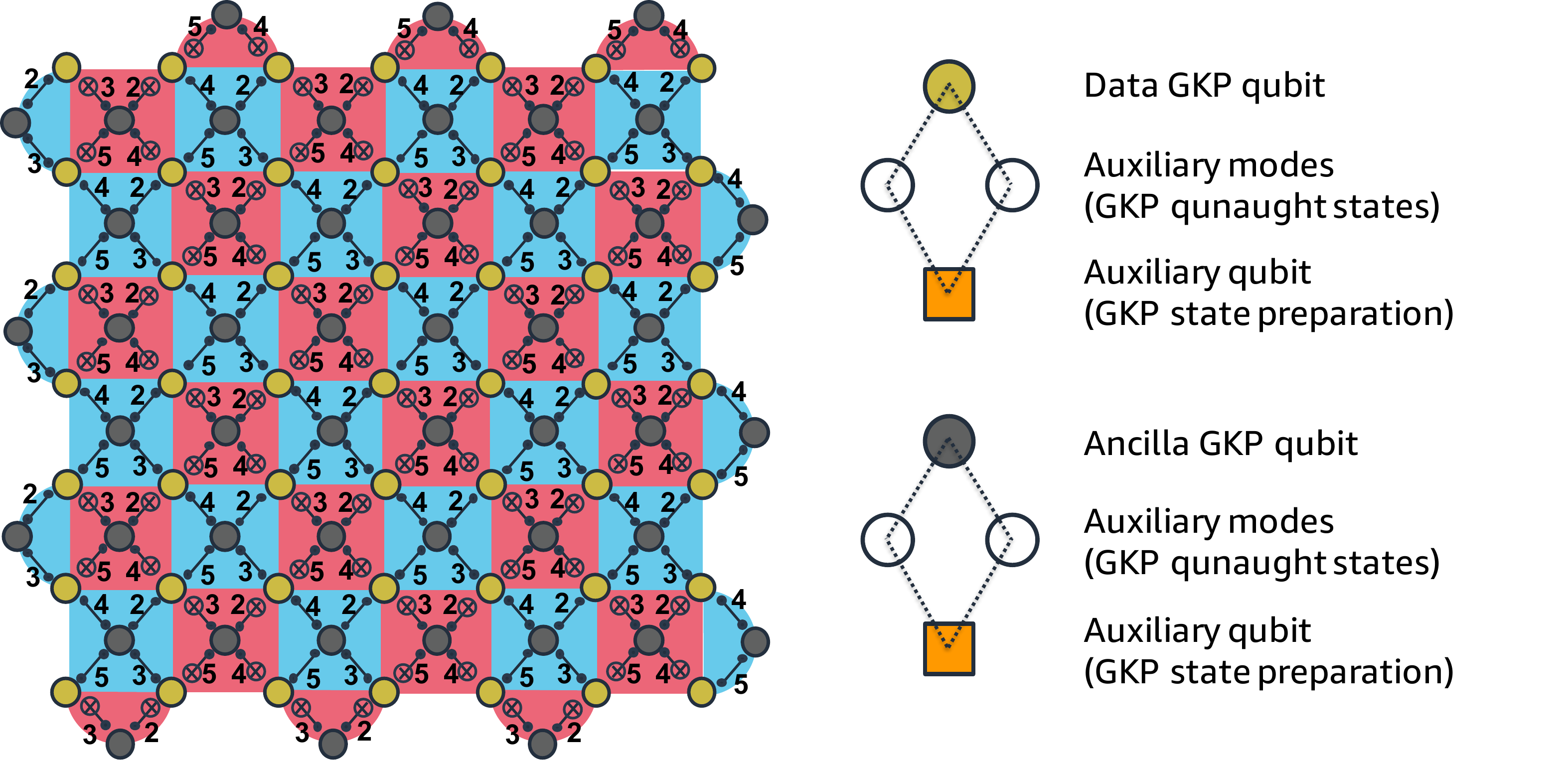}
    \caption{Lattice for a surface code of distance $d_x=d_z=7$. The logical $\overline{X}$-operator has minimum weight $d_x$ and is represented by a vertical string whereas the logical $\overline{Z}$ operator has minimum weight $d_z$ and is represented by a horizontal string. Red plaquettes correspond to $X$-type stabilizers and blue plaquettes to $Z$-type stabilizers. Data qubits correspond to yellow vertices and ancilla qubits to grey vertices. We assume that all data qubits are square-lattice GKP qubits but consider the use of rectangular-lattice GKP qubits with $0.8\le \lambda \le 1.2$ for the ancilla qubits. All ancilla qubits are prepared in the $|+_{\lambda}\rangle$ state and measured in the $X$-basis via a momentum homodyne measurement. Numbers beside the CNOT and CZ gates indicate the time-step in which the corresponding gate is applied. Each CNOT or CZ gate is an error-corrected logical CNOT or CZ gate between two GKP qubits. That is, the accumulated shift errors during each two-qubit gate are corrected at the end of the gate via a (teleportation-based) GKP error correction protocol and using the maximum likelihood decoding method described in \cref{subsection:Maximum likelihood decoding for GKP}. Due to the need to implement the GKP error correction, data (yellow) and ancilla (grey) GKP qubits should be constantly provided with two GKP qunaught states for every two-qubit gate. If the GKP qunaught states are supplied in a non-multiplexed fashion, each GKP qubit requires two auxiliary oscillator modes (white vertices) to host two GKP qunaught states and one auxiliary qubit (orange square) to prepare the GKP qunaught states. Connectivity between a GKP qubit, auxiliary modes, and an auxiliary qubit is represented by the dashed lines.     }
    \label{fig:SurfaceCodeLattice}
\end{figure*}

\subsection{Example experimental setup and decoding methods}
\label{subsection:Example experimental setup and decoding methods}

The surface code lattice for a distance $d_x=d_z=7$ surface code (encoding one logical qubit) is illustrated in \cref{fig:SurfaceCodeLattice}. Yellow vertices correspond to the data qubits where the logical information is stored, and grey vertices are the ancilla qubits used to measure the surface code stabilizers. The $X$-type stabilizers are represented by red-plaquettes and detect $Z$ errors whereas $Z$-type stabilizers are represented by blue plaquettes and detect $X$ errors. The logical $\overline{X}$ operator has minimum support on $d_x$ qubits along each column of the lattice. The logical $\overline{Z}$ operator has minimum support on $d_z$ qubits along each row of the lattice. We assume that all the data qubits are encoded in the square-lattice GKP qubit but consider the use of noise-biased rectangular-lattice GKP qubits (i.e., $0.8\le \lambda \le 1.2 $) for the ancilla qubits. All ancilla qubits in the lattice of \cref{fig:SurfaceCodeLattice} are prepared in the $| +_{\lambda} \rangle$ state and measured in the $X$ basis via a momentum homodyne measurement. The numbers adjacent to the CNOT and CZ gates indicate the time step at which such gates are applied. We choose the same two-qubit gate scheduling as in Ref.~\cite{TS14}. 

We apply a (teleportation-based) GKP error correction at the end of each CNOT or CZ gate. That is, each CNOT (or CZ) gate is an error-corrected logical CNOT (or CZ) gate between two GKP qubits as described in \cref{subsection:Maximum likelihood decoding for GKP}. Also importantly, since GKP error correction is performed after every two-qubit gate, extra analog information \cite{Fukui2017_analog,Fukui2018_high_threshold,Vuillot2019_quantum,Noh2020_fault_tolerant,Yamasaki2020_polylog_overhead,Larsen2021_fault_tolerant,Bourassa2021_blueprint} is gathered from the GKP error correction for each two-qubit gate. This analog information can then be used to compute the conditional probabilities of Pauli error rates for each error-corrected CNOT or CZ gate between two GKP qubits. The same applies to the ancilla state preparation, idling, and measurements as well (see \cref{subappedix:Noise model} for more details). Note that due to the need to implement the GKP error correction to all GKP qubits, each GKP qubit should be assisted by two auxiliary modes (white vertices in \cref{fig:SurfaceCodeLattice}) that host two GKP qunaught states and one auxiliary qubit (orange square in \cref{fig:SurfaceCodeLattice}) used to prepare the GKP qunaught states. We also emphasize that we use the maximum likelihood decoding instead of the simple closest-integer decoding to correct for shift errors in error-corrected CNOT and CZ gates between two GKP qubits.

\begin{figure}
	\centering
	\subfloat[\label{fig:GKPl1AlldNoAnalog}]{%
		\includegraphics[width=0.475\textwidth]{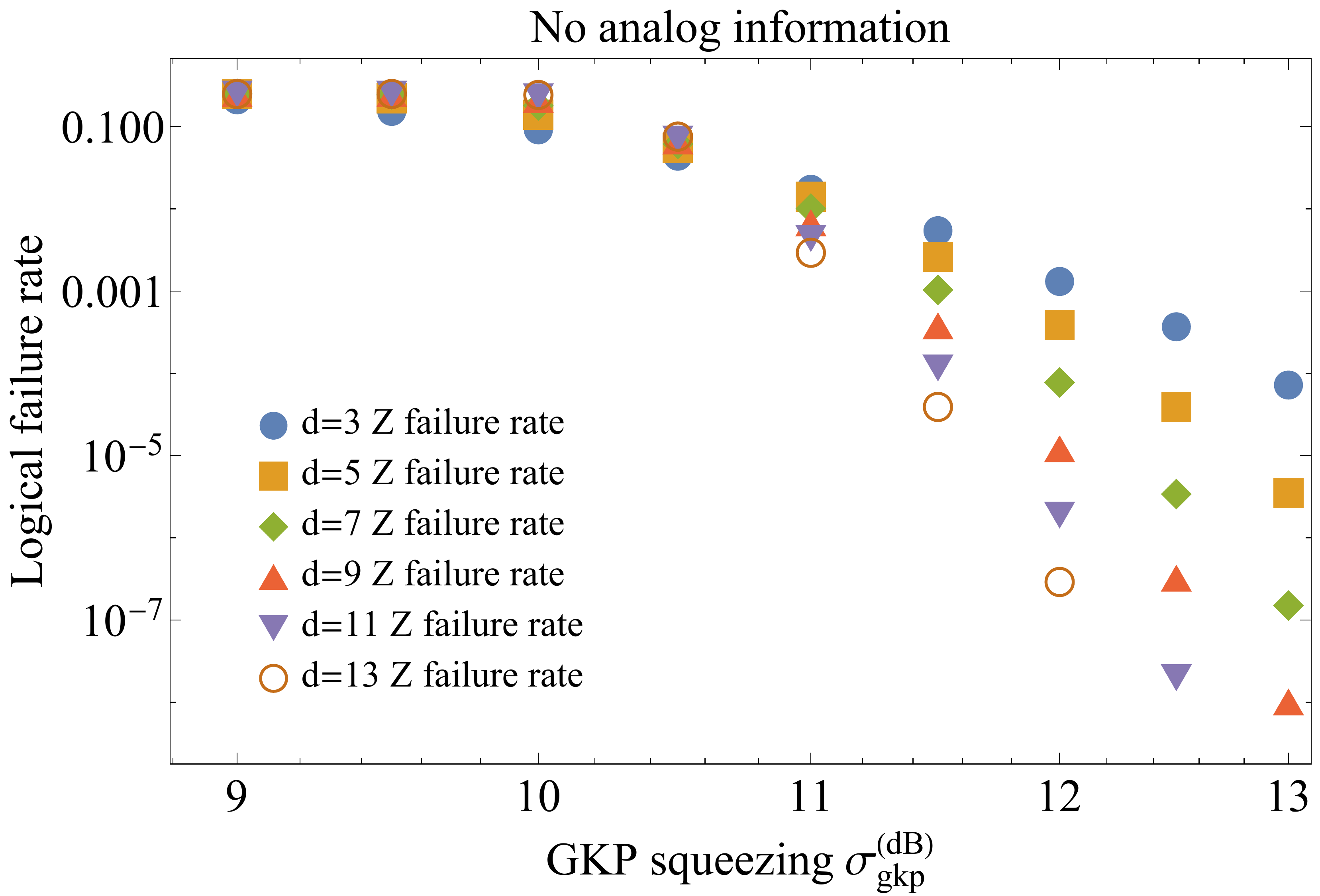}
	}
	\vfill
	\subfloat[\label{fig:GKPl1AlldAnalog}]{%
		\includegraphics[width=0.475\textwidth]{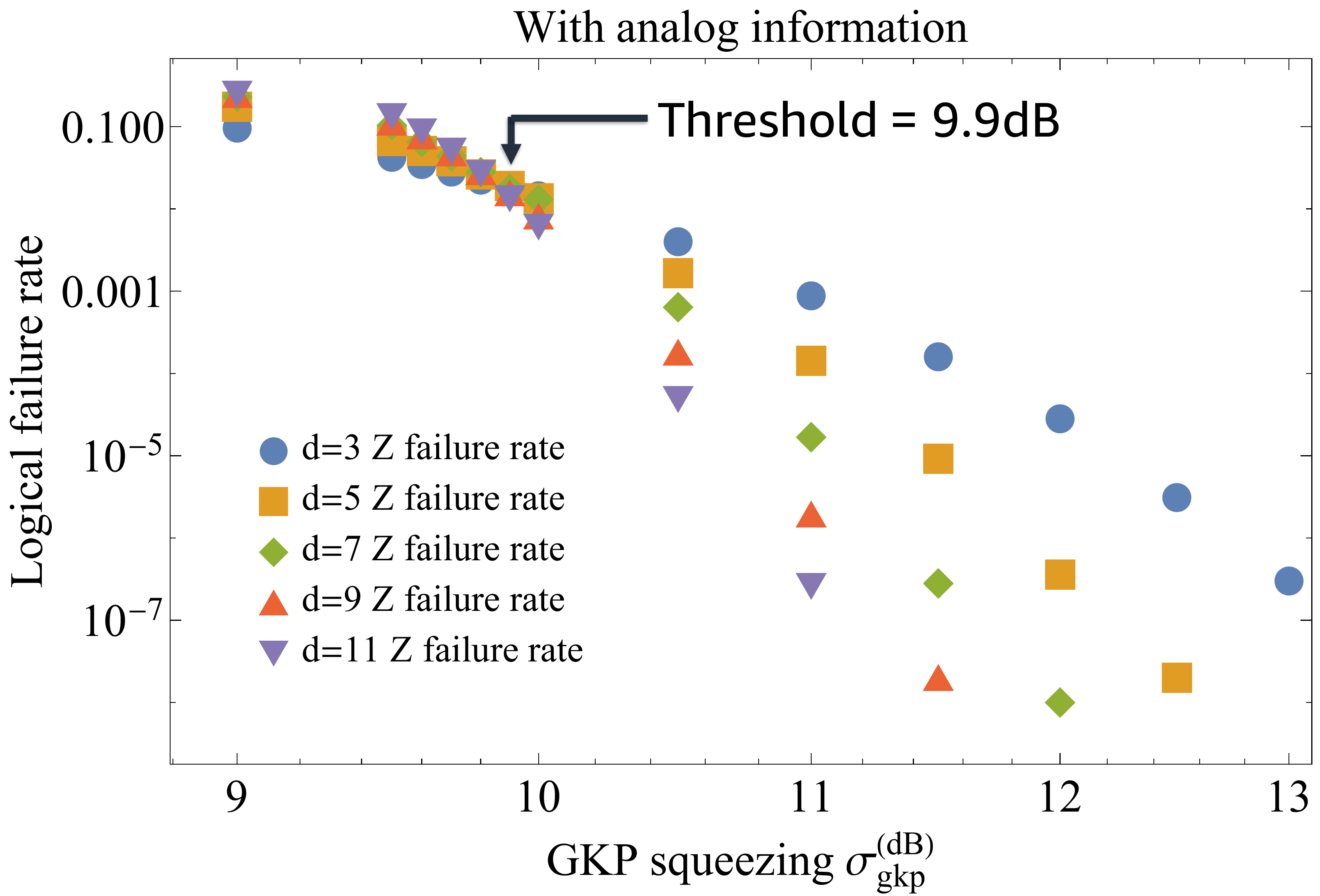}
	}

	\caption{\label{fig:NoAnalogAnalogFailSurfaceCode} (a) Logical $Z$ failure rates for the surface-GKP code that does not use the analog information from the GKP error correction as a function of the GKP squeezing. Logical failure rate curves are plotted for distances $d_x=d_z=d \in \{3,5, \cdots, 13 \}$. (b) Same as in (a) but where the analog information from the GKP error correction is taken into account in computing the edge weights of the matching graphs. All results were obtained by performing $10^{8}$ Monte-Carlo simulations and setting $\lambda = 1$. In (b), we limit the surface-GKP code distance to $d \le 11$ due to the very low logical failure rates. Logical $X$ failure rates are not plotted since their values are almost identical to the logical $Z$ failure rates shown in the above plots, although a very small discrepancy arises due to the differing boundary edge weights of $X$ and $Z$-type stabilizers.}
\end{figure}

In order to ensure a fault-tolerant error correction protocol, measurements of the stabilizers must be repeated at least $d = \max{ \{d_x,d_z \}}$ times \cite{FMMC12, FowlerAutotune}. Errors are then corrected using a Minimum Weight Perfect Matching (MWPM) decoder \cite{Edmonds65} applied to three-dimensional matching graphs with weighted edges which incorporates the conditional probabilities of the GKP error correction for the entire syndrome history. In \cref{subappedix:Edge weights in the matching graphs}, we provide the matching graphs for $X$ and $Z$-type stabilizers and show how edge weights can be computed dynamically using analog information from the GKP error correction. It is important to note here that when using the analog information for computing edge weights in the matching graphs, at every time step and in each syndrome measurement round, every fault location (which includes two-qubit gates, idle, state-preparation and ancilla measurements) can have different conditional probabilities associated with a particular Pauli error at the given location. As such, edge weights cannot be pre-computed and must be updated dynamically. Our methods for efficient dynamical updates in the edge weights are described in \cref{subappedix:Edge weights in the matching graphs}. 

\subsection{Performance of the surface-GKP code}
\label{subsection:Performance of the surface-GKP code}

Having described all the methods, we now present the main results, i.e., performance of the surface-GKP code under our highly optimized error correction and decoding procedures. In \cref{fig:GKPl1AlldNoAnalog}, we plot the logical $Z$ failure rates of the surface-GKP code when analog information from the GKP error correction is omitted. The results are obtained for square surface code patches (i.e. $d_x = d_z = d$) with $3 \le d \le 13$ by performing $10^8$ Monte Carlo simulations using the full circuit level noise model described in \cref{appendix:Simulation details} and setting $\lambda = 1$. In \cref{fig:GKPl1AlldAnalog}, we plot the logical $Z$ failure rates of the surface-GKP code when analog information from the GKP error correction is taken into account, and edge weights are dynamically updated following the techniques presented in \cref{subappedix:Edge weights in the matching graphs}. Note that for the analog simulations, since the logical $Z$ failure rates drop very quickly with increasing GKP squeezing, we limit the code distance to $d \le 11$ to avoid large statistical errors from the Monte-Carlo simulations. We omitted plots for the logical $X$ error rates since for both cases with and without the use of analog information, logical $X$ error rates are nearly identical to the logical $Z$ error rates shown. 

Comparing the plots of \cref{fig:GKPl1AlldNoAnalog,fig:GKPl1AlldAnalog}, it can be seen that using the analog information in the GKP decoding scheme results in smaller logical failure rates by several orders of magnitude. For instance, for $d=9$ and $\sigma_{\mathrm{gkp}}^{(\mathrm{dB})} = 11\mathrm{dB}$, without the analog information the logical failure rate is found to be $6.72 \times 10^{-3}$ compared to $1.94 \times 10^{-6}$ when using the analog information, an improvement by a factor of $3461$. Such a large improvement is possible since all edge weights of the matching graphs used by the MWPM decoder are determined based on the conditional probabilities of all types of errors in the full history of syndrome measurement rounds given the extra analog information. If GKP qubits are sufficiently squeezed (i.e., $\sigma_{\mathrm{gkp}}^{(\mathrm{dB})} \gtrsim 9\mathrm{dB}$), a Pauli error on GKP qubits occurs due to a shift that is barely larger than the largest correctable shift (e.g., $\sqrt{\pi}/2$ in the case of an idling square-lattice GKP qubit). Hence, most Pauli errors on GKP qubits occur near the decision boundary of a GKP error correction and hence come with a high conditional probability because decisions made near the decision boundary are less reliable than the ones made deep inside the decision boundary (see, e.g., Ref.\ \cite{Noh2020_fault_tolerant} for a more detailed and quantitative explanation). 

For instance, in the case of the distance-three surface code, there are many cases where two faults happened during the three syndrome measurement rounds. These two faults come with high conditional probabilities as they are caused by shifts close the the decision boundary. If this extra analog information (i.e., high conditional probabilities) are not taken into account, the MWPM decoder will choose a wrong path consistent with the syndrome history when pairing highlighted vertices. The correction will then result in a logical Pauli error in the surface-GKP code. However, since the two faults that did happen are likely to have high conditional probabilities, the paths correcting the resulting errors are favored by the MWPM decoder when the analog information is incorporated in the decoding protocol. Thus, even though the distance of the surface code is only three (and hence can only correct at most one fault in the standard surface code setting with bare two-level qubits), many two-fault events are correctable in the surface-GKP code setting with the help of the extra analog information. 

We point out that the opposite can also be true in principle. That is, there may be cases where only a single fault occurred but two other candidate fault locations have much higher conditional probabilities despite the fact that no errors were introduced at these other two locations (i.e., false alarms). In this case, the MWPM decoder favors edges of two data qubits not afflicted by an error (due to the large conditional probabilities). The applied correction would then result in a logical Pauli error for the surface-GKP code even though only one fault actually happened (see \cref{subappedix:FaultToleranceSurfaceGKP} for a more detailed discussion). Such possibilities might seem problematic from a fault-tolerance perspective. However, such events occur with very small probability (smaller than the computed logical failure rates) and the benefit of using the extra analog information strongly outweighs the side effects due to false alarms. This can be seen from the significantly reduced logical failure rates in \cref{fig:GKPl1AlldAnalog} compared to those in \cref{fig:GKPl1AlldNoAnalog}.

\begin{figure}
    \centering
    \includegraphics[width = 0.475\textwidth]{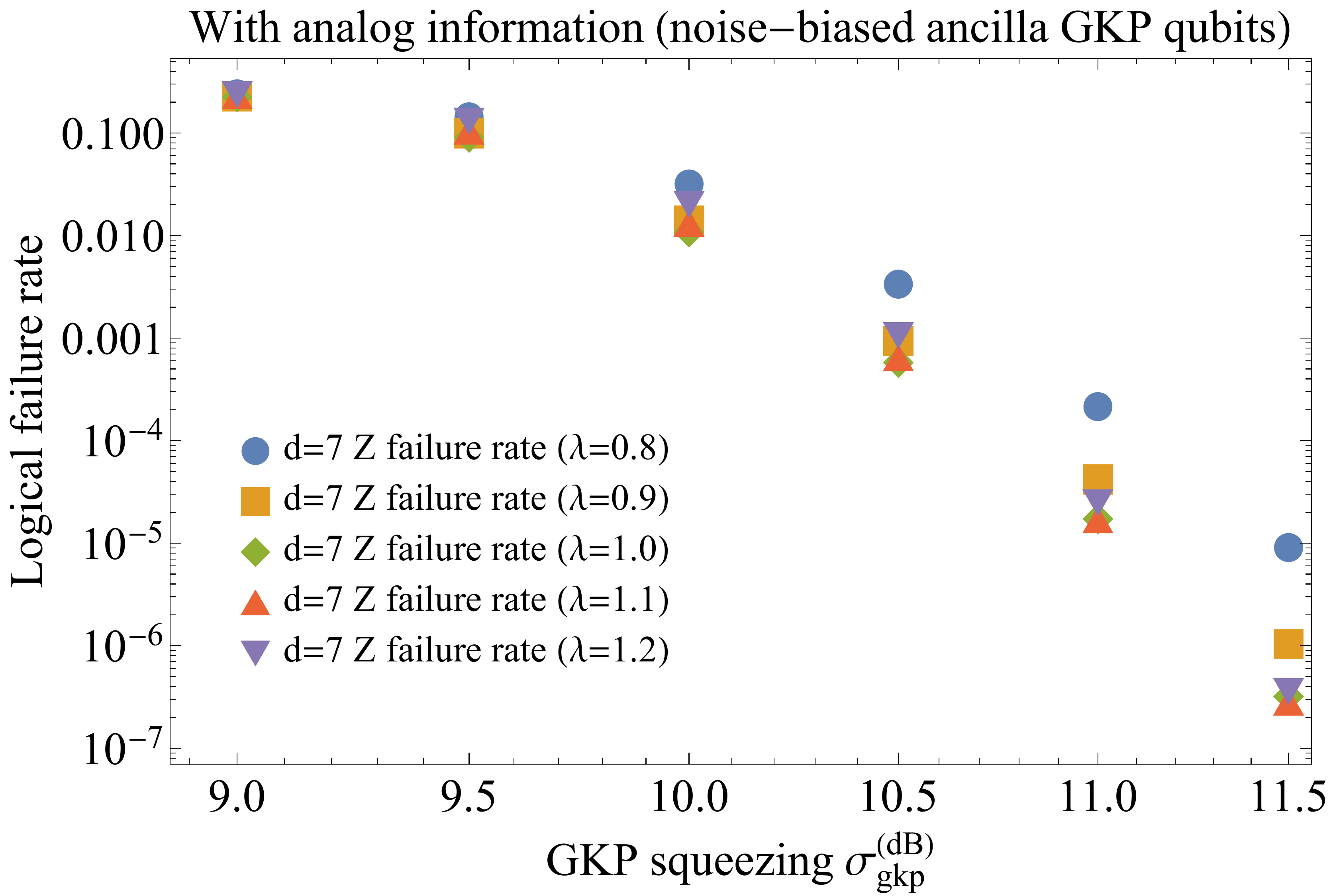}
    \caption{Logical $Z$ failure rates for a $d_x = d_z = d = 7$ surface-GKP code using analog information for the different values of $\lambda$ shown in the legend. Below $\sigma_{\mathrm{gkp}}^{(\mathrm{dB})}  = 11\mathrm{dB}$, the best logical failure rates are achieved by setting $\lambda = 1$. Above $\sigma_{\mathrm{gkp}}^{(\mathrm{dB})}  = 11\mathrm{dB}$, $\lambda = 1.1$ provides a very minor improvement. }
    \label{fig:CompareLambdas}
\end{figure}

From the intersection of the logical failure rate curves in \cref{fig:GKPl1AlldAnalog}, the threshold can be seen to be approximately $\sigma_{\mathrm{gkp}}^{(\mathrm{dB})} \approx 9.9\mathrm{dB}$, which represents the lowest threshold to date (see for instance the results in Ref.~\cite{Larsen2021_fault_tolerant,Bourassa2021_blueprint} where the threshold for the same noise model was found to be $\sigma_{\mathrm{gkp}}^{(\mathrm{dB})} = 10.5-10.6\mathrm{dB}$). More importantly, however, it can be observed that for some parameter regimes, the logical failure rate curve in \cref{fig:GKPl1AlldAnalog} is about three orders of magnitude lower than the one in Ref.~\cite{Larsen2021_fault_tolerant} (e.g., improved from $5\times 10^{-3}$ to $1.94\times 10^{-6}$ at $\sigma_{\mathrm{gkp}}^{(\mathrm{dB})} = 11\mathrm{dB}$ and $d=9$).

Note that there are two reasons for the large reduction in the logical failure rate compared to the results of Ref.\ \cite{Larsen2021_fault_tolerant}. First, we use the maximum likelihood decoding for error-corrected CNOT and CZ gates between two GKP qubits. Hence, the total failure rates of the two-qubit gates are significantly reduced at the GKP qubit level (e.g., from $2.33\%$ to $1.18\%$ for the CNOT gate at $\sigma_{\mathrm{gkp}}^{(\mathrm{dB})} = 11\mathrm{dB}$; see \cref{tab:CNOT logical failure rates}). Furthermore, since the shift errors are corrected at the end of every two-qubit gates, each two-qubit gate may be afflicted by a Pauli error if the shift errors are incorrectly inferred and countered. Hence, while it was not crucial to use space-time correlated edges in the embodiment of the surface-GKP codes as considered in Refs.\ \cite{Vuillot2019_quantum,Noh2020_fault_tolerant} (since the GKP error correction is only performed once after all four two-qubit gates per each surface code syndrome measurement), not using space-time correlated edges in the matching graphs in our case (as well as in Ref.\ \cite{Larsen2021_fault_tolerant}) would result in much higher logical failure rates. Such improved performance with space-time correlated edges is primarily due to the fact that GKP error correction is performed after each two-qubit gate, sometimes resulting in an error pattern correlated in both space and time at the surface code level. We can thus conclude that the use of maximum likelihood decoding in the GKP error correction and the space-time correlated edges are the main reasons why the logical error rates are much lower in \cref{fig:NoAnalogAnalogFailSurfaceCode} compared to what was found in Ref.\ \cite{Larsen2021_fault_tolerant}.

Recall that one can change the aspect ratio of the ancilla GKP qubit (i.e., $\lambda$) to bias the noise towards a certain Pauli error. When $\lambda > 1$, the two-qubit gates in the surface code lattice of \cref{fig:SurfaceCodeLattice} introduce $Z$ errors on the ancilla qubits with higher probability resulting in increased measurement error rates, while simultaneously reducing the probability of $X$ errors on the ancilla qubits thus reducing the rate of space-time correlated errors (a subset of which are hook errors \cite{DKLP02}). The opposite is true for $\lambda < 1$ (see \cref{tab:GKP CNOT CZ detailed error rates}). In \cref{fig:CompareLambdas}, we compare logical $Z$ failure rate curves of $d=7$ surface codes for $0.8 \le \lambda \le 1.2$ in increments of $0.1$. As can be seen, $\lambda < 1$ always has the worst performance given the impact of increased space-time correlated errors. For $\sigma_{\mathrm{gkp}}^{(\mathrm{dB})}  < 11\mathrm{dB}$, logical failure rates are always minimized at $\lambda =1$ given the same GKP squeezing $\sigma_{\mathrm{gkp}}^{(\mathrm{dB})}$, in part due to the fact that the total failure rates of the two-qubit gates are minimized when $\lambda = 1$, as shown in \cref{tab:GKP CNOT CZ detailed error rates}. For $\sigma_{\mathrm{gkp}}^{(\mathrm{dB})}  \ge 11\mathrm{dB}$, $\lambda = 1.1$ offers slightly lower failure rates than $\lambda = 1.0$. For instance, at $\sigma_{\mathrm{gkp}}^{(\mathrm{dB})}  = 11\mathrm{dB}$ and $\lambda = 1$, the logical $Z$ error rate is $1.722 \times 10^{-5}$ compared to $1.679 \times 10^{-5}$ when $\lambda = 1.1$. Due to the negligible increase in performance, however, we only plot the logical failure rates for $\lambda = 1$ (i.e., square-lattice GKP qubits for all data and ancilla qubits) in \cref{fig:NoAnalogAnalogFailSurfaceCode}. 

\subsection{Resource overhead comparison}
\label{subsection:Resource overhead comparison}

\begin{table*}[t!]
    \begin{center}
    \begin{tabular}{c  c  c  c  c }
            \toprule
            $\sigma_{\mathrm{gkp}}^{(\mathrm{dB})} = 11.5$dB  & \multirow{2}{*}{ $d_{\min}$ for $p_{L} < 10^{-7}$ } & \multirow{2}{*}{ $p_{L}$ at $d= d_{\min}$ } &   \# surface code  & \multirow{2}{*}{ hardware requirements } \\
            ($\leftrightarrow p= 6.71\times 10^{-3}$) &  & & qubits &  \\
            \midrule
            Surface code ($p$)  & $d_{\min} = 69$ & $8.62\times 10^{-8}$ & $9521$ & $9521$ qubits \\
            Surface-GKP code ($\sigma_{\mathrm{gkp}}^{(\mathrm{dB})}$)  &  $d_{\min} = 9$ & $2\times 10^{-8}$ & $161$ & $483$ modes \& $161$ qubits \\ 
            \bottomrule 
        \end{tabular}
        \begin{tabular}{c  c  c  c  c }
            \toprule
            $\sigma_{\mathrm{gkp}}^{(\mathrm{dB})} = 12$dB  & \multirow{2}{*}{ $d_{\min}$ for $p_{L} < 10^{-7}$ } & \multirow{2}{*}{ $p_{L}$ at $d= d_{\min}$ } &  \# surface code   & \multirow{2}{*}{ hardware requirements } \\
            ($\leftrightarrow p= 3.61\times 10^{-3}$) &  & & qubits & \\
            \midrule
            Surface code ($p$)  & $d_{\min} = 27$ & $6.38 \times 10^{-8}$ & $1457$ & $1457$ qubits \\
            Surface-GKP code ($\sigma_{\mathrm{gkp}}^{(\mathrm{dB})}$)  &  $d_{\min} = 7$ & $1\times 10^{-8}$ & $97$ & $291$ modes \& $97$ qubits  \\
            \bottomrule
        \end{tabular}
        \begin{tabular}{c  c  c  c  c }
            \toprule
            $\sigma_{\mathrm{gkp}}^{(\mathrm{dB})} = 12.5$dB  & \multirow{2}{*}{ $d_{\min}$ for $p_{L} < 10^{-7}$ } & \multirow{2}{*}{ $p_{L}$ at $d= d_{\min}$ } &  \# surface code  & \multirow{2}{*}{ hardware requirements } \\
            ($\leftrightarrow p= 1.82\times 10^{-3}$) &  & & qubits & \\
            \midrule
            Surface code ($p$)  & $d_{\min} = 17$ & $2.19\times 10^{-8}$ & $577$ & $577$ qubits  \\
            Surface-GKP code ($\sigma_{\mathrm{gkp}}^{(\mathrm{dB})}$)  &  $d_{\min} = 5$ & $2\times 10^{-8}$ & $49$ & $147$ modes \& $49$ qubits  \\
            \bottomrule
        \end{tabular}
    \end{center}
    \caption{Resource overheads for achieving a logical failure rate $p_{L} < 10^{-7}$ with the standard (rotated) surface code and the surface-GKP code. For a given value of GKP squeezing (e.g., $\sigma_{\mathrm{gkp}}^{(\mathrm{dB})} = 12$dB), we take the failure rate of the error-corrected CNOT (or CZ) gate between two square-lattice GKP qubits (e.g., $p=3.61\times 10^{-3}$) from \cref{tab:CNOT logical failure rates} using the maximum likelihood decoding. To compare the resource overheads between the standard surface code and the surface-GKP code, we assume that each element in the standard surface code fails with the probability $p$ found above. Then, by using the widespread estimate of the surface code logical failure rate $p_{L} = 0.1(100p)^{(d+1)/2}$ \cite{FMMC12} (obtained from a toy depolarizing noise model), we estimate the minimum distance of the surface code $d_{\min}$ needed to achieve $p_{L} < 10^{-7}$ and compute the number of qubits in the surface code using $n_{\text{tot}} = 2d_{\min}^{2} - 1$. For the surface-GKP code, we find $d_{\min}$ needed to achieve $p_{L}<10^{-7}$ from \cref{fig:GKPl1AlldAnalog} and estimate the total number of required modes using $n_{\text{tot}} = 3(2d_{\min}^{2} - 1)$. The extra factor of 3 comes from the fact that we assume the GKP qunaught states are supplied in a non-multiplexed manner for the GKP error correction. Note that additional $2d_{\min}^{2} - 1$ auxiliary qubits are needed for the surface-GKP code to prepare GKP qunaught states in the auxiliary modes. The experimental feasibility of a large GKP squeezing (i.e., $\sigma_{\mathrm{gkp}}^{(\mathrm{dB})} \gtrsim 12$dB) is discussed in \cref{section:Experimental realization of a highly squeezed GKP qunaught state}.    }
    \label{tab:overhead comparison}
\end{table*}

Lastly, to put our results into perspective, we briefly compare the resource overhead costs of our surface-GKP code scheme to that of the standard surface code approach using regular two-level qubits (e.g., transmons or internal states of trapped ions). Note that at the GKP squeezing $\sigma_{\mathrm{gkp}}^{(\mathrm{dB})} = 12$dB, the total failure rate of the error-corrected CNOT or CZ gate between two square-lattice GKP qubits is given by $3.61\times 10^{-3}$ using the maximum likelihood decoding (see \cref{tab:CNOT logical failure rates}). In the case of the standard surface code schemes, the logical failure rate of the surface code is roughly given by $p_{L} = 0.1(100p)^{(d+1)/2}$ \cite{FMMC12} for a toy depolarizing noise model where $p$ is the failure rate of each circuit element (including two-qubit gates). In this case, by plugging in $p= 3.61\times 10^{-3} $, we find that the surface code distance $d= 27$ is needed to reach a logical failure rate $p_{L} < 10^{-7}$. The distance-$27$ surface code requires $729$ data qubits, $728$ ancilla qubits (a total of $1457$ qubits), and more hardware elements to support these qubits.

On the other hand, as shown in \cref{fig:GKPl1AlldAnalog}, the surface-GKP code with $\sigma_{\mathrm{gkp}}^{(\mathrm{dB})} = 12$dB (corresponding to the two-qubit error rate $3.61\times 10^{-3}$) achieves a very low logical failure rate $<10^{-7}$ with only $d=7$, which requires $49$ data GKP qubits and $48$ ancilla GKP qubits, a massive reduction compared to the case of the standard surface code based on two-level qubits. However, shift errors in each GKP qubit are constantly corrected by a teleportation-based GKP error correction protocol. Note that two fresh GKP qunaught states must be supplied to realize the teleportation-based GKP error correction. Thus, each data or ancilla GKP qubit requires one mode to encode the quantum information and, assuming no multiplexing, two auxiliary modes (represented by white vertices in \cref{fig:SurfaceCodeLattice}) are needed to provide fresh GKP qunaught states for the GKP error correction. In this case, the number of required oscillator modes is three times the number of GKP qubits. Moreover, for each GKP qubit, one auxiliary qubit is needed to prepare the GKP qunaught states in the auxiliary modes. Such auxiliary qubits are represented by orange squares in \cref{fig:SurfaceCodeLattice}. Hence, the surface-GKP code requires a total of $291$ physical modes, $97$ qubits and hardware elements to support these modes. Note that if the GKP qunaught states are supplied in a multiplexed fashion, the extra overhead associated with the GKP error correction may be further reduced. However, despite the extra two auxiliary modes and one auxiliary qubit for each surface code qubit, the surface-GKP code requires fewer physical elements than the standard surface code consisting of bare qubits.   

Comparisons between resource costs for standard surface codes and surface-GKP codes for other values of the GKP squeezing (i.e., $\sigma_{\mathrm{gkp}}^{(\mathrm{dB})} = 11.5$dB and $\sigma_{\mathrm{gkp}}^{(\mathrm{dB})} = 12.5$dB) are provided in \cref{tab:overhead comparison}. Note that at $\sigma_{\mathrm{gkp}}^{(\mathrm{dB})} =11.5$dB (corresponding to $p=0.67\%$), the standard surface code is estimated to require more than $9521$ qubits to achieve a logical error rate $p_{L}<10^{-7}$, whereas the surface-GKP only requires $483$ oscillator modes and $161$ qubits. Such a large difference is due to the fact that the error rate $p=0.67\%$ is too close to the standard surface code threshold $1\%$ (based on the rough estimate $p_{L} = 0.1(100p)^{(d+1)/2}$; this estimate, however, tends to underestimate the logical error rate and overestimate the threshold error rate near the threshold). On the other hand, in the case of the surface-GKP code, $\sigma_{\mathrm{gkp}}^{(\mathrm{dB})} =11.5$dB is decently away from the threshold value $\sigma_{\mathrm{gkp}}^{(\mathrm{dB})} =9.9$dB. To put it in another way, at $\sigma_{\mathrm{gkp}}^{(\mathrm{dB})} = 9.9$dB, the failure rate of the error-corrected CNOT and CZ gate between two square-lattice GKP qubits is given by $p = 3.4\%$ (see also \cref{tab:CNOT logical failure rates}). Such a high CNOT (or CZ) failure rate at the threshold is possible through the help of extra analog information gathered from GKP error corrections (compare \cref{fig:GKPl1AlldNoAnalog} with \cref{fig:GKPl1AlldAnalog}). Thus, the use of extra analog information is particularly crucial for reducing the resource overheads in the high (but below threshold) noise regime, which is most experimentally relevant today. In the following section, we show that high GKP squeezing $\sigma_{\mathrm{gkp}}^{(\mathrm{dB})} \gtrsim 12$dB can in principle be achieved by assuming reasonable experimental parameters.

\section{Experimental realization of a highly squeezed GKP qunaught state}
\label{section:Experimental realization of a highly squeezed GKP qunaught state} 

\begin{figure*}
    \centering
    \includegraphics[width=0.8\textwidth]{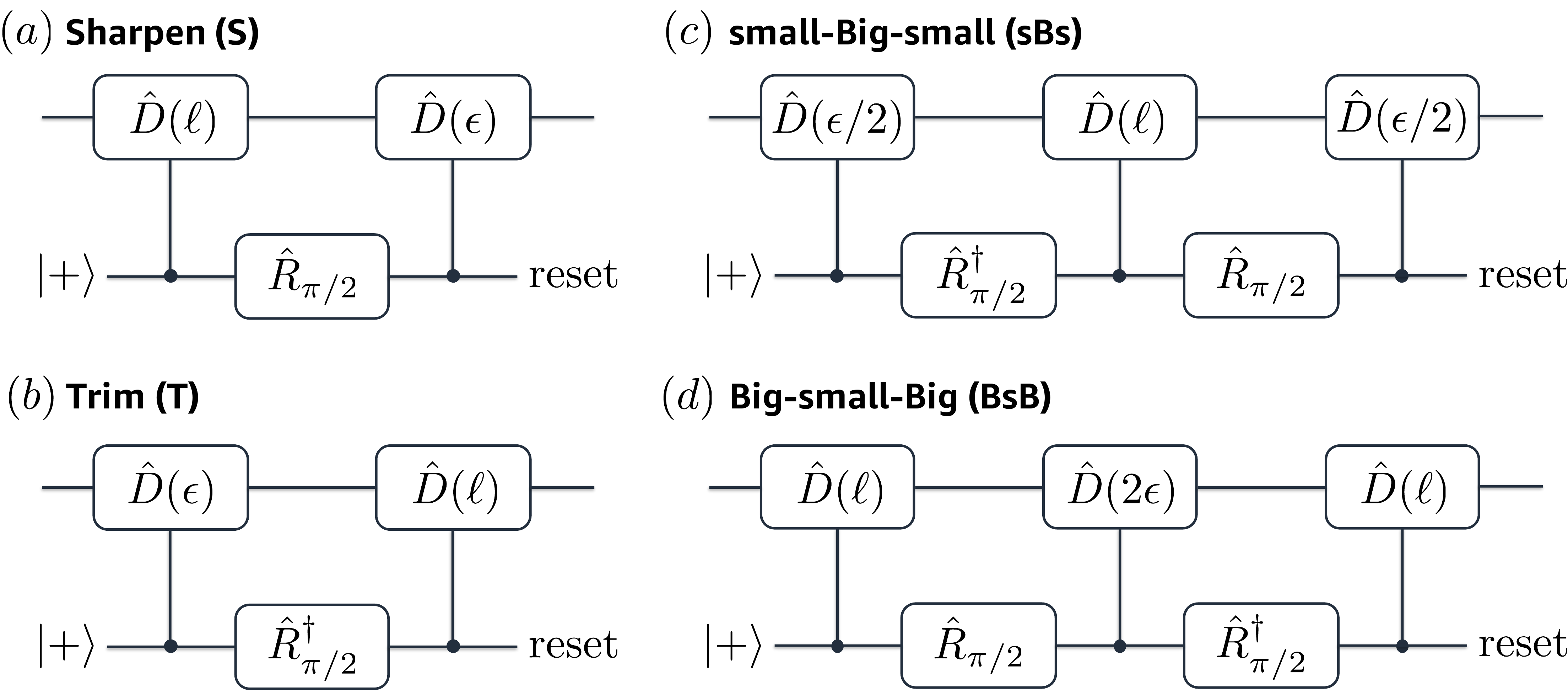}
    \caption{Circutis for (a) Sharpen, (b) Trim, (c) small-Big-small, and (d) Big-small-Big schemes used to stabilize a finitely-squeezed square-lattice GKP qunaught state $|\emptyset_{\Delta}\rangle \propto \exp[ -2\Delta^{2}\hat{n} ]|\emptyset\rangle$. Here, $|\emptyset\rangle \equiv |\emptyset_{\lambda = 1}\rangle$ is the square-lattice GKP qunaught state. These circuits dissipatively stabilize the exact stabilizers of a finitely-squeezed GKP qunaught state, i.e., $\hat{S}_{q}^{[\Delta]} \equiv  \exp[ -2\Delta^{2}\hat{n} ]  \hat{S}_{q} \exp[ 2\Delta^{2}\hat{n} ] $ and $\hat{S}_{p}^{[\Delta]} \equiv  \exp[ -2\Delta^{2}\hat{n} ]  \hat{S}_{p} \exp[ 2\Delta^{2}\hat{n} ] $, where $\hat{S}_{q} \equiv \exp[ i\sqrt{2\pi}\hat{q} ]$ and $\hat{S}_{p} \equiv \exp[ -i\sqrt{2\pi}\hat{p} ]$ are the stabilizers of an infinitely-squeezed square-lattice GKP qunaught state. Here, $\Delta^{2}$ is approximately equal to the noise variance of a GKP qunaught state $\sigma_{\mathrm{gkp}}^{2}$, i.e., $\sigma_{\mathrm{gkp}}^{2} = \Delta^{2} + \mathcal{O}(\Delta^{4})$ (see \cref{subappendix:Finitely squeezed ancilla GKP states}). In the circuit for stabilizing $\hat{S}_{q}^{[\Delta]}$, $\ell$ and $\epsilon$ are given by $\ell  = \ell_{q} \equiv i\sqrt{2\pi}\cosh(2\Delta^{2})$ and $\epsilon = \epsilon_{q} \equiv \sqrt{2\pi}\sinh(2\Delta^{2})$. In the other circuit stabilizing $\hat{S}_{p}^{[\Delta]}$, $\ell$ and $\epsilon$ are given by $\ell  = \ell_{p} \equiv \sqrt{2\pi}\cosh(2\Delta^{2})$ and $\epsilon = \epsilon_{p} \equiv -i\sqrt{2\pi}\sinh(2\Delta^{2})$. Note that in our convention, the displacement operator is defined as $\hat{D}(\xi) \equiv \exp[ i(\xi_{p}\hat{q} - \xi_{q}\hat{p}) ]$. The ancilla input state is given by $|+\rangle = (|0\rangle + |1\rangle)/\sqrt{2}$ and the single qubit rotation $\hat{R}_{\theta}$ is defined as $\hat{R}_{\theta} \equiv [ -i\theta \hat{\sigma}_{x}/2 ]$, where $\sigma_{x} \equiv |0\rangle\langle 1| + |1\rangle\langle 0| $. The conditional displacement operator $C\hat{D}(\xi)$ is defined as $C\hat{D}(\xi) \equiv \exp[ -i(\xi_{p}\hat{q}-\xi_{q}\hat{p})\otimes \sigma_{z}/2 ] = \hat{D}(\xi/2)\otimes |1\rangle\langle 1| + \hat{D}(-\xi/2)\otimes |0\rangle\langle 0|$, where $\sigma_{z} \equiv |0\rangle\langle 0| - |1\rangle\langle 1|$.           }
    \label{fig:GKP dissipative stabilization}
\end{figure*}

Here, we study the experimental feasibility of a highly squeezed GKP qunaught state with $\sigma_{\mathrm{gkp}}^{(\mathrm{dB})} \gtrsim 12$dB. Recall that the GKP squeezing $\sigma_{\mathrm{gkp}}^{(\mathrm{dB})} \gtrsim 12$dB is sufficient for achieving a low logical failure rate of the surface-GKP code with a small surface code distance $3\le d \le 11$ (see \cref{fig:GKPl1AlldAnalog}). There have been many theoretical proposals for preparing a finitely-squeezed GKP state \cite{Gottesman2001_encoding,travaglione2002_preparing,pirandola2004_constructing,pirandola2006_generating,vasconcelos2010_alloptical,Terhal2016_encoding,motes2017_encoding,weigand2018_generating,arrazola2019_machine,Shi2019_fault_tolerant,su2019_conversion,eaton2019_nongaussian,hastrup2019_measurementfree,weigand2020_realizing,hastrup2020_improved,Royer2020_stabilization,Conrad2021_twirling}. In recent years, GKP qubits have been experimentally realized in trapped ion systems \cite{Fluhmann2019_encoding,Fluhmann2020_direct,deNeeve2020_error} as well as in a circuit QED system \cite{CampagneIbarcq_2020_quantum}. The achieved GKP squeezing $\sigma_{\mathrm{gkp}}^{(\mathrm{dB})}$ in these experiments ranges from $5.5$dB to $9.5$dB. Notably, two of the recent experiments \cite{CampagneIbarcq_2020_quantum,deNeeve2020_error} have used an autonomous (or dissipative) stabilization method, namely the Sharpen-and-Trim (ST) scheme \cite{CampagneIbarcq_2020_quantum} and the small-Big-small (sBs) scheme \cite{deNeeve2020_error}, to prepare and stabilize a finitely-squeezed GKP state by using an ancilla qubit. Note also that a recent theoretical work \cite{Royer2020_stabilization} has identified the ST scheme as a first-order Trotter method and proposed two new methods, small-Big-small (sBs) and Big-small-Big (BsB) schemes, by using the second-order Trotter formula (see \cref{fig:GKP dissipative stabilization}; the sBs scheme is also independently discovered in Ref.\ \cite{deNeeve2020_error}).    

In this section, we demonstrate that the BsB scheme is particularly well suited for preparing a highly squeezed GKP qunaught state (with a GKP squeezing $\sigma_{\mathrm{gkp}}^{(\mathrm{dB})} > 10$dB) in the presence of realistic imperfections such as photon loss and ancilla qubit decay and dephasing. We also propose a modification of the existing schemes by using a three-level ancilla instead of a two-level ancilla qubit. More specifically, we show that the stabilization schemes can be made robust against single ancilla decay events by using a three-level ancilla with a properly engineered conditional displcement operation. In particular, we demonstrate that a highly squeezed GKP qunaught state of GKP squeezing $12$dB can in principle be realized in circuit QED systems assuming reasonable experimental parameters. 

\subsection{Dissipative preparation of a GKP qunaught state}
\label{subsection:Dissipative stabilization of a GKP qunaught state}

Given the optimal performance of the square-lattice GKP code in the surface-GKP code as demonstrated in \cref{fig:CompareLambdas}, we focus on stabilizing the square-lattice GKP qunaught state. Let $|\emptyset\rangle\equiv |\emptyset_{\lambda = 1}\rangle$ be the ideal, infinitely-squeezed square-lattice GKP qunaught state. A finitely-squeezed GKP qunaught state (with a Gaussian envelope) is given by $|\emptyset^{\Delta}\rangle \equiv \exp[ -2\Delta^{2}\hat{n} ] |\emptyset\rangle$. Here, $\Delta^{2}$ approximately equals the noise variance $\sigma_{\mathrm{gkp}}^{2}$ of the finitely-squeezed GKP state, i.e., $\sigma_{\mathrm{gkp}}^{2} = \Delta^{2} + \mathcal{O}(\Delta^{4})$ (see \cref{subappendix:Finitely squeezed ancilla GKP states} for more details). Note that the finitely-squeezed GKP qunaught state $|\emptyset^{\Delta}\rangle$ is only approximately stabilized by the stabilizers of the infinitely-squeezed GKP qunaught state $\hat{S}_{q} \equiv \exp[i\sqrt{2\pi}\hat{q}]$ and $\hat{S}_{p} \equiv \exp[-i\sqrt{2\pi}\hat{p}]$. However, as observed in Ref.\ \cite{Royer2020_stabilization}, it is possible to define exact stabilizers of a finitely-squeezed GKP state. For instance, the finitely-squeezed GKP qunaught state $|\emptyset^{\Delta}\rangle$ is exactly stabilized by the following stabilizers: 
\begin{align}
    \hat{S}_{q}^{[\Delta]} &\equiv \exp[ -2\Delta^{2}\hat{n} ]  \hat{S}_{q} \exp[ 2\Delta^{2}\hat{n} ] , 
    \nonumber\\
    \hat{S}_{p}^{[\Delta]} &\equiv \exp[ -2\Delta^{2}\hat{n} ]  \hat{S}_{p} \exp[ 2\Delta^{2}\hat{n} ].   
\end{align}
In two recent experiments \cite{deNeeve2020_error,CampagneIbarcq_2020_quantum}, these stabilizers are stabilized by using the Sharpen-and-Trim (ST) method shown in \cref{fig:GKP dissipative stabilization} (a) and (b). The two new schemes proposed in Ref.\ \cite{Royer2020_stabilization}, namely the sBs and BsB schemes, are shown in \cref{fig:GKP dissipative stabilization} (c) and (d), respectively. Note that unlike phase estimation methods \cite{Terhal2016_encoding,Shi2019_fault_tolerant}, the dissipative schemes in \cref{fig:GKP dissipative stabilization} do not require active monitoring of the ancilla qubit state and real-time feedback control based on the ancilla measurement outcome. Instead, it suffices to simply reset the ancilla qubit to $|+\rangle$ regardless of the ancilla state at the end of the circuit. 

\begin{figure*}
    \centering
    \includegraphics[width=0.8\textwidth]{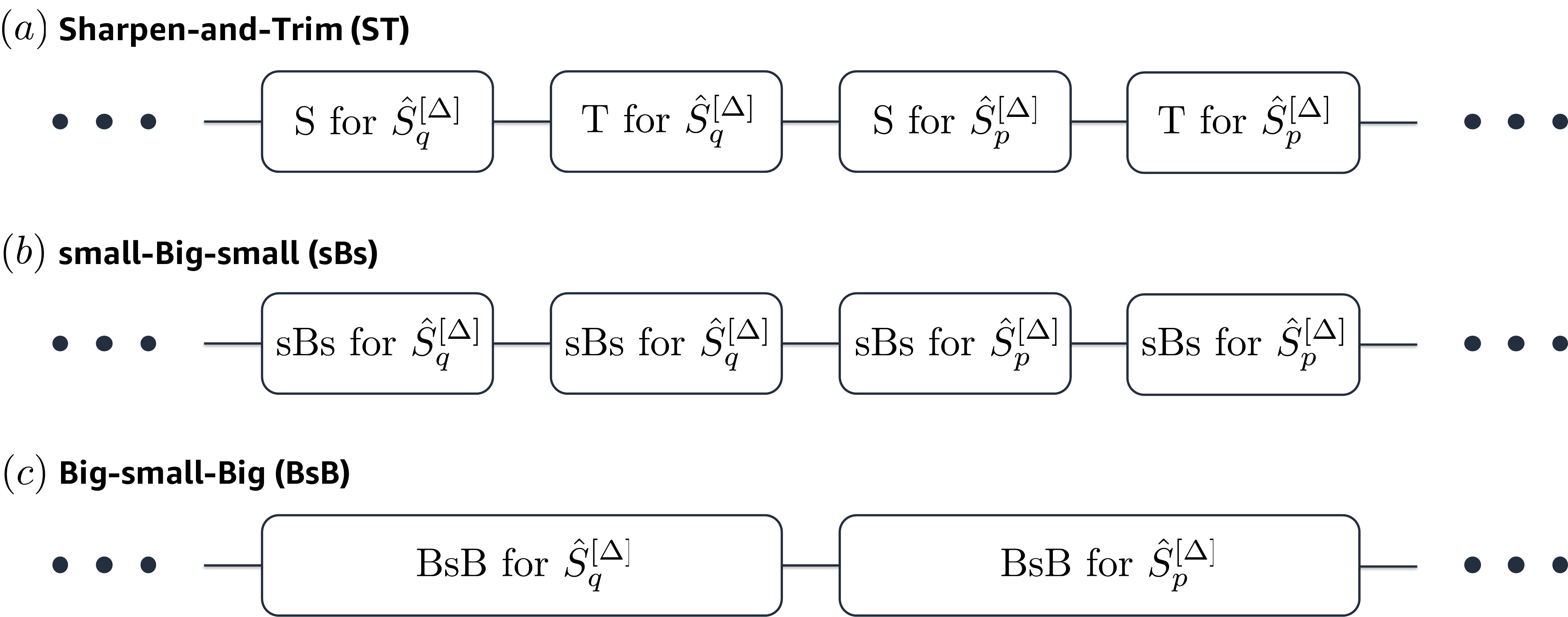}
    \caption{Single round of stabilization of the finitely-squeezed GKP qunaught state by using (a) ST, (b) sBs, and (c) BsB schemes. Note that in the sBs scheme, stabilization of $\hat{S}_{q}^{[\Delta]}$ ($\hat{S}_{p}^{[\Delta]}$) is repeated twice before moving on to the stabilization of $\hat{S}_{p}^{[\Delta]}$ ($\hat{S}_{q}^{[\Delta]}$). We used this specific sequence because we observed that a GKP qunaught state cannot be properly prepared from the vacuum state if the sBs stabilization for each quadrature is not repeated at least twice before moving on to the stabilization of another quadrature. Note also that while S, T, and sBs contain only one big conditional displacement (i.e., $C\hat{D}(\ell)$), BsB contains two big conditional displacements (see \cref{fig:GKP dissipative stabilization}). Thus, each BsB scheme takes twice as long as the other schemes if the conditional displacement is the slowest process, hence the longer width in our schematic diagram.    }
    \label{fig:GKP dissipative stabilization one round}
\end{figure*}

Every scheme in \cref{fig:GKP dissipative stabilization} requires conditional displacement operations $C\hat{D}(\ell)$ and $C\hat{D}(\epsilon)$ (or $C\hat{D}(2\epsilon)$ or $C\hat{D}(\epsilon/2)$), single-qubit rotations $\hat{R}_{\pi/2}$ and $\hat{R}^{\dagger}_{\pi/2}$, and the ancilla qubit reset (to the $|+\rangle$ state). Here, the conditional displacement operator $C\hat{D}(\xi)$ is defined as $C\hat{D}(\xi) \equiv \exp[ -i(\xi_{p}\hat{q}-\xi_{q}\hat{p})\otimes \sigma_{z}/2 ] = \hat{D}(\xi/2)\otimes |1\rangle\langle 1| + \hat{D}(-\xi/2)\otimes |0\rangle\langle 0|$ and the single-qubit rotation $\hat{R}_{\theta}$ is defined as $\hat{R}_{\theta} \equiv [ -i\theta \hat{\sigma}_{x}/2 ]$, where $\sigma_{z} \equiv |0\rangle\langle 0| - |1\rangle\langle 1|$ and $\sigma_{x} \equiv |0\rangle\langle 1| + |1\rangle\langle 0|$. To stabilize $\hat{S}_{q}^{[\Delta]}$ (i.e., position quadrature), we choose $\ell  = \ell_{q} \equiv i\sqrt{2\pi}\cosh(2\Delta^{2})$ and $\epsilon = \epsilon_{q} \equiv \sqrt{2\pi}\sinh(2\Delta^{2})$. Also, to stabilize $\hat{S}_{p}^{[\Delta]}$ (i.e., momentum quadrature), we choose $\ell  = \ell_{p} \equiv \sqrt{2\pi}\cosh(2\Delta^{2})$ and $\epsilon = \epsilon_{p} \equiv -i\sqrt{2\pi}\sinh(2\Delta^{2})$.

To prepare and stabilize a finitely-squeezed GKP qunaught state, we start from the vacuum state $|\hat{n}=0\rangle$ and apply multiple rounds of the disspative stabilization circuits in \cref{fig:GKP dissipative stabilization}. The exact sequence of the stabilization is shown in \cref{fig:GKP dissipative stabilization one round}. Each sequence consists of stabilization of the position quadrature which is followed by stabilization of the momentum quadrature. In the case of the sBs scheme, we observe that a GKP qunaught state cannot be properly prepared from the vacuum state if the position (momentum) stabilization is not repeated at least twice. Thus, we choose to repeat the stabilization of each quadrature twice before moving on to the stabilization of the other quadrature (see \cref{fig:GKP dissipative stabilization one round} (b)). Note also that each BsB scheme contains two big conditional displacements ($\hat{D}(\ell)$), whereas all the other schemes (i.e., S, T, sBs) contain only one big conditional displacement. Thus, if the conditional displacement is the slowest process, each BsB scheme takes twice as long as the other schemes.     

\begin{figure*}
    \centering
    \includegraphics[width=0.9\textwidth]{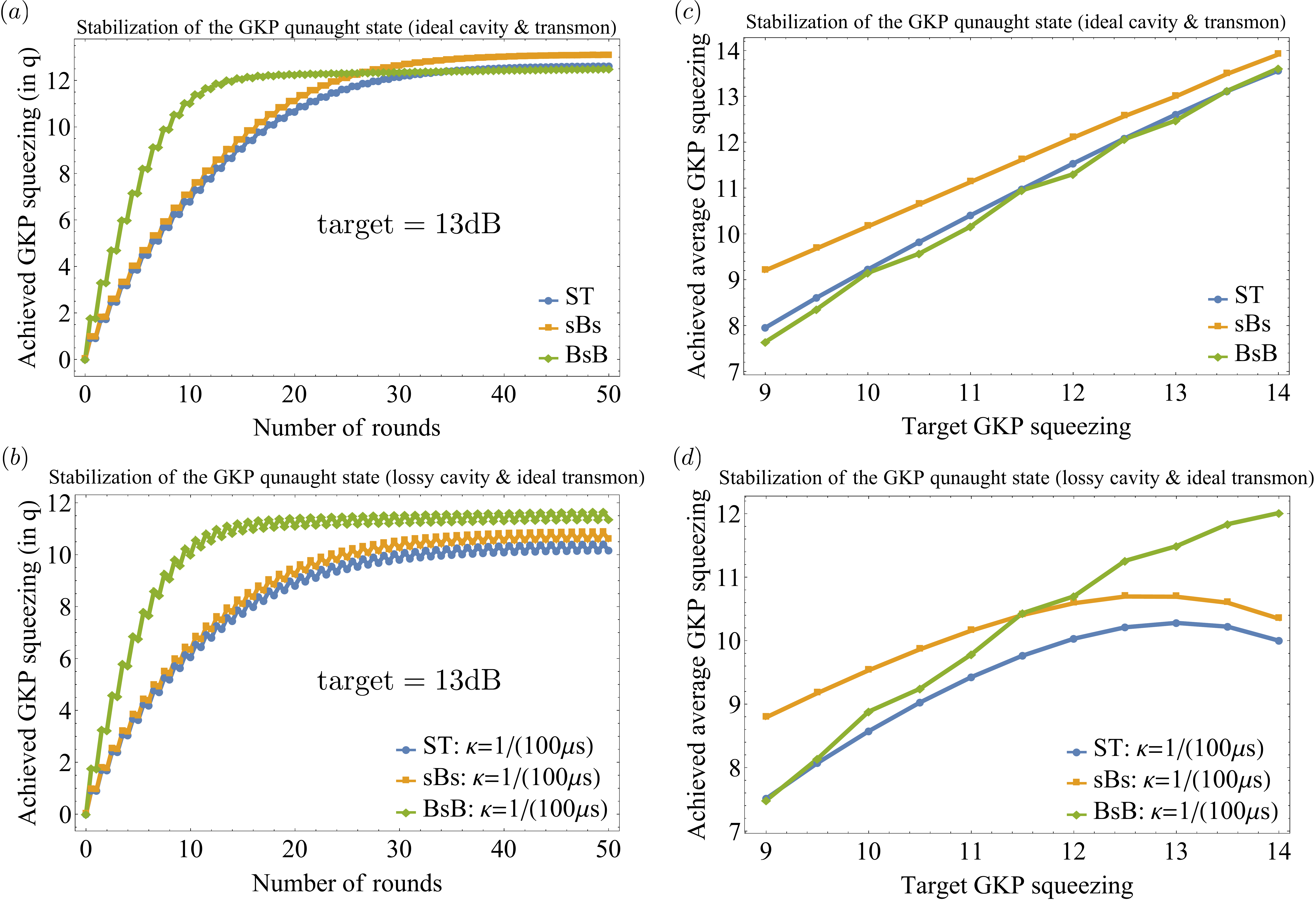}
    \caption{(a), (b) Achieved GKP squeezing $\Delta_{\mathrm{eff},q}^{(\mathrm{dB})}$ in the position quadrature as a function of the number of stabilization rounds, starting from the vacuum state $|\hat{n}=0\rangle$. In both (a) and (b), the target GKP squeezing is chosen to be $\Delta^{(\mathrm{dB})} = 13$dB. The achieved GKP squeezing in the momentum quadrature $\Delta_{\mathrm{eff},p}^{(\mathrm{dB})}$ is essentially the same as $\Delta_{\mathrm{eff},q}^{(\mathrm{dB})}$ except for the fact that $\Delta_{\mathrm{eff},p}^{(\mathrm{dB})}$ temporarily drops when $\Delta_{\mathrm{eff},q}^{(\mathrm{dB})}$ peaks (and vice versa) in the final steady-state oscillation. The saturated values of the GKP squeezing are plotted as a function of the target GKP squeezing in (c) and (d). In (a) and (c), we assume that both the cavity and the ancilla transmon are noiseless. In (b) and (d), we assume that the ancilla transmon is noiseless but the cavity has a photon loss rate $\kappa = 1/(100\mu s)$. Also, we assume that the coupling strength $g$ for the conditional displacement $g\hat{q}\otimes \hat{\sigma}_{z}$ (or $g\hat{p}\otimes \hat{\sigma}_{z}$) is given by $g = 2\pi \times   1\mathrm{MHz}$. For the target GKP squeezings from $9$dB to $12.5$dB, we used $100$-dimensional Hilbert space for the cavity mode. For the target GKP squeezing of $13$dB, $13.5$dB, and $14$dB, we used $120$, $120$, $160$-dimensional Hilbert spaces for the cavity mode, respectively.      }
    \label{fig:Achieved GKP squeezing}
\end{figure*}

Note that we can freely choose the parameter $\Delta$ which determines the size (and thus squeezing) of the output GKP state. Since $\Delta^{2}$ and the noise variance of the finitely-squeezed GKP state $\sigma_{\mathrm{gkp}}^{2}$ are approximately identical to each other in the $\Delta \ll 1$ (or high squeezing) regime, we define the target GKP squeezing $\Delta^{(\mathrm{dB})}$ analogously as in \cref{eq:GKP squeezing}. Hence, given the target GKP squeezing $\Delta^{(\mathrm{dB})}$, we choose $\Delta$ to be 
\begin{align}
    \Delta = \sqrt{ \frac{1}{2}10^{- \Delta^{(\mathrm{dB})} / 10} } .
\end{align}
Then, suppose that the stabilization circuit outputs a state $\hat{\rho}$. Unlike in Ref.\ \cite{Royer2020_stabilization} where the goal was to stabilize a GKP qubit, we only aim to stabilize a finitely-squeezed GKP qunaught state. Hence, there is no notion of logical qubit fidelity, which was the main focus of the study in Ref.\ \cite{Royer2020_stabilization}, because there is no logical information encoded in the GKP qunaught state. Instead, we are only interested in how much the output GKP qunaught state is squeezed in both the position and the momentum quadratures. Thus, similarly as in Refs.\ \cite{Duivenvoorden2017_single_mode,Terhal2020_towards}, we define the effective GKP squeezings (in position and momentum quadratures) of the state $\hat{\rho}$ as follows and use them as the key figure of merit: \begin{align}
    \Delta_{\mathrm{eff},q}^{(\mathrm{dB})} &\equiv 10\log_{10}\Big{(} \frac{ 1/2 }{  \log( 1/|\mathrm{Tr}[ \hat{S}_{q} \hat{\rho} ]| ) / \pi  } \Big{)} , 
    \nonumber\\
    \Delta_{\mathrm{eff},p}^{(\mathrm{dB})} &\equiv 10\log_{10}\Big{(} \frac{ 1/2 }{  \log( 1/|\mathrm{Tr}[ \hat{S}_{p} \hat{\rho} ]| ) / \pi  } \Big{)} , \label{eq:effective GKP squeezing}
\end{align}
Here, $\hat{S}_{q} \equiv \exp[i\sqrt{2\pi}\hat{q}]$ and $\hat{S}_{p} \equiv \exp[-i\sqrt{2\pi}\hat{p}]$. This definition of the GKP squeezing is motivated by the following observation. 
\begin{align}
    &\mathbb{E}_{\xi_{q} \sim \mathcal{N}( 0, \sigma^{2} )} [ \exp[ i\sqrt{2\pi} \xi_{q}  ] ] 
    \nonumber\\
    &= \mathbb{E}_{\xi_{p} \sim \mathcal{N}( 0, \sigma^{2} )} [ \exp[ -i\sqrt{2\pi} \xi_{p}  ] ] = \exp[ -\pi \sigma^{2}  ].   
\end{align}
That is, one can infer the effective noise variance $\sigma_{\mathrm{gkp}}^{2}$ from the expectation values of the stabilizers $\hat{S}_{q}$ and $\hat{S}_{p}$ (of the infinitely-squeezed GKP qunaught state) and then convert them into the unit of decibel as in \cref{eq:effective GKP squeezing}.  

\subsection{Performance of the dissipative preparation methods}
\label{section:Performance of the dissipative preparation methods}

\subsubsection{Ideal ancilla qubit}
\label{subsubsection:Ideal ancilla qubit}

We now show the performance of the three dissipative GKP-state preparation methods (i.e., ST, sBs, and BsB). For now, we consider an ideal case where the ancilla qubit (e.g., transmon) is noiseless. In \cref{fig:Achieved GKP squeezing} (a) and (b), we show the achieved GKP squeezing (in the position quadrature, i.e., $\Delta_{\mathrm{eff},q}^{(\mathrm{dB})}$) as a function of the number of rounds, starting from the vacuum state $|\hat{n}=0\rangle$. In particular, we set the target GKP squeezing to be $\Delta^{(\mathrm{dB})} = 13$dB. With the circuit QED system in mind, we refer to the bosonic mode (where the GKP state is prepared) as a cavity and the ancilla qubit as a transmon. To demonstrate that all three schemes (ST, sBs, BsB) achieve the intended goal, we assume that the cavity mode and the transmon are noiseless in \cref{fig:Achieved GKP squeezing} (a). In this case, after sufficiently many rounds of stabilization, the ST, sBs, and BsB schemes output a finitely-squeezed GKP qunaught state with an effective GKP squeezing of $12.6$dB, $13.0$dB, and $12.5$dB, respectively. Notably, the BsB scheme reaches the steady state configuration with the fewest number of stabilization rounds.

Then, we consider the adverse effects of photon loss in the cavity mode and transmon relaxation and dephasing. In particular, we focus on the effects of photon loss and transmon errors that occur during the conditional displacement operations. That is, while we assume that single-qubit rotations are noiseless and instantaneous, we simulate the conditional displacement operations via a Lindblad master equation 
\begin{align}
    \frac{d\hat{\rho}(t)}{dt} &= -i[\hat{H} , \hat{\rho}(t)] 
    \nonumber\\
    &+ \kappa\mathcal{D}[\hat{a}] \hat{\rho}(t) + \Gamma_{\downarrow}\mathcal{D}[\hat{b}]\hat{\rho}(t) + \Gamma_{\phi}\mathcal{D}[\hat{b}^{\dagger}\hat{b}] \hat{\rho}(t),
\end{align}
where $\hat{H} = g\hat{q}\otimes \hat{\sigma_{z}}$ or $\hat{H} = g\hat{p}\otimes \hat{\sigma_{z}}$ and we assume $g = 2\pi \times 1\mathrm{MHz}$. Here, $\hat{a}$ is the annihilation operator of the cavity mode and $\hat{b}$ is the annihilation operator of the ancilla transmon, i.e., $\hat{b} = |0\rangle\langle 1|$. For now, we assume that the ancilla transmon is free from relaxation and dephasing to understand the limitations imposed by the non-zero photon loss rate $\kappa$. 

In \cref{fig:Achieved GKP squeezing} (b), we take $\kappa  = 1/(100\mu s)$ and $\Gamma_{\downarrow} = \Gamma_{\phi}=0$. In this case, the achieved GKP squeezings in the final steady state configuration are given by $10.2-10.4$dB, $10.6-10.8$dB, and $11.4-11.7$dB for ST, sBs, and BsB schemes, respectively. The temporary drop in the output GKP squeezing (e.g., in the position quadrature) is due to the photon loss during the stabilization of the other quadrature (e.g., the momentum quadrature). Note that the BsB scheme is most resilient against the cavity photon loss as it experiences a decrease in the achieved GKP squeezing by only $1$dB whereas the ST and sBs schemes see a decrease by about $2$dB.

\begin{figure*}
    \centering
    \includegraphics[width=0.95\textwidth]{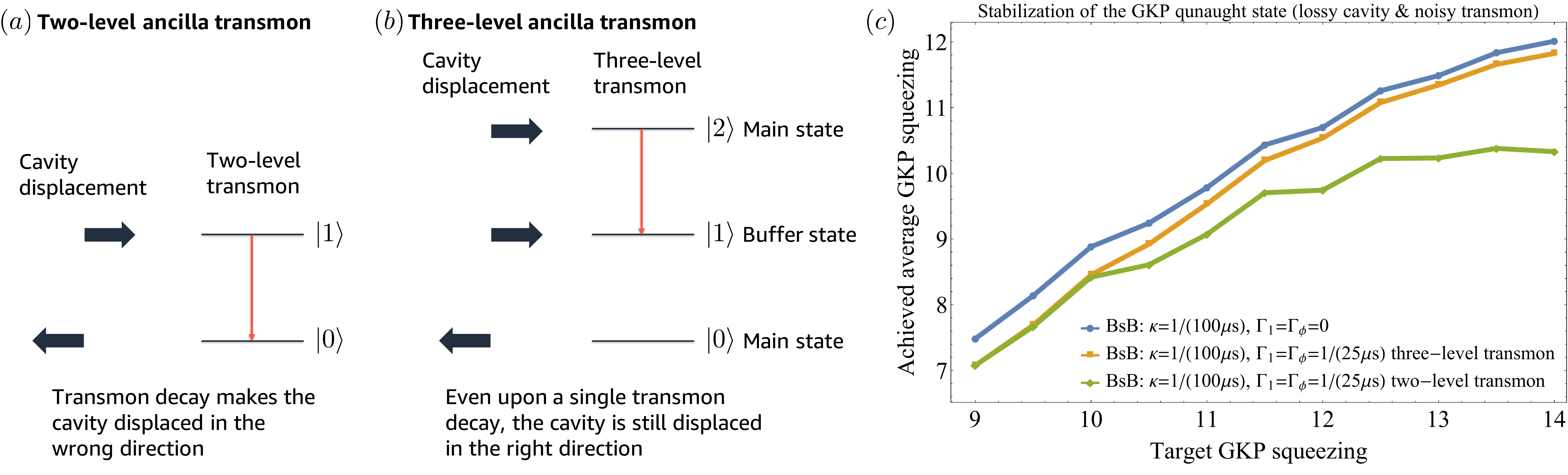}
    \caption{Conditional displacement operation (a) for the two-level ancilla transmon, i.e., $\hat{H} \propto \hat{q}\otimes(-|0\rangle\langle 0| + |1\rangle\langle 1|)$ or $\hat{H} \propto \hat{p}\otimes(-|0\rangle\langle 0| + |1\rangle\langle 1|)$ and (b) for the three-level ancilla transmon, i.e., $\hat{H} \propto \hat{q}\otimes(-|0\rangle\langle 0| + |1\rangle\langle 1| + |2\rangle\langle 2|)$ or $\hat{H} \propto \hat{p}\otimes(-|0\rangle\langle 0| + |1\rangle\langle 1| + |2\rangle\langle 2|)$. In the three-level transmon scheme, the two states $|0\rangle$ and $|2\rangle$ form a main ancilla qubit and $|1\rangle$ is a buffer state for possible single transmon decay events. Note that the conditional displacement operation in the three-level scheme is carefully chosen such that even upon a single transmon decay (from $|2\rangle$ to $|1\rangle$), the cavity mode is still displaced in the right direction by the same amount. (c) Achieved GKP squeezing as a function of the target GKP squeezing for the BsB scheme in the presence of photon loss only (blue line: $\kappa = 1/(100\mu s)$, $\Gamma_{\downarrow} = \Gamma_{\phi} = 0$) and in the presence of photon loss and transmon decay and dephasing (orange and green lines: $\kappa = 1/(100\mu s)$, $\Gamma_{\downarrow} = \Gamma_{\phi} = 1/(25\mu s)$). The three-level transmon scheme (represented by the orange line) nearly saturates the limit set by the photon loss only (blue line) despite strong transmon decay and dephasing. The two-level transmon scheme (green line), on the other hand, achieves a significantly lower GKP squeezing compared to the case where the transmon is assumed to be noiseless as well as to the case where the third level of the transmon is also utilized.         }
    \label{fig:three level ancilla transmon for GKP stabilization}
\end{figure*}

In \cref{fig:Achieved GKP squeezing} (c) and (d), we vary the target GKP squeezing from $9$dB to $14$dB and study the saturated value of the GKP squeezing as a function of the target GKP squeezing. Simiarly as in \cref{fig:Achieved GKP squeezing} (a), we assume that both cavity and transmon are noiseless in \cref{fig:Achieved GKP squeezing} (c). In this case, not so surprisingly, the achieved GKP squeezing increases as we increase the target GKP squeezing. This is not necessarily the case, however, when the cavity mode suffers from photon loss. As shown in \cref{fig:Achieved GKP squeezing} (d), if the cavity photon loss rate is given by $\kappa = 1/(100\mu s)$ (which should be compared with $g  = 2\pi \times 1\mathrm{MHz}$), the achieved GKP squeezing peaks at a certain value of the target GKP squeezing. For instance, the maximum achievable GKP squeezing with the ST scheme is given by $10.3$dB which is achived when the target GKP squeezing is $13$dB. Similarly, the optimal performance of the sBs scheme is achieved when the target GKP squeezing is $12.5$dB with which a GKP squeezing of $10.7$dB is achieved. The existence of a peak is due to the fact that, given the same coupling strength $g$ for the conditional displacement operation, the effective stabilization rate decreases as we increase the target GKP squeezing (see Ref.\ \cite{Royer2020_stabilization} for more details), whereas the photon loss rate remains constant.

In the case of the BsB scheme, the achieved GKP squeezing continues to increase as we increase the target GKP squeezing from $9$dB to $14$dB. In particular, a GKP squeezing of $12.0$dB can be achieved by setting the target GKP squeezing to be $14$dB. We expect, however, that the achieved GKP squeezing will eventually peak at a certain value of the target GKP squeezing just like in the case of the ST and sBs schemes for the reason described above. 

The result in \cref{fig:Achieved GKP squeezing} (d) has an important implication: even if the ancilla transmon is assumed to be noiseless, the maximum achievable GKP squeezing is eventually limited by the photon loss rate $\kappa$ in proportion to the coupling strength $g$ for the conditional displacement operations. Hence, regardless of whether we mitigate the ancilla transmon errors (which we will discuss shortly), the GKP squeezing cannot be made arbitrarily large under a non-vanishing photon loss rate. Nevertheless, the BsB scheme is shown to be most robust against photon loss errors and achieves a GKP squeezing of $12$dB assuming realistic parameters for the coupling strength $g = 2\pi \times 1\mathrm{MHz}$ and the photon loss rate $\kappa = 1/(100\mu s)$. Thus, we mainly focus on the BsB scheme from now on.

\subsubsection{Noisy ancilla qubit}
\label{subsubsection:Noisy ancilla qubit}

We now consider the adverse effects due to transmon decay and dephasing. As previously noted in Ref.\ \cite{Royer2020_stabilization}, ancilla transmon dephasing does not limit the performance of all three stabilization methods (i.e., ST, sBs, BsB methods). This is because the transmon dephasing commutes with the conditional displacement operations and hence does not cause any undesired displacement in the cavity mode. However, transmon decay can significantly compromise the performance of the GKP state stabilization methods. This is because transmon decay in the middle of a conditional displacement operation can cause a large undesirable shift to the cavity mode. That is, if the transmon decays from $|1\rangle$ to $|0\rangle$ during a conditional displacement (generated by $\hat{H} \propto \hat{q}\otimes (-|0\rangle\langle 0| + |1\rangle\langle 1|)$ or $\hat{H} \propto \hat{p}\otimes (-|0\rangle\langle 0| + |1\rangle\langle 1|)$), the cavity state is displaced in wrong direction, possibly by a large amount depending on when the transmon decayed (see \cref{fig:three level ancilla transmon for GKP stabilization} (a)). Indeed, as shown in \cref{fig:three level ancilla transmon for GKP stabilization}, if the transmon decay and dephasing rates are given by $\Gamma_{\downarrow} = \Gamma_{\phi} = 1/(25\mu s)$ and the cavity photon loss rate is $\kappa = 1/(100\mu s)$, the achievable GKP squeezing is significantly lower ($\le 10.4$dB) compared to the case when the cavity photon loss rate remains the same but the transmon is assumed to be noiseless ($>12$dB).

Here, to mitigate the adverse effects of transmon decay, we propose to use the third level of the ancilla transmon. That is, we propose to use the ground state $|0\rangle$ and the second excited state $|2\rangle$ of the transmon as the main ancilla qubit basis states. Thus, for instance, the ancilla transmon is initialized to a state $|+'\rangle  = (|0\rangle + |2\rangle)/\sqrt{2}$ instead of $|+\rangle  = (|0\rangle + |1\rangle)/\sqrt{2}$. Also, the single-qubit rotation $\hat{R}_{\theta} = [ -i\theta (|0\rangle\langle 1| + |1\rangle\langle 0|)/2 ]$ is replaced by $\hat{R}'_{\theta} = [ -i\theta (|0\rangle\langle 2| + |2\rangle\langle 0|)/2 ]$, and the annihilation operator of the ancilla transmon is replaced by $\hat{b} = |0\rangle\langle 1| + \sqrt{2}|1\rangle\langle 2|$ (note that we are thus assuming the second excited state of the transmon decays twice faster than the first excited state). The key feature of this three-level ancilla scheme is to use the first excited state $|1\rangle$ of the transmon as a buffer state for possible single transmon decay events (from $|2\rangle$ to $|1\rangle$). Most importantly, to make sure that the cavity mode is displaced in the right direction by the same amount even upon a single transmon decay, we carefully engineer the conditional displacement operation and propose to realize it via the following Hamiltonian 
\begin{align}
    \hat{H} &\propto \hat{q} \otimes ( -|0\rangle\langle 0| + |1\rangle\langle 1| + |2\rangle\langle 2|  ),   \label{eq:conditional displacement three level error transparent}
\end{align} 
or $\hat{H} \propto \hat{p} \otimes ( -|0\rangle\langle 0| + |1\rangle\langle 1| + |2\rangle\langle 2|  )$. In this case, even if the second excited state $|2\rangle$ of the transmon decays to its first excited state $|1\rangle$, the above Hamiltonian still generates the same displacement in the cavity mode. Thus, regardless of when the transmon decays from $|2\rangle$ to $|1\rangle$, the cavity mode is still displaced by $\ell_{q}$ or $\ell_{p}$ (relative to the case where the transmon is in the ground state $|0\rangle$) which is approximately equal to $\sqrt{2\pi}$, hence a trivial shift to the square-lattice GKP qunaught state. However, in a much less likely event where the transmon experiences two decay events ($|2\rangle \rightarrow |1\rangle \rightarrow |0\rangle$), the cavity mode may be shifted by an undesirable amount, leading to an increased noise variance in the output GKP qunaught state. Note that the design principle behind the interaction in \cref{eq:conditional displacement three level error transparent} is also related to the ideas of error transparency \cite{Vy2013_error_transparent,Kapit2018_error_transparent,Rosenblum2018,Reinhold2020_error_corrected} and path independence \cite{Ma2020_path_independent}.    

The effectiveness of the three-level transmon scheme is demonstrated in \cref{fig:three level ancilla transmon for GKP stabilization} (c). As indicated by the orange line, despite the strong transmon decay and dephasing rates $\gamma_{\downarrow} = \Gamma_{\phi} = 1/(25\mu s)$, the three-level transmon scheme nearly achieves as high GKP squeezing as in the case where the transmon is assumed to be noiseless (i.e., blue line). In particular, while the two-level transmon scheme can only achieve a GKP squeezing of $10.4$dB at most, the three-level scheme can achieve a GKP squeezing of $11.8$dB choosing the target GKP squeezing to be $14$dB. Moreover, it can potentially realize a higher GKP squeezing by further increasing the target GKP squeezing. 

Thus, we show that a highly squeezed GKP qunaught state of GKP squeezing $\sigma_{\mathrm{gkp}}^{(\mathrm{dB})} \gtrsim 12$dB can in principle be achieved via the BsB scheme with reasonable experimental parameters $g = 2\pi \times 1\mathrm{MHz}$, $\kappa = 1/(100\mu s)$, and $\Gamma_{\downarrow} = \Gamma_{\phi} = 1/(25\mu s)$ by using the three-level transmon scheme. Note that if the transmon relaxation time is longer, the third level of the transmon may not be needed and the two-level transmon scheme may suffice to reach the ultimate limit set by $\kappa/g$. For instance, if the photon loss rate and the transmon dephasing rates remain the same ($\kappa = 1/(100\mu s)$ and $\Gamma_{\phi} = 1/(25\mu s)$) but the transmon decay rate is smaller ($\Gamma_{\phi} = 1/(100\mu s)$), the two-level transmon scheme achieves a GKP squeezing of $11.5$dB, close to the limit $12.0$dB set by the photon loss, by choosing the target GKP squeezing to be $14$dB.    

\section{Discussion and outlook}
\label{section:Discussion and outlook}

In this work, we delivered three main results: first, we introduced a maximum likelihood decoding for correcting shift errors in two-qubit gates between GKP qubits and showed that it significantly outperforms the simple closest-integer decoding scheme (see \cref{tab:CNOT logical failure rates}). Secondly, using space-time correlated edges in the matching graphs of the surface code decoder, we carefully took into account all types of errors arising from every error-corrected two-GKP-qubit gates throughout the full syndrome history. By doing so, we were able to achieve a low logical failure rate of the surface-GKP code with only moderate hardware requirements and a reasonable GKP squeezing $\sigma_{\mathrm{gkp}}^{(\mathrm{dB})} \gtrsim 12$dB (see \cref{fig:GKPl1AlldAnalog} and \cref{tab:overhead comparison}). Lastly, we demonstrated that a highly-squeezed GKP state of GKP squeezing $\sigma_{\mathrm{gkp}}^{(\mathrm{dB})} \gtrsim 12$dB can in principle be realized by a dissipative stabilization method, namely the Big-small-Big method, with reasonable experimental parameters. In particular, we showed that the ancilla decay errors can be effectively mitigated through a suitably engineered three-level ancilla in the dissipative stabilization scheme.        

Several remarks are in order. Recall that throughout the analysis of the surface-GKP code, we made a twirling approximation and treated the shift errors due to finite GKP squeezing as incoherent random shift errors. This is mostly for simulation purposes and it is not desirable to physically implement the shift twirling. This is because the shift twirling increases the energy of an encoded GKP state, making it susceptible to uncontrolled non-linear interactions. Thus, to have a refined understanding of the realistic, finitely-squeezed GKP qubits, one needs to analyze them without making the twirling approximation and instead assuming coherent shift errors. A recently proposed subsystem decomposition method \cite{Pantaleoni2020_modular} has shown to be useful for analyzing finitely-squeezed GKP states exactly \cite{Tzitrin2020_progress,Wan2020_memory_assisted,Pantaleoni2021_subsystem}. In this work, however, we have not performed such an analysis partly because one would then get a generic noise channel for a GKP qubit after each GKP error correction. In this case, the Gottesman-Knill simulation methods \cite{Gottesman1998_heisenberg,Aaronson2004_improved} no longer apply in the analysis of the surface code. As such, large-distance surface codes could not be simulated efficiently using the exact noise model for finitely squeezed GKP states. Nevertheless, analyzing the effects of the coherence is still an important task and we leave it as a future work. Related to this, we remark that recent work, which uses a teleportation-based GKP error correction protocol which is similar to the one presented in this manuscript, has demonstrated that approximate results based on shift twirling agree well with the exact numerical results. The results apply to the single GKP qubit setting \cite{Hillmann2021_performance}.   

Recall that we have neglected photon loss and heating since finite squeezing of the GKP qunaught states is the dominant source of error (at least in the near term). The addition of other noise sources such as photon loss and heating will modify the shift distribution and correlation structure. Hence, maximum-likelihood decoding for two-GKP-qubit gates should be adapted to the new shift distribution. Similarly, edge weights in the surface-code matching graphs should also be adjusted to account for modified conditional probabilities of various error types. 

Note also that in this work we focused on fault-tolerant quantum error correction for building a logical quantum memory and have not discussed schemes for universal fault-tolerant quantum computation. Since we conclude that biasing the noise of the GKP qubits does not offer significant advantages in reducing the logical error rate (especially for the data qubits which will be used for carrying out the computation), the Toffoli magic state preparation schemes \cite{Chamberland2020_building} and piece-wise fault-tolerant Toffoli gates \cite{Guillaud2019_repetition,Guillaud2020_error} that are tailored to noise-biased qubits would be sub-optimal for GKP encoded qubits. A more promising avenue would be to pursue the more widely used $|H\rangle$-type magic states \cite{Bravyi2005_universal}. In doing so, we can take advantage of the fact every operation between GKP qubits is error corrected and thus comes with extra analog information that can be used to determine the reliability of the gate. Hence, an interesting research direction is to see if the direct, fault-tolerant $|H\rangle$-type magic state preparation schemes in Ref.\ \cite{Chamberland2019_fault_tolerant,Chamberland2020_very} (which use flag-qubit techniques \cite{CR17v1,CR17v2,CB17,TCD18Flag,ChamberlandPRX,CKYZ20,ReichardtFlag18,ChaoAnyFlag20} and redundant ancilla encoding) can be combined with the GKP qubits to significantly reduce the resource overheads for fault-tolerant quantum computing compared to the cases where bare two-level qubits are used. We also remark that a different decoder other than the MWPM decoder may be used in analyzing such fault-tolerant computing schemes to either reduce logical failure rates further (e.g., by using a tensor-network decoder \cite{Bravyi2014_efficient,Chubb2018_statistical}) or to speed up the decoding at the expense of decreased performance (e.g., by using a Union Find decoder \cite{Delfosse2017_almost_linear,Newman2020_generating,Huang2020_fault_tolerant}).

Regarding the experimental realization of the GKP qubits, we emphasize that for the scope of this work, we only focused on achieving a high GKP squeezing $\sigma_{\mathrm{gkp}}^{(\mathrm{dB})} \gtrsim 12$dB and have not considered how long it takes to prepare such a highly-squeezed GKP state. In particular, the dissipative stabilization schemes, including the Big-small-Big scheme, require tens of stabilization rounds to prepare a GKP state from the vaccum state (corresponding to a preparation time $\gtrsim 10\mu s$ assuming the coupling strength of the conditional displacement is $2\pi\times 1\mathrm{MHz}$). Such long preparation times can potentially be problematic because it increases the idling time and hence makes the data qubits of the surface-GKP code susceptible to photon loss. Alternatively, idling times can be reduced by using more than two modes for each GKP qubit in a (teleportation-based) GKP error correction protocol. This is due to the fact that a subset of modes can begin preparing a GKP qunaught state well before it needs to be supplied to the circuit. In this case, the total number of oscillator modes will be larger than simply three times the number of GKP qubits in the surface-GKP code. As such, reducing the GKP state preparation time is important to maintain the hardware resource estimates obtained in this work.   

One possible way to speed up the GKP state preparation protocol is to increase the coupling strength of the conditional displacement operations by driving the cavity \cite{CampagneIbarcq_2020_quantum} or transmon qubit \cite{Touzard2019_gated} more strongly. The state preparation protocol can also be accelerated by starting from a squeezed state in one quadrature (e.g., $\hat{q}$) \cite{Fluhmann2019_encoding,Fluhmann2020_direct,deNeeve2020_error} instead of the vacuum state so that we can mostly focus on squeezing the other quadrature (e.g., $\hat{p}$). Also, it is important to realize that the dissipative stabilization method we considered in this work (e.g., the Big-small-Big scheme) does not utilize information gained by transmon as it is a completely autonomous process (see \cref{fig:GKP dissipative stabilization}). However, there are other methods, namely phase estimation methods, that do take advantage of information gained by transmon \cite{Terhal2016_encoding,Shi2019_fault_tolerant}. In this case, fewer rounds of measurements will be needed to achieve a high GKP squeezing with the help of feedback control based on the transmon measurement outcomes. As shown in Ref.\ \cite{Shi2019_fault_tolerant}, however, transmon measurement errors should be carefully post-processed via a majority-voting type of scheme. This is due to the fact that an incorrect feedback based on a single faulty measurement outcome can cause a large shift error and hence introduce extra noise thereby destroying the desired fault-tolerant properties of the state-preparation protocol. 

We note that a long state preparation time $\gtrsim 5\mu s$ will not be a limiting factor if the cavity lifetime is given by $1ms$ instead of $100\mu s$: at the GKP squeezing $\sigma_{\mathrm{gkp}}^{(\mathrm{dB})} = 12$dB, the noise variance due to the finite GKP squeezing is given by $2\sigma_{\mathrm{gkp}}^{2} = 0.063$. If the state preparation time is $t_{\mathrm{prep}} = 5\mu s$ and the cavity lifetime is $\kappa = 100\mu s$, an additional noise variance $\kappa t_{\mathrm{prep}} = 0.05$ (comparable to $2\sigma_{\mathrm{gkp}}^{2} = 0.063$) is introduced. However, if the cavity lifetime is $1 ms$, the extra noise variance is reduced to $0.005$ and is not appreciable compared to the noise variance due to the finite GKP squeezing. 

Note also that we only considered preparation of a GKP qunaught state via an auxiliary qubit. Thus if we want to prepare a GKP-Bell state, which is needed for the teleportation-based GKP error correction, two GKP qunaught states have to be prepared sequentially using the same auxiliary qubit and then a beam-splitter interaction should be applied to them to turn them into a GKP-Bell state. In principle, this procedure can be simplified as we can use an auxiliary qubit to directly prepare a GKP-Bell state. More specifically, instead of measuring the two stabilizers of a GKP qunaught state for each mode, we can measure the four stabilizers of a GKP-Bell state jointly on the two modes. We leave a more detailed study of such a direct GKP-Bell state preparation as a future work.    

Lastly, we remark that reducing the strength of non-linear interactions is especially crucial when working with highly-squeezed GKP states since these highly-energetic states are susceptible to non-linear interactions such as the self-Kerr and cross-Kerr interactions. In circuit QED systems, non-linear interactions have been minimized by coupling the cavity mode (where the GKP state resides in) much more weakly to an ancilla transmon than in the standard regime \cite{CampagneIbarcq_2020_quantum}. The desired conditional displacement operation is then selectively enhanced either by driving the cavity mode \cite{CampagneIbarcq_2020_quantum} or the transmon \cite{Touzard2019_gated}. To go beyond preparing a single GKP state and be able to manipulate many GKP qubits, it is also essential to couple cavity modes via Gaussian operations. In circuit QED systems, Gaussian operations between cavity modes have been realized by using a four-wave mixing element \cite{Gao2018_programmable,Zhang2019_engineering,Gao2019_entanglement}. In the context of manipulating GKP qubits, however, it is more desirable to use a three-wave mixing element (e.g., SNAIL \cite{Frattini2017} or ATS \cite{Lescanne2020_exponential}) such that higher-order non-linear interactions are minimized. A holistic and detailed architecture analysis incorporating all these considerations, including the effects of photon loss, is left as a future work.         

\section*{Acknowledgments}


We would like to acknowledge the AWS EC2 resources which were used for part of the simulations performed in this work.

\appendix

\section{GKP error correction}
\label{appendix:GKP error correction}

In this appendix, we derive \cref{eq:ideal output state of the GKP error correction Steane,eq:ideal output state of the GKP error correction teleportation}, i.e., ideal post-measurement state after a Steane-type and a teleportation-based GKP error correction, respectively. We also discuss how the envelope operators of finitely-squeezed ancilla GKP states propagate during a teleportation-based GKP error correction. 


\subsection{Steane-type GKP error correction}
\label{subappendix:Steane-type GKP error correction}

The Steane-type GKP error correction scheme can be decomposed into two stages. In the first stage, the GKP qubit in the second mode (prepared in the $|+_{\lambda}\rangle \propto \sum_{n \in\mathbb{Z} } |\hat{q}_{2} = n\sqrt{\pi}\lambda\rangle$) measures the position of the data mode modulo $\sqrt{\pi}\lambda$. Let $|\psi\rangle = \int dq_{1} \psi(q_{1})|\hat{q}_{1} = q_{1}\rangle$ be the input state in the data mode. After applying the SUM gate $\mathrm{SUM}_{1 \rightarrow 2} \equiv \exp[-i\hat{q}_{1}\hat{p}_{2}]$, we get 
\begin{align}
    &|\psi_{1}\rangle \equiv \mathrm{SUM}_{1 \rightarrow 2} |\psi\rangle |+_{\lambda}\rangle 
    \nonumber\\
    &\propto \mathrm{SUM}_{1 \rightarrow 2}  \int dq_{1} \psi(q_{1})|\hat{q}_{1} = q_{1}\rangle \sum_{n \in\mathbb{Z} } |\hat{q}_{2} = n\sqrt{\pi}\lambda\rangle
    \nonumber\\
    &= \sum_{n \in\mathbb{Z} } \int dq_{1}\psi(q_{1})|\hat{q}_{1} = q_{1}\rangle|\hat{q}_{2} = n\sqrt{\pi}\lambda + q_{1}\rangle, 
\end{align}
where we used 
\begin{align}
    \mathrm{SUM}_{1 \rightarrow 2} |\hat{q}_{1} = q_{1}\rangle|\hat{q}_{2} = q_{2}\rangle = |\hat{q}_{1} = q_{1}\rangle|\hat{q}_{2} = q_{2} + q_{1} \rangle .     
\end{align}
Then, conditioned on measuring $q_{m}$ in the second mode via a position homodyne measurement, we get 
\begin{align}
    &|\psi'\rangle \equiv \langle \hat{q}_{2} = q_{m} |\psi_{1}\rangle 
    \nonumber\\
    &\propto \sum_{n \in\mathbb{Z} } \int dq_{1}\psi(q_{1})|\hat{q}_{1} = q_{1}\rangle \delta\Big{(} q_{m} - (n\sqrt{\pi}\lambda + q_{1}) \Big{)}  
    \nonumber\\
    &= \sum_{n \in\mathbb{Z} }  \psi( q_{m} - n\sqrt{\pi}\lambda )|\hat{q}_{1} = q_{m} - n\sqrt{\pi}\lambda \rangle  
    \nonumber\\
    &= \sum_{n \in\mathbb{Z} } |\hat{q}_{1} = q_{m} - n\sqrt{\pi}\lambda \rangle  \langle \hat{q}_{1} = q_{m} - n\sqrt{\pi}\lambda|  \psi\rangle 
    \nonumber\\
    &= \hat{\Pi}_{q,\lambda}(q_{m}) |\psi\rangle, \label{eq:Steane error correction after position measurement}
\end{align}
where the projection operator $\hat{\Pi}_{q,\lambda}(q_{m})$ is given by
\begin{align}
    \hat{\Pi}_{q,\lambda}(q_{m}) &= \sum_{n \in\mathbb{Z} } |\hat{q}_{1} = q_{m} + n\sqrt{\pi}\lambda \rangle  \langle \hat{q}_{1} = q_{m} + n\sqrt{\pi}\lambda| ,  
\end{align}
where the sign of the integer $n$ is flipped for a cosmetic reason without changing the results. Note that $\hat{\Pi}_{q,\lambda}(q_{m})$ projects an input state to the space of states that have $\hat{q}_{1} = q_{m}$ modulo $\sqrt{\pi}\lambda$. 

In the second stage, the GKP qubit in the third mode (prepared in $|0_{\lambda}\rangle \propto \sum_{n \in \mathbb{Z}} |\hat{p}_{3} = n\sqrt{\pi}/\lambda\rangle$) measures the momentum of the data mode modulo $\sqrt{\pi}/\lambda$. This time, we expand the state in the data mode in the momentum basis, i.e., $|\psi'\rangle = \int dp_{1} \psi'(p_{1})|\hat{p}_{1} = p_{1}\rangle$. Then, after applying the inverse SUM gate $\mathrm{SUM}_{3\rightarrow 1}^{-1} \equiv \exp[ i\hat{p}_{1}\hat{q}_{3} ]$, we get 
\begin{align}
    &|\psi'_{1}\rangle \equiv \mathrm{SUM}_{3\rightarrow 1}^{-1} |\psi'\rangle |0_{\lambda}\rangle
    \nonumber\\
    &\propto \mathrm{SUM}_{3\rightarrow 1}^{-1} \int dp_{1} \psi'(p_{1})|\hat{p}_{1} = p_{1}\rangle \sum_{n \in \mathbb{Z}} |\hat{p}_{3} = n\sqrt{\pi}/\lambda\rangle 
    \nonumber\\
    &= \sum_{n \in \mathbb{Z}} \int dp_{1} \psi'(p_{1}) |\hat{p}_{1} = p_{1}\rangle |\hat{p}_{3} = n\sqrt{\pi}/\lambda + p_{1} \rangle  , 
\end{align}
where we used 
\begin{align}
    \mathrm{SUM}_{3 \rightarrow 1}^{-1} |\hat{p}_{1} = p_{1}\rangle|\hat{p}_{3} = p_{3}\rangle = |\hat{p}_{1} = p_{1}\rangle|\hat{p}_{3} = p_{3} + p_{1} \rangle .  
\end{align}
Thus, similarly as above, the post-measurement state upon measuring $\hat{p}_{3} = p_{m}$ is given by 
\begin{align}
    |\psi''\rangle \equiv \langle \hat{p}_{3} = p_{m} |\psi'_{1}\rangle &\propto  \hat{\Pi}_{p,\lambda}(p_{m}) |\psi'\rangle, \label{eq:Steane error correction after momentum measurement}
\end{align}
where 
\begin{align}
    \hat{\Pi}_{p,\lambda}(p_{m}) &\equiv \sum_{n \in\mathbb{Z} } |\hat{p}_{1} = p_{m} + n\sqrt{\pi}/\lambda \rangle  \langle \hat{p}_{1} = p_{m} + n\sqrt{\pi}/\lambda|  
\end{align}
is the projection operator to the space of states that have $\hat{p}_{1} = p_{m}$ modulo $\sqrt{\pi}/\lambda$. 

Combining \cref{eq:Steane error correction after position measurement,eq:Steane error correction after momentum measurement}, we find that the ideal output state after the Steane-type GKP error correction is given by 
\begin{align}
    |\psi(q_{m},p_{m})\rangle_{\mathrm{Steane}} &\propto \hat{\Pi}_{p,\lambda}(p_{m}) \hat{\Pi}_{q,\lambda}(q_{m}) |\psi\rangle . 
\end{align}
Thus, to derive \cref{eq:ideal output state of the GKP error correction Steane}, it remains to show 
\begin{align}
    \hat{\Pi}_{p,\lambda}(p_{m}) \hat{\Pi}_{q,\lambda}(q_{m}) \propto \hat{D}(q_{m}+ip_{m})\hat{\Pi}_{\lambda}\hat{D}(-q_{m}-ip_{m}). \label{eq:Steane projection}
\end{align}
Note that 
\begin{align}
    &\hat{\Pi}_{p,\lambda}(p_{m}) \hat{\Pi}_{q,\lambda}(q_{m}) 
    \nonumber\\
    &= \sum_{n_{p} \in\mathbb{Z} } |\hat{p} = p_{m} + n_{p}\sqrt{\pi}/\lambda \rangle  \langle \hat{p} = p_{m} + n_{p}\sqrt{\pi}/\lambda|  
    \nonumber\\
    &\quad \times \sum_{n_{q} \in\mathbb{Z} } |\hat{q} = q_{m} + n_{q}\sqrt{\pi}\lambda \rangle  \langle \hat{q} = q_{m} + n_{q}\sqrt{\pi}\lambda| 
    \nonumber\\
    &\propto \sum_{n_{q}, n_{p} \in\mathbb{Z} } \hat{D}(ip_{m}) |\hat{p} =  n_{p}\sqrt{\pi}/\lambda \rangle \langle \hat{q} = n_{q}\sqrt{\pi}\lambda| \hat{D}(-q_{m}) 
    \nonumber\\
    &\qquad\quad  \times \exp[ -i( p_{m} + n_{p}\sqrt{\pi}/\lambda ) ( q_{m} + n_{q}\sqrt{\pi}\lambda ) ]  , 
\end{align}
where we used $\langle \hat{p} = p | \hat{q} = q\rangle \propto \exp[-ipq]$ and the convention $\hat{D}(\xi_{q} + i\xi_{p}) \equiv \exp[ i(\xi_{p}\hat{q} - \xi_{q}\hat{p}) ]$ (see the caption of \cref{fig:GKP error correction}). Carrying on, we find 
\begin{align}
    &\hat{\Pi}_{p,\lambda}(p_{m}) \hat{\Pi}_{q,\lambda}(q_{m}) 
    \nonumber\\
    &\propto \sum_{n_{q}, n_{p} \in\mathbb{Z} } (-1)^{n_{q}n_{p}} \hat{D}(ip_{m}) \exp[ -iq_{m}\hat{p} ] |\hat{p} =  n_{p}\sqrt{\pi}/\lambda \rangle 
    \nonumber\\
    &\qquad\quad \times \langle \hat{q} = n_{q}\sqrt{\pi}\lambda| \exp[ -ip_{m}\hat{q} ] \hat{D}(-q_{m}) 
    \nonumber\\
    &\propto \hat{D}(q_{m} + ip_{m}) 
    \nonumber\\
    &\quad\times \sum_{n_{q}, n_{p} \in\mathbb{Z} } (-1)^{n_{q}n_{p}} |\hat{p} =  n_{p}\sqrt{\pi}/\lambda \rangle \langle \hat{q} = n_{q}\sqrt{\pi}\lambda| 
    \nonumber\\
    &\quad \times \hat{D}(-q_{m} - ip_{m})  . 
\end{align}
Lastly, since 
\begin{align}
    &\sum_{n_{q}, n_{p} \in\mathbb{Z} } (-1)^{n_{q}n_{p}} |\hat{p} =  n_{p}\sqrt{\pi}/\lambda \rangle \langle \hat{q} = n_{q}\sqrt{\pi}\lambda|   
    \nonumber\\
    &= \underbrace{ \sum_{n_{p} \in\mathbb{Z} }|\hat{p} =  n_{p}\sqrt{\pi}/\lambda \rangle}_{ = \sqrt{2}|0_{\lambda}\rangle } \underbrace{  \sum_{n_{q}\in 2\mathbb{Z} }\langle \hat{q} = n_{q}\sqrt{\pi}\lambda|   }_{ =\langle 0_{\lambda}| }
    \nonumber\\
    &\quad +  \underbrace{  \sum_{n_{p} \in\mathbb{Z} } (-1)^{n_{p}} |\hat{p} =  n_{p}\sqrt{\pi}/\lambda \rangle }_{ = \sqrt{2} |1_{\lambda}\rangle }  \underbrace{ \sum_{n_{q}\in 2\mathbb{Z} + 1 }\langle \hat{q} = n_{q}\sqrt{\pi}\lambda|   }_{ =\langle 1_{\lambda}|  }
    \nonumber\\
    &\propto |0_{\lambda}\rangle\langle 0_{\lambda}| + |1_{\lambda}\rangle\langle 1_{\lambda}| \equiv \hat{\Pi}_{\lambda},  
\end{align}
\cref{eq:Steane projection} and hence \cref{eq:ideal output state of the GKP error correction Steane} follow.  
  
\begin{figure*}[t!]
    \centering
    \includegraphics[width = 0.7\textwidth]{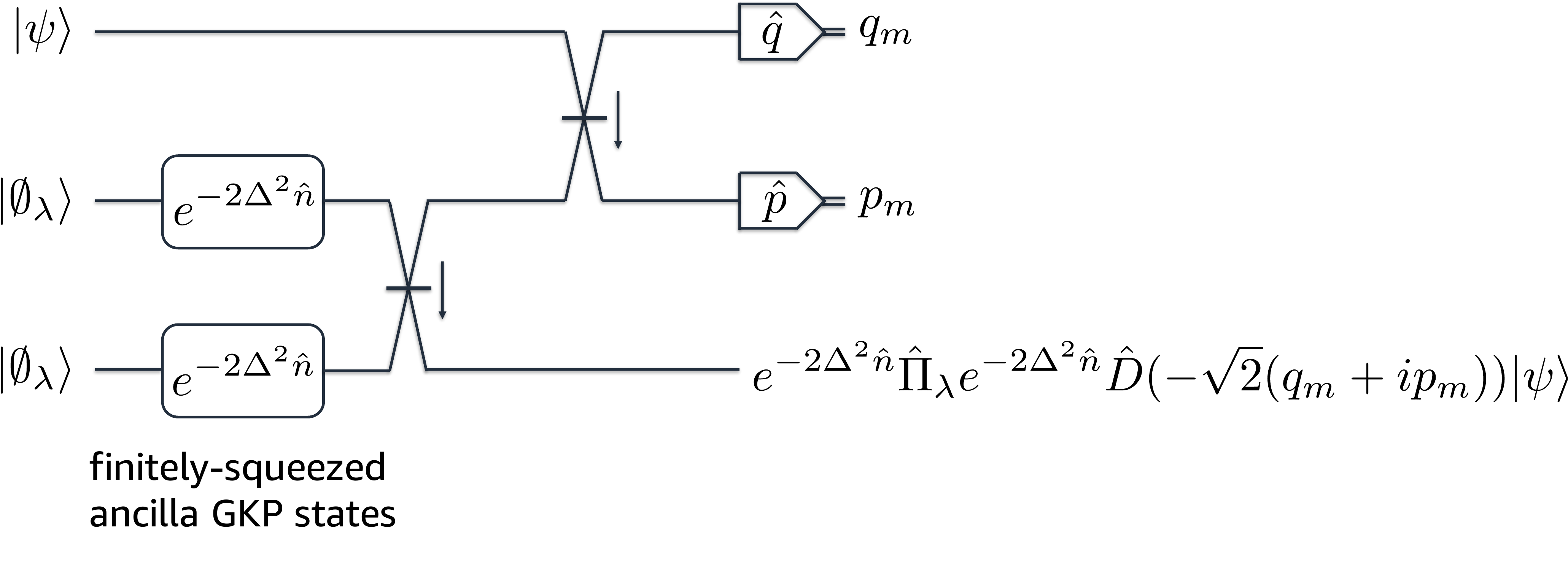}
    \caption{Teleportation-based GKP error correction with finitely-squeezed ancilla GKP states. Conditioned on measuring $q_{m}$ and $p_{m}$, the output state is given by $\exp[-2\Delta^{2}\hat{n}]\hat{\Pi}_{\lambda} \exp[-2\Delta^{2}\hat{n}]\hat{D}(-\sqrt{2}(q_{m} + ip_{m}))|\psi\rangle$. Due to the Gaussian envelope operator $\exp[-2\Delta^{2}\hat{n}]$ after the projection $\hat{\Pi}_{\lambda}$, the output state is guaranteed to be in the finitely-squeezed GKP code space and hence have a bounded energy. Moreover, this envelope operator corresponds to the post-measurement shifts $\xi_{q}^{(+)}$ and $\xi_{p}^{(+)}$ in \cref{fig:GKP error correction noisy ancilla} (b). Another envelope operator $\exp[-2\Delta^{2}\hat{n}]$ before the projection $\hat{\Pi}_{\lambda}$, on the other hand, corresponds to the pre-measurement shifts $-\xi_{q}^{(-)}$ and $\xi_{p}^{(-)}$ in \cref{fig:GKP error correction noisy ancilla} (b).   }
    \label{app_fig:teleportation based GKP error correction finite squeezing}
\end{figure*}

\subsection{Teleportation-based GKP error correction}
\label{subappendix:Teleportation-based GKP error correction}

Recall that the teleportation-based GKP error correction is based on beam-splitter interactions. Under the beam-splitter interaction $\hat{B}_{j\rightarrow k}(\pi/4)$, position eigenstates are transformed as follows: 
\begin{align}
    &\hat{B}_{j\rightarrow k}\Big{(} \frac{ \pi}{4}\Big{)} |\hat{q}_{j} = q_{j}\rangle|\hat{q}_{k} = q_{k}\rangle
    \nonumber\\
    &= |\hat{q}_{j} = ( q_{j} - q_{k} )/\sqrt{2}  \rangle|\hat{q}_{k} =  ( q_{j} + q_{k} )/\sqrt{2}\rangle . 
\end{align}
Applying the balanced beam-splitter interaction to two GKP qunaught states, we get a GKP-Bell state.  
\begin{align}
    &\hat{B}_{2 \rightarrow 3}\Big{(}\frac{\pi}{4}\Big{)}|\emptyset_{\lambda}\rangle_{2}|\emptyset_{\lambda}\rangle_{3} 
    \nonumber\\
    &\propto \hat{B}_{2 \rightarrow 3} \sum_{n_{2},n_{3} \in \mathbb{Z} } |\hat{q}_{2} = n_{2} \sqrt{2\pi}\lambda \rangle |\hat{q}_{3} = n_{3} \sqrt{2\pi}\lambda \rangle 
    \nonumber\\
    &= \sum_{n_{2},n_{3} \in \mathbb{Z} } |\hat{q}_{2} = (n_{2}-n_{3}) \sqrt{\pi}\lambda \rangle |\hat{q}_{3} = (n_{2}+n_{3}) \sqrt{\pi}\lambda \rangle 
    \nonumber\\
    &= \sum_{n_{2},n_{3} \in \mathbb{Z} } |\hat{q}_{2} = n_{2} \sqrt{\pi}\lambda \rangle |\hat{q}_{3} = (n_{2}+2n_{3}) \sqrt{\pi}\lambda \rangle 
    \nonumber\\
    &= |0_{\lambda}\rangle_{2}|0_{\lambda}\rangle_{3} + |1_{\lambda}\rangle_{2}|1_{\lambda}\rangle_{3} \propto |\Phi^{+}_{\lambda}\rangle_{2,3} . 
\end{align}
Then, applying a balanced beam-splitter interaction $\hat{B}_{1\rightarrow 2}(\pi/4)$ to $|\psi\rangle_{1}|\Phi^{+}_{\lambda}\rangle_{2,3}$, we find 
\begin{align}
    &\hat{B}_{1 \rightarrow 2}\Big{(}\frac{\pi}{4}\Big{)} |\psi\rangle_{1}|\Phi^{+}_{\lambda}\rangle_{2,3} 
    \nonumber\\
    &\propto \hat{B}_{1 \rightarrow 2}\Big{(}\frac{\pi}{4}\Big{)} |\psi\rangle_{1} ( |0_{\lambda}\rangle_{2}|0_{\lambda}\rangle_{3} + |1_{\lambda}\rangle_{2}|1_{\lambda}\rangle_{3} )
    \nonumber\\
    &= \sum_{ \mu\in \lbrace 0,1\rbrace }\hat{B}_{1 \rightarrow 2}\Big{(}\frac{\pi}{4}\Big{)} \int dq_{1}\psi(q_{1}) |\hat{q}_{1} = q_{1}\rangle 
    \nonumber\\
    &\quad \times \sum_{n\in 2\mathbb{Z} + \mu } |\hat{q}_{2} = n\sqrt{\pi}\lambda\rangle |\mu_{\lambda}\rangle_{3}, 
\end{align}
and hence 
\begin{align}
    &\hat{B}_{1 \rightarrow 2}\Big{(}\frac{\pi}{4}\Big{)} |\psi\rangle_{1}|\Phi^{+}_{\lambda}\rangle_{2,3} 
    \nonumber\\
    &= \sum_{ \mu\in \lbrace 0,1\rbrace } \sum_{n\in 2\mathbb{Z} + \mu }  \int dq_{1}\psi(q_{1}) \Big{|} \hat{q}_{1} = \frac{ q_{1} }{\sqrt{2}} -n\sqrt{ \frac{\pi}{2} }\lambda \Big{ \rangle} 
    \nonumber\\
    &\quad \times \Big{|} \hat{q}_{2} = \frac{ q_{1} }{\sqrt{2}} + n\sqrt{ \frac{\pi}{2} }\lambda \Big{ \rangle}  |\mu_{\lambda}\rangle_{3} . 
\end{align}
Thus, after measuring $\hat{q}_{1} = q_{m}$ and $\hat{p}_{2} = p_{m}$ via homodyne measurements, we get 
\begin{align}
    &|\psi(q_{m},p_{m})\rangle_{\mathrm{Teleport}} 
    \nonumber\\
    &\equiv \langle \hat{q}_{1} = q_{m}|\hat{p}_{2} = p_{m}| \hat{B}_{1 \rightarrow 2}\Big{(}\frac{\pi}{4}\Big{)} |\psi\rangle_{1}|\Phi^{+}_{\lambda}\rangle_{2,3} 
    \nonumber\\
    &\propto \sum_{ \mu\in \lbrace 0,1\rbrace } \sum_{n\in 2\mathbb{Z} + \mu } \psi( \sqrt{2}q_{m} + n\sqrt{\pi}\lambda  )
    \nonumber\\
    &\quad \times \exp \Big{[} -ip_{m} \Big{(}  q_{m} + n\sqrt{ 2\pi }\lambda \Big{)} \Big{]}  |\mu_{\lambda}\rangle 
    \nonumber\\
    &\propto \sum_{ \mu\in \lbrace 0,1\rbrace } \sum_{n\in 2\mathbb{Z} + \mu } \exp \Big{[} -ip_{m} \Big{(}  q_{m} + n\sqrt{ 2\pi }\lambda \Big{)} \Big{]} 
    \nonumber\\
    &\quad \times  |\mu_{\lambda}\rangle \langle \hat{q} = \sqrt{2}q_{m} +  n\sqrt{\pi}\lambda  | \psi \rangle 
    \nonumber\\
    &\propto \sum_{ \mu\in \lbrace 0,1\rbrace } \sum_{n\in 2\mathbb{Z} + \mu } \exp \Big{[} -i\sqrt{2} p_{m}  n\sqrt{ \pi }\lambda \Big{]} 
    \nonumber\\
    &\quad \times  |\mu_{\lambda}\rangle \langle \hat{q} =   n\sqrt{\pi}\lambda | \hat{D}(-\sqrt{2}q_{m}) | \psi \rangle \nonumber\\
    &= \sum_{ \mu\in \lbrace 0,1\rbrace } \sum_{n\in 2\mathbb{Z} + \mu }  |\mu_{\lambda}\rangle \langle \hat{q} =   n\sqrt{\pi}\lambda | e^{-i\sqrt{2}p_{m}\hat{q} } \hat{D}(-\sqrt{2}q_{m}) | \psi \rangle . 
\end{align}
Carrying on, we obtain the desired result in \cref{eq:ideal output state of the GKP error correction teleportation}: 
\begin{align}
    &|\psi(q_{m},p_{m})\rangle_{\mathrm{Teleport}} 
    \nonumber\\
    &\propto  \sum_{ \mu\in \lbrace 0,1\rbrace }  |\mu_{\lambda}\rangle \underbrace{ \sum_{n\in 2\mathbb{Z} + \mu }  \langle \hat{q} =   n\sqrt{\pi}\lambda | }_{ =\langle \mu_{\lambda}| } \hat{D}(-\sqrt{2}(q_{m} +i p_{m} ) ) | \psi \rangle
    \nonumber\\
    &= \underbrace{ \sum_{ \mu\in \lbrace 0,1\rbrace }  |\mu_{\lambda}\rangle \langle \mu_{\lambda} | }_{ =\hat{\Pi}_{\lambda} } \hat{D}(-\sqrt{2}(q_{m} +i p_{m} ) ) | \psi \rangle
    \nonumber\\
    &= \hat{\Pi}_{\lambda} \hat{D}(-\sqrt{2}(q_{m} +i p_{m} ) ) | \psi \rangle. 
\end{align}

\subsection{Finitely-squeezed ancilla GKP states}
\label{subappendix:Finitely squeezed ancilla GKP states}

A finitely-squeezed GKP state can be understood as a state resulting from applying a Gaussian envelope operator $\exp[-2\Delta^{2}\hat{n}]$ to an ideal GKP state. As shown in Ref.\ \cite{Noh2020_fault_tolerant}, the envelope operator can also be expressed as a coherent superposition of Gaussian shift errors, i.e., 
\begin{align}
    \exp[-2\Delta^{2}\hat{n}] &\propto \int d^{2}\xi \exp\Big{[} -\frac{|\xi|^{2}}{4\sigma_{\mathrm{gkp}}^{2}} \Big{]} \hat{D}(\xi) , 
\end{align}
where $\xi = \xi_{q} + i\xi_{p}$ and
\begin{align}
    \sigma_{\mathrm{gkp}}^{2} = \frac{ 1-e^{ -2\Delta^{2} } }{ 1 + e^{-2\Delta^{2}} }  = \Delta^{2} + \mathcal{O}( \Delta^{4} ) . 
\end{align}
For instance, a finitely-squeezed GKP qunaught state is given by
\begin{align}
    |\emptyset_{\lambda}^{\Delta}\rangle &\propto  \exp[-2\Delta^{2}\hat{n}] |\emptyset_{\lambda}\rangle . 
\end{align}
At the continuous-variable level, we can apply random shifts $\hat{D} ( n_{q}\sqrt{2\pi}\lambda + in_{p}\sqrt{2\pi}/\lambda  )$ with random integers $n_{q}$ and $n_{p}$, and convert the coherent Gaussian shifts into incoherent Gaussian shifts, i.e., 
\begin{align}
    |\emptyset_{\lambda}^{\Delta}\rangle \rightarrow \int  \frac{d^{2}\xi}{ 2\pi\sigma_{\mathrm{gkp}}^{2} }\exp\Big{[} -\frac{|\xi|^{2}}{2\sigma_{\mathrm{gkp}}^{2}} \Big{]} \hat{D}(\xi) |\emptyset_{\lambda}\rangle \langle \emptyset_{\lambda}| \hat{D}^{\dagger}(\xi)
\end{align}
if $\sigma_{\mathrm{gkp}} \ll \sqrt{2\pi}\lambda, \sqrt{2\pi}/\lambda $. See Appendix A in Ref.\ \cite{Noh2020_fault_tolerant} for derivation. In our simulation, we use this incoherent shift noise model because otherwise the simulation cannot scale to large system sizes.

In realistic experimental setups, however, it is not desirable to physically implement the random shifts $\hat{D} ( n_{q}\sqrt{2\pi}\lambda + in_{p}\sqrt{2\pi}/\lambda  )$ with random integers $n_{q}$ and $n_{p}$ because it will increase the energy of the encoded GKP states. Thus, it is also important to understand the outcome of teleportation-based GKP error correction with finitely-squeezed ancilla GKP states $|\emptyset_{\lambda,\Delta}\rangle$ without making the twirling approximation. Note that the product of Gaussian envelope operators $\exp[ -2\Delta^{2}\hat{n}_{2} ]$ and $ \exp[ -2\Delta^{2}\hat{n}_{3} ]$ commutes with the balanced beam-splitter interaction $\hat{B}_{2\rightarrow 3}(\pi/4)$. Thus, after the balanced beam-splitter interaction $\hat{B}_{2\rightarrow 3}(\pi/4)$, we have a finitely-squeezed GKP-Bell state $\exp[ -2\Delta^{2}(\hat{n}_{2} + \hat{n}_{3}) ] |\Phi^{+}_{\lambda}\rangle_{2,3}$ Following the same derivation in \cref{subappendix:Teleportation-based GKP error correction}, we then find (see \cref{app_fig:teleportation based GKP error correction finite squeezing})
\begin{align}
    &|\psi(q_{m},p_{m};\Delta)\rangle_{\mathrm{Teleport}} 
    \nonumber\\
    &\equiv \langle \hat{q}_{1} = q_{m}|\hat{p}_{2} = p_{m}|  \hat{B}_{1 \rightarrow 2}\Big{(}\frac{\pi}{4}\Big{)} |\psi\rangle_{1} e^{ -2\Delta^{2}(\hat{n}_{2} + \hat{n}_{3}) } |\Phi^{+}_{\lambda}\rangle_{2,3} 
    \nonumber\\
    &\propto e^{-2\Delta^{2}\hat{n}}\hat{\Pi}_{\lambda} e^{-2\Delta^{2}\hat{n}} \hat{D}(-\sqrt{2}(q_{m} + ip_{m}))|\psi\rangle . 
\end{align}
See also Ref.\ \cite{Walshe2020_continuous_variable}. Due to the Gaussian envelope operator $\exp[-2\Delta^{2}\hat{n}]$ after the projection $\hat{\Pi}_{\lambda}$, the output state is guaranteed to be in the finitely-squeezed GKP code space and hence have a bounded energy. If we are not concerned about the energy of the state and are only interested in shift errors, the action of the envelope operator can be understood as the post-measurement shifts $\xi_{q}^{(+)}$ and $\xi_{p}^{(+)}$ shown in \cref{fig:GKP error correction noisy ancilla} (b). Similarly, another envelope operator $\exp[-2\Delta^{2}\hat{n}]$ before the projection $\hat{\Pi}_{\lambda}$ guarantee that the measurement outcomes $q_{m}$ and $p_{m}$ are finite whenever the input state $|\psi\rangle$ has a bounded energy. However, if we are only interested in the shift errors, this envelope operator can be understood as the pre-measurement shifts $-\xi_{q}^{(-)}$ and $\xi_{p}^{(-)}$ shown in \cref{fig:GKP error correction noisy ancilla} (b). 

\section{Maximum likelihood decoding of the logical two-qubit gates between two GKP qubits}
\label{appendix:Maximum likelihood decoding of the logical two-qubit gates between two GKP qubits}

Recall the optimization problem in \cref{eq:CNOT qq optimization}: 
\begin{align}
    &(n^{\star(1)}_{q} , n^{\star(2)}_{q} ) = \argmin_{ n_{1} , n_{2}} \Big{(} (2 + \frac{1}{\lambda^{2}})x_{1}^{2} + 2x_{2}^{2} -\frac{ 2}{ \lambda } x_{1}x_{2}  \Big{)} , 
    \nonumber\\
    &\mathrm{where}\,\,\, x_{1} \equiv  \sqrt{2}q_{m}^{(1)} - n_{1}\sqrt{\pi}\lambda , \,\,\, x_{2} \equiv  \sqrt{2}q_{m}^{(2)} - n_{2}\sqrt{\pi},  
\end{align}
given $\sqrt{2}q_{m}^{(1)}$ and $\sqrt{2}q_{m}^{(2)}$. To solve this optimization, we tackle a closely related but slightly different problem: That is, we seek to find conditions on $q_{1} \equiv \sqrt{2}q_{m}^{(1)}$ and $q_{2} \equiv \sqrt{2}q_{m}^{(2)}$ such that $(n^{\star(1)}_{q} , n^{\star(2)}_{q} )$ is given by $(0,0)$. Define $f_{\mathrm{CNOT}}^{[qq]}(x_{1},x_{2})$ as 
\begin{align}
    f_{\mathrm{CNOT}}^{[qq]}(x_{1},x_{2}) \equiv \Big{(} (2 + \frac{1}{\lambda^{2}})x_{1}^{2} + 2x_{2}^{2} -\frac{ 2}{ \lambda } x_{1}x_{2}  \Big{)} . 
\end{align}
Then, for $q_{1}$ and $q_{2}$ to have $(n^{\star(1)}_{q} , n^{\star(2)}_{q} ) = (0,0)$ as the solution of the above optimization problem, they must satisfy
\begin{align}
    f_{\mathrm{CNOT}}^{[qq]}( q_{1} , q_{2} ) &< f_{\mathrm{CNOT}}^{[qq]}( q_{1} \pm \sqrt{\pi}\lambda , q_{2} ), 
    \nonumber\\
    f_{\mathrm{CNOT}}^{[qq]}( q_{1} , q_{2} ) &< f_{\mathrm{CNOT}}^{[qq]}( q_{1}  , q_{2} \pm \sqrt{\pi} ), \nonumber\\
    f_{\mathrm{CNOT}}^{[qq]}( q_{1} , q_{2} ) &< f_{\mathrm{CNOT}}^{[qq]}( q_{1} \pm \sqrt{\pi}\lambda , q_{2} \pm \sqrt{\pi} ). 
\end{align}
These conditions are reduced to 
\begin{align}
    &\Big{(}2 + \frac{1}{\lambda^{2}}\Big{)}q_{1}^{2} + 2q_{2}^{2} -\frac{ 2}{ \lambda } q_{1}q_{2} 
    \nonumber\\
    &< \Big{(}2 + \frac{1}{\lambda^{2}}\Big{)}(q_{1} \pm \sqrt{\pi}\lambda)^{2} + 2q_{2}^{2} -\frac{ 2}{ \lambda } (q_{1} \pm \sqrt{\pi}\lambda)q_{2}
    \nonumber\\
    &\leftrightarrow  \Big{|} q_{2} - \Big{(}2\lambda + \frac{1}{\lambda}\Big{)}q_{1} \Big{|} <   \frac{(2\lambda^{2} + 1)}{2}\sqrt{\pi} , 
\end{align}

\begin{align}
    &\Big{(}2 + \frac{1}{\lambda^{2}}\Big{)}q_{1}^{2} + 2q_{2}^{2} -\frac{ 2}{ \lambda } q_{1}q_{2}
    \nonumber\\
    &< \Big{(}2 + \frac{1}{\lambda^{2}}\Big{)}q_{1}^{2} + 2(q_{2}\pm\sqrt{\pi})^{2} -\frac{ 2}{ \lambda } q_{1} (q_{2}\pm\sqrt{\pi}) 
    \nonumber\\
    &\leftrightarrow  \Big{|} q_{2} - \frac{1}{2\lambda}q_{1} \Big{|} <   \frac{\sqrt{\pi}}{2} , 
\end{align} 
and 
\begin{align}
    &\Big{(}2 + \frac{1}{\lambda^{2}}\Big{)}q_{1}^{2} + 2q_{2}^{2} -\frac{ 2}{ \lambda } q_{1}q_{2}
    \nonumber\\
    &< \Big{(}2 + \frac{1}{\lambda^{2}}\Big{)}(q_{1} \pm \sqrt{\pi}\lambda)^{2} + 2(q_{2}\pm\sqrt{\pi})^{2}
    \nonumber\\
    &\quad -\frac{ 2}{ \lambda } (q_{1} \pm \sqrt{\pi}\lambda) (q_{2}\pm\sqrt{\pi}) 
    \nonumber\\
    &\leftrightarrow  \Big{|} q_{2} + 2\lambda q_{1} \Big{|} <   \frac{(2\lambda^{2}+1)}{2}\sqrt{\pi} . 
\end{align} 
One can verify that other conditions such as $f_{\mathrm{CNOT}}^{[qq]}( q_{1} , q_{2} ) < f_{\mathrm{CNOT}}^{[qq]}( q_{1} \pm \sqrt{\pi}\lambda , q_{2} \mp \sqrt{\pi} )$ are trivially satisfied if these three conditions hold. Thus, we can define Voronoi cells $\mathrm{Vor}_{\mathrm{CNOT}}^{[qq]}(n_{1} , n_{2})$ as follows: 
\begin{align}
    &\mathrm{Vor}_{\mathrm{CNOT}}^{[qq]}(n_{1} , n_{2}) \equiv \Big{\lbrace } (q_{1} , q_{2}) :  
    \nonumber\\
    &\Big{|} q_{2} - n_{2}\sqrt{\pi} - \Big{(}2\lambda + \frac{1}{\lambda}\Big{)} (q_{1} -n_{1}\sqrt{\pi}\lambda ) \Big{|} <   \frac{(2\lambda^{2} + 1)}{2}\sqrt{\pi} , 
    \nonumber\\
    & \Big{|} q_{2} - n_{2}\sqrt{\pi} - \frac{1}{2\lambda}(q_{1} -n_{1}\sqrt{\pi}\lambda) \Big{|} <   \frac{\sqrt{\pi}}{2}, 
    \nonumber\\
    & \Big{|} q_{2} - n_{2}\sqrt{\pi} + 2\lambda (q_{1} -n_{1}\sqrt{\pi}\lambda) \Big{|} <   \frac{(2\lambda^{2}+1)}{2}\sqrt{\pi}  \Big{ \rbrace } . \label{eq:CNOT qq Voronoi cell}
\end{align}
Thus, to solve the optimization problem in \cref{eq:CNOT qq optimization}, we can simply ask in which Voronoi cell $(q_{1} , q_{2}) \equiv (\sqrt{2}q_{m}^{(1)} , \sqrt{2}q_{m}^{(2)})$ lies in. This can be efficiently done by using \cref{alg:CNOTnq1nq2}.

\begin{algorithm}[t!]
\SetAlgoLined
\KwResult{$n^{\star(1)}_{q}$, $n^{\star(2)}_{q}$}
 Input: $\sqrt{2}q_{m}^{(1)}$, $\sqrt{2}q_{m}^{(2)}$\;
$\bar{n}_{q}^{(1)} \leftarrow   \Big{\lfloor} \frac{ \sqrt{2}q_{m}^{(1)} }{\sqrt{\pi}\lambda}+\frac{1}{2} \Big{\rfloor}$\,\,\, and\,\,\,
$\bar{n}_{q}^{(2)} \leftarrow  \Big{\lfloor} \frac{ \sqrt{2}q_{m}^{(2)} }{\sqrt{\pi}}+\frac{1}{2} \Big{\rfloor}$\;
$v_{1} \leftarrow \frac{1}{2\lambda}( \sqrt{2}q_{m}^{(1)}  - \bar{n}_{q}^{(1)}\sqrt{\pi}\lambda) - \frac{1}{2}\sqrt{\pi}$\;
$v_{2} \leftarrow \sqrt{2}q_{m}^{(2)} - \bar{n}_{q}^{(2)}\sqrt{\pi}$\;
$v_{3} \leftarrow \frac{1}{2\lambda}( \sqrt{2}q_{m}^{(1)} - \bar{n}_{q}^{(1)}\sqrt{\pi}\lambda) + \frac{1}{2}\sqrt{\pi}$\;
$v_{4} \leftarrow (2\lambda + \frac{1}{\lambda}) ( \sqrt{2}q_{m}^{(1)} - \bar{n}_{q}^{(1)}\sqrt{\pi}\lambda) - \frac{2\lambda^{2}+1}{2}\sqrt{\pi}$\; 
$v_{5} \leftarrow (2\lambda + \frac{1}{\lambda}) ( \sqrt{2}q_{m}^{(1)} - \bar{n}_{q}^{(1)}\sqrt{\pi}\lambda) + \frac{(2\lambda^{2}+1)}{2}\sqrt{\pi}$\; 
$v_{6} \leftarrow -2\lambda( \sqrt{2}q_{m}^{(1)} - (\bar{n}_{q}^{(1)} -1 ) \sqrt{\pi}\lambda) + \frac{(2\lambda^{2}+1)}{2}\sqrt{\pi}$\; 
$v_{7} \leftarrow -2\lambda( \sqrt{2}q_{m}^{(1)} - (\bar{n}_{q}^{(1)} +1 )\sqrt{\pi}\lambda) - \frac{(2\lambda^{2}+1)}{2}\sqrt{\pi}$\; 
\uIf{ $ v_{1} < v_{2} < v_{3} $ }{
   
  \uIf{ $ v_{4} <  v_{2} < v_{5}$ }{
  $n^{\star(1)}_{q} \leftarrow \bar{n}_{q}^{(1)}$ and $n^{\star(2)}_{q} \leftarrow \bar{n}_{q}^{(2)}$\;
  }

	\uElseIf{ $v_{2} \ge v_{5}$ }{
	 $n^{\star(1)}_{q} \leftarrow \bar{n}_{q}^{(1)}-1$ and $n^{\star(2)}_{q} \leftarrow \bar{n}_{q}^{(2)}$\;
	}   
   
  \Else{
  $n^{\star(1)}_{q} \leftarrow \bar{n}_{q}^{(1)}+1$ and $n^{\star(2)}_{q} \leftarrow \bar{n}_{q}^{(2)}$\;
  }
   
  }{

  \uElseIf{ $v_{2} \ge v_{3}$ }{
      
    \uIf{ $ v_{2} > v_{6}$ }{
  $n^{\star(1)}_{q} \leftarrow \bar{n}_{q}^{(1)}$ and $n^{\star(2)}_{q} \leftarrow \bar{n}_{q}^{(2)}+1$\;
  }
  \Else{
  $n^{\star(1)}_{q} \leftarrow \bar{n}_{q}^{(1)}-1$ and $n^{\star(2)}_{q} \leftarrow \bar{n}_{q}^{(2)}$\;
  }
   
  }
   
  \Else{
   
	\uIf{ $v_{2} > v_{7}$ }{
  $n^{\star(1)}_{q} \leftarrow \bar{n}_{q}^{(1)}+1$ and $n^{\star(2)}_{q} \leftarrow \bar{n}_{q}^{(2)}$\;
  }
  \Else{
  $n^{\star(1)}_{q} \leftarrow \bar{n}_{q}^{(1)}$ and $n^{\star(2)}_{q} \leftarrow \bar{n}_{q}^{(2)}-1$\;
  }

  }
}
 \caption{Maximum likelihood decoding of the position shifts in the logical CNOT gate between a rectangular-lattice (labeled as 1) and a square-lattice (labeled as 2) GKP qubits}
 \label{alg:CNOTnq1nq2}
\end{algorithm}

\begin{algorithm}[t!]
\SetAlgoLined
\KwResult{$n^{\star(1)}_{p}$, $n^{\star(2)}_{p}$}
 Input: $ \sqrt{2}p_{m}^{(1)}$, $\sqrt{2}p_{m}^{(2)}$\;
$\bar{n}_{p}^{(1)} \leftarrow   \Big{\lfloor} \frac{  \sqrt{2}p_{m}^{(1)} }{\sqrt{\pi}/ \lambda}+\frac{1}{2} \Big{\rfloor}$ \,\,\, and\,\,\, $\bar{n}_{p}^{(2)} \leftarrow  \Big{\lfloor} \frac{ \sqrt{2}p_{m}^{(2)} }{\sqrt{\pi}}+\frac{1}{2} \Big{\rfloor}$\;
$v_{1} \leftarrow -\frac{\lambda}{2\lambda^{2}+1}( \sqrt{2}p_{m}^{(1)} - \bar{n}_{p}^{(1)} \sqrt{\pi}/\lambda) - \frac{1}{2} \sqrt{\pi}$\; 
$v_{2}\leftarrow \sqrt{2}p_{m}^{(2)} - \bar{n}_{p}^{(2)}\sqrt{\pi}$\; 
$v_{3} \leftarrow -\frac{\lambda}{2\lambda^{2}+1}( \sqrt{2}p_{m}^{(1)} - \bar{n}_{p}^{(1)}\sqrt{\pi}/\lambda) + \frac{1}{2} \sqrt{\pi}$\; 
$v_{4} \leftarrow -2\lambda( \sqrt{2}p_{m}^{(1)} - \bar{n}_{p}^{(1)}\sqrt{\pi}/\lambda) - \sqrt{\pi}$\; 
$v_{5} \leftarrow -2\lambda( \sqrt{2}p_{m}^{(1)} - \bar{n}_{p}^{(1)}\sqrt{\pi}/\lambda) + \sqrt{\pi}$\; 
$v_{6} \leftarrow \frac{1}{2\lambda}( \sqrt{2}p_{m}^{(1)} - (\bar{n}_{p}^{(1)}+1)\sqrt{\pi}/\lambda) + \frac{(2\lambda^{2}+1)}{4\lambda^{2}} \sqrt{\pi}$\; 
$v_{7} \leftarrow \frac{1}{2\lambda}( \sqrt{2}p_{m}^{(1)} - (\bar{n}_{p}^{(1)} -1 )\sqrt{\pi}/\lambda) - \frac{(2\lambda^{2}+1)}{4\lambda^{2}} \sqrt{\pi}$\;
\uIf{ $ v_{1} < v_{2} < v_{3}$ }{
   
   \uIf{ $ v_{4} < v_{2} < v_{5} $ }{
   $n^{\star(1)}_{p} \leftarrow \bar{n}_{p}^{(1)}$ and $n^{\star(2)}_{p} \leftarrow \bar{n}_{p}^{(2)}$\;
   }

	\uElseIf{ $v_{2} \ge  v_{5} $ }{
	 $n^{\star(1)}_{p} \leftarrow \bar{n}_{p}^{(1)}+1$ and $n^{\star(2)}_{p} \leftarrow \bar{n}_{p}^{(2)}$\;
	}   
   
   \Else{
   $n^{\star(1)}_{p} \leftarrow \bar{n}_{p}^{(1)}-1$ and $n^{\star(2)}_{p} \leftarrow \bar{n}_{p}^{(2)}$\;
  }
   
   }{

   \uElseIf{ $v_{2} \ge v_{3}$ }{
      
    \uIf{ $v_{2} > v_{6}$ }{
   $n^{\star(1)}_{p} \leftarrow \bar{n}_{p}^{(1)}$ and $n^{\star(2)}_{p} \leftarrow \bar{n}_{p}^{(2)}+1$\;
   }
   \Else{
   $n^{\star(1)}_{p} \leftarrow \bar{n}_{p}^{(1)} +1$ and $n^{\star(2)}_{p} \leftarrow \bar{n}_{p}^{(2)}$\;
  }
   
   }
   
   \Else{
   
	\uIf{ $v_{2} > v_{7}$ }{
   $n^{\star(1)}_{p} \leftarrow \bar{n}_{p}^{(1)}-1$ and $n^{\star(2)}_{p} \leftarrow \bar{n}_{p}^{(2)}$\;
   }
   \Else{
   $n^{\star(1)}_{p} \leftarrow \bar{n}_{p}^{(1)}$ and $n^{\star(2)}_{p} \leftarrow \bar{n}_{p}^{(2)}-1$\;
  }

   }
}
 \caption{Maximum likelihood decoding of the momentum shifts in the logical CNOT gate between a rectangular-lattice (labeled as 1) and a square-lattice (labeled as 2) GKP qubits}
 \label{alg:CNOTnp1np2}
\end{algorithm}

Similarly for the other optimization problem in \cref{eq:CNOT pp optimization} for the momentum quadratures, i.e.,
\begin{align}
    &(n^{\star(1)}_{p} , n^{\star(2)}_{p} ) = \argmin_{ n_{1} , n_{2}} \Big{(} 2x_{1}^{2} + ( 2+\frac{1}{\lambda^{2}} )x_{2}^{2} + \frac{2}{\lambda} x_{1}x_{2}  \Big{)} , 
    \nonumber\\
    &\mathrm{where}\,\,\, x_{1} \equiv  \sqrt{2}p_{m}^{(1)} - n_{1}\sqrt{\pi}/\lambda , \,\,\, x_{2} \equiv  \sqrt{2}p_{m}^{(2)} - n_{2}\sqrt{\pi}, 
\end{align}
given $p_{1} \equiv \sqrt{2}p_{m}^{(1)}$ and $p_{2} \equiv \sqrt{2}p_{m}^{(2)}$, we can identify the set of $(p_{1} , p_{2})$ for which the solution is given by $(n^{\star(1)}_{p} , n^{\star(2)}_{p} )  = (0,0)$ by imposing the following conditions: 
\begin{align}
    f_{\mathrm{CNOT}}^{[pp]}( p_{1} , p_{2} ) &< f_{\mathrm{CNOT}}^{[pp]}( p_{1} \pm \sqrt{\pi}/\lambda , p_{2} ), 
    \nonumber\\
    f_{\mathrm{CNOT}}^{[pp]}( p_{1} , p_{2} ) &< f_{\mathrm{CNOT}}^{[pp]}( p_{1}  , p_{2} \pm \sqrt{\pi} ), \nonumber\\
    f_{\mathrm{CNOT}}^{[pp]}( p_{1} , p_{2} ) &< f_{\mathrm{CNOT}}^{[pp]}( p_{1} \pm \sqrt{\pi} / \lambda , p_{2} \mp \sqrt{\pi} ), 
\end{align}
where $f_{\mathrm{CNOT}}^{[pp]}( p_{1} , p_{2} )$ is defined as 
\begin{align}
    f_{\mathrm{CNOT}}^{[pp]}( p_{1} , p_{2} ) &\equiv 2p_{1}^{2} + \Big{(} 2+\frac{1}{\lambda^{2}} \Big{)}p_{2}^{2} + \frac{2}{\lambda} p_{1}p_{2}  . 
\end{align}
In this case, the right hand side of the third condition is given by $f_{\mathrm{CNOT}}^{[pp]}( p_{1} \pm \sqrt{\pi} / \lambda , p_{2} \mp \sqrt{\pi} )$, instead of $f_{\mathrm{CNOT}}^{[pp]}( p_{1} \pm \sqrt{\pi} / \lambda , p_{2} \pm \sqrt{\pi} )$. This is related to the fact that the momentum noise $\xi_{p}^{(1)}$ and $\xi_{p}^{(2)}$ are negatively correlated. The corresponding Voronoi cells are given by 
\begin{align}
    &\mathrm{Vor}_{\mathrm{CNOT}}^{[pp]}(n_{1} , n_{2}) \equiv \Big{\lbrace } (p_{1} , p_{2}) :  
    \nonumber\\
    &\Big{|} p_{2} - n_{2}\sqrt{\pi} +2\lambda (p_{1} -n_{1}\sqrt{\pi}/\lambda ) \Big{|} <   \sqrt{\pi}, 
    \nonumber\\
    & \Big{|} p_{2} - n_{2}\sqrt{\pi} + \frac{\lambda}{2\lambda^{2}+1}(p_{1} -n_{1}\sqrt{\pi}/\lambda) \Big{|} <   \frac{\sqrt{\pi}}{2}, 
    \nonumber\\
    & \Big{|} p_{2} - n_{2}\sqrt{\pi} -\frac{1}{2\lambda}(p_{1} -n_{1}\sqrt{\pi}/\lambda) \Big{|} <   \frac{(2\lambda^{2}+1)}{4\lambda^{2}}\sqrt{\pi}  \Big{ \rbrace } . \label{eq:CNOT pp Voronoi cell}
\end{align}
Hence, the optimization problem in \cref{eq:CNOT pp optimization} can be solved by finding in which Voronoi cell $(p_{1} , p_{2}) \equiv (\sqrt{2}p_{m}^{(1)} , \sqrt{2}p_{m}^{(2)})$ is in. This search can be efficiently done by using \cref{alg:CNOTnp1np2}. Note that the decision boundaries defined by the Voronoi cells in \cref{eq:CNOT qq Voronoi cell,eq:CNOT pp Voronoi cell} are visualized in \cref{fig:GKP CNOT hexagon decoding} (c). 

\section{Simulation details}
\label{appendix:Simulation details}

In \cref{subappedix:Noise model}, we discuss the noise model used in our surface-GKP code simulation. We also explain how compute the conditional failure probabilities of various circuit elements based on extra analog information obtained during GKP error corrections. Then, in \cref{subappedix:Edge weights in the matching graphs}, we describe how the edge weights in the surface-code matching graphs are dynamically computed by using the conditional failure probabilities of circuit elements.  

\subsection{Noise model}
\label{subappedix:Noise model}

Here, we discuss how to simulate $|+_{\lambda}\rangle$ state preparation, Pauli $X$ measurement, idling, CNOT, and CZ gates of the GKP qubits which are necessary to implement the surface-GKP code.  

\subsubsection{$|+_{\lambda}\rangle$ state preparation}

To prepare a $|+_{\lambda}\rangle$ state in an ancilla mode of the surface-GKP code, we assume that the ancilla mode is initially prepared in a squeezed state, close to a momentum eigenstate $|\hat{p}=0\rangle$. In particular, we assume that the variance of the momentum quadrature is given by $\sigma_{\mathrm{gkp}}^{2}$. We then apply (noisy) teleportation-based GKP error-correction by using finitely-squeezed GKP qunaught states with the same noise variance $\sigma_{\mathrm{gkp}}^{2}$. Since the GKP error correction is noisy due to the finite GKP squeezing, a momentum noise of variance $\sigma_{\mathrm{gkp}}^{2}$ is added to the ancilla mode (see \cref{fig:GKP error corrected two qubit gates} (b)). Then the net momentum noise variance is given by $2\sigma_{\mathrm{gkp}}^{2}$. 

Hence, we simulate the $|+_{\lambda}\rangle$ state preparation as follows: we first sample a random shift $\xi_{p}$ from the Gaussian distribution $\mathcal{N}(0,2\sigma_{\mathrm{gkp}}^{2})$ of zero mean and variance $2\sigma_{\mathrm{gkp}}^{2}$. Then, compute an integer $n_{p}$ 
\begin{align}
    n_{p}  &= \Big{\lfloor} \frac{\xi_{p}}{\sqrt{\pi}/\lambda}+\frac{1}{2} \Big{\rfloor} . 
\end{align}
Then, if $n_{p}$ is even (i.e., $n_{p}\in 2\mathbb{Z}$), assume that $|+_{\lambda}\rangle$ is correctly prepared. On the other hand, if $n_{p}$ is odd (i.e., $n_{p}\in 2\mathbb{Z} + 1$), apply a Pauli $Z$ error, or equivalently, assume that $|-_{\lambda}\rangle$ is prepared instead of $|+_{\lambda}\rangle$.  

To compute the edge weights in the matching graphs of the surface code, we need to know the failure probability $p_{ |+_{\lambda}\rangle \rightarrow |-_{\lambda}\rangle }$ of the $|+_{\lambda}\rangle$ state preparation. In the case where we ignore the extra analog information, we simply set $p_{ |+_{\lambda}\rangle \rightarrow |-_{\lambda}\rangle }$ to be the probability of $n_{p}$ being odd: 
\begin{align}
    &p_{ |+_{\lambda}\rangle \rightarrow |-_{\lambda}\rangle } 
    \nonumber\\
    &= \sum_{n \in 2\mathbb{Z} + 1} \int_{( n-\frac{1}{2} )\sqrt{\pi}/\lambda}^{( n+\frac{1}{2} )\sqrt{\pi}/\lambda}  \frac{d\xi}{\sqrt{2\pi ( 2\sigma_{\mathrm{gkp}}^{2} )}}  \exp\Big{[} -\frac{\xi^{2}}{ 2 ( 2\sigma_{\mathrm{gkp}}^{2} ) } \Big{]} . 
\end{align}
In practice, for the parameters we consider in this work ($9\mathrm{dB} \le \sigma_{\mathrm{gkp}}^{(\mathrm{dB})} \le 13\mathrm{dB}$), it suffices to approximate $\sum_{n \in 2\mathbb{Z} + 1}$ by $\sum_{n \in \lbrace -1,1 \rbrace }$. 

On the other hand, if the analog information is utilized, we first compute 
\begin{align}
    \bar{\xi}_{p} \equiv \xi_{p} - n_{p}\sqrt{\pi}/\lambda . 
\end{align}
Then, given this analog information, we compute the conditional probability $p_{ |+_{\lambda}\rangle \rightarrow |-_{\lambda}\rangle } $ as 
\begin{align}
    p_{ |+_{\lambda}\rangle \rightarrow |-_{\lambda}\rangle }  = \frac{b}{a+b}, 
\end{align}
where $a$ and $b$ are given by
\begin{align}
    a &= \sum_{n \in 2\mathbb{Z}} \exp\Big{[} -\frac{( \bar{\xi}_{p} + n\sqrt{\pi}/\lambda )^{2}}{ 2 ( 2\sigma_{\mathrm{gkp}}^{2} ) } \Big{]} , 
    \nonumber\\
    b &= \sum_{n \in 2\mathbb{Z} + 1} \exp\Big{[} -\frac{( \bar{\xi}_{p} + n\sqrt{\pi}/\lambda )^{2}}{ 2 ( 2\sigma_{\mathrm{gkp}}^{2} ) } \Big{]} . 
\end{align}
Similarly as in the no-analog information case, it suffices to approximate $\sum_{n \in 2\mathbb{Z} }$ by $\sum_{n \in \lbrace 0 \rbrace }$ and $\sum_{n \in 2\mathbb{Z} + 1}$ by $\sum_{n \in \lbrace -1,1 \rbrace }$.  

\subsubsection{Pauli $X$ measurement}

To simulate the Pauli $X$ measurement of an ancilla qubit of the surface-GKP code (which is encoded in a rectangular-lattice GKP code with $\lambda$), we assume that it follows a noisy GKP error correction. This is because all Pauli $X$ measurements are preceded by an error-corrected two-qubit gate between two GKP qubits which ends with a noisy GKP error correction. Thus, the ancilla GKP qubit inherits a momentum shift $\xi_{p} \sim \mathcal{N}(0, \sigma_{\mathrm{gkp}}^{2})$ from the noisy GKP error correction (see \cref{fig:GKP error corrected two qubit gates} (b)) before it is measured in the $X$ basis. The Pauli $X$ measurement itself is performed via a momentum homodyne measurement which we assume to be noiseless for simplicity. Assuming that the resolution of the homodyne measurement is not much worse than the noise standard deviation $\sigma_{\mathrm{gkp}}$ of the GKP qunaught states, the errors in the two-qubit gates (to be discussed below) are the dominant noise sources in the surface-GKP code. Thus, if this is the case, how precisely we model the homodyne measurement does not make an appreciable overall impact to the performance of the surface-GKP code.

Under this noise model, the rest of the noise simulation goes similarly as in the case of the $|+_{\lambda}\rangle$ state preparation. We first sample a random shift $\xi_{p}$ from the Gaussian distribution $\mathcal{N}( 0, \sigma_{\mathrm{gkp}}^{2})$ and compute an integer $n_{p}$ as follows:   
\begin{align}
    n_{p}  &= \Big{\lfloor} \frac{\xi_{p}}{\sqrt{\pi}/\lambda}+\frac{1}{2} \Big{\rfloor} . 
\end{align}
If $n_{p}\in 2\mathbb{Z}$, we assume that the $X$ measurement is performed faithfully. If $n_{p}\in 2\mathbb{Z} + 1$, however, we assume that the Pauli $X$ measurement is preceded by a Pauli $Z$ error, i.e., the measurement outcome is flipped. 

Then, we compute the probability (or conditional probability) $p_{+\leftrightarrow -}$ of having a measurement error. If we ignore the analog information, we simply compute the probability of $n_{p}$ being odd:   
\begin{align}
    &p_{ + \leftrightarrow - }  = \sum_{n \in 2\mathbb{Z} + 1} \int_{( n-\frac{1}{2} )\sqrt{\pi}/\lambda}^{( n+\frac{1}{2} )\sqrt{\pi}/\lambda}  \frac{d\xi}{\sqrt{2\pi \sigma_{\mathrm{gkp}}^{2} }}  \exp\Big{[} -\frac{\xi^{2}}{ 2 \sigma_{\mathrm{gkp}}^{2}  } \Big{]} , 
\end{align}
where $\sum_{n \in 2\mathbb{Z} + 1} \rightarrow \sum_{n \in \lbrace -1,1 \rbrace }$ suffices in practice for $9\mathrm{dB} \le \sigma_{\mathrm{gkp}}^{(\mathrm{dB})} \le 13\mathrm{dB}$.  

On the other hand, if we exploit the extra analog information, we compute 
\begin{align}
    \bar{\xi}_{p} \equiv \xi_{p} - n_{p}\sqrt{\pi}/\lambda  
\end{align}
and then set the conditional probability $p_{+\leftrightarrow -}$ to be 
\begin{align}
    p_{+\leftrightarrow -} = \frac{b}{a+b}, 
\end{align}
where $a$ and $b$ are given by
\begin{align}
    a &= \sum_{n \in 2\mathbb{Z}} \exp\Big{[} -\frac{( \bar{\xi}_{p} + n\sqrt{\pi}/\lambda )^{2}}{ 2 \sigma_{\mathrm{gkp}}^{2} } \Big{]} , 
    \nonumber\\
    b &= \sum_{n \in 2\mathbb{Z} + 1} \exp\Big{[} -\frac{( \bar{\xi}_{p} + n\sqrt{\pi}/\lambda )^{2}}{ 2  \sigma_{\mathrm{gkp}}^{2}  } \Big{]} . 
\end{align}
Again, it suffices to use $\sum_{n \in \lbrace 0 \rbrace }$ instead of $\sum_{n \in 2\mathbb{Z} }$ and $\sum_{n \in \lbrace -1,1 \rbrace }$ instead of $\sum_{n \in 2\mathbb{Z} + 1}$.  

\subsubsection{Idling}

The idling scenario is already discussed in great detail in \cref{subsection:Teleportation-based GKP error correction}. To briefly remind ourselves, we assume that an idling is always preceded by a noisy GKP error correction and thus the idling mode inherits position and momentum shifts of noise variance $\sigma_{\mathrm{gkp}}^{2} $ from the previous noisy GKP error correction. At the end of the idling, another noisy GKP error correction is performed and hence extra position and momentum shifts errors of the same variance $\sigma_{\mathrm{gkp}}^{2} $ are added (see \cref{fig:GKP error correction noisy ancilla} (b)). Hence, the noise variance of the net accumulated position and momentum shift errors is given by $2\sigma_{\mathrm{gkp}}^{2}$.  

Thus, to simulate an idling event, we first sample position and momentum shifts $\xi_{q}$ and $\xi_{p}$ independently from the Gaussian distribution $\mathcal{N}( 0, 2\sigma_{\mathrm{gkp}}^{2} )$, i.e., $(\xi_{q}.\xi_{p}) \sim_{\mathrm{iid}} \mathcal{N}( 0, 2\sigma_{\mathrm{gkp}}^{2} )$. Then, compute the two integers  
\begin{align}
    n_{q}  &= \Big{\lfloor} \frac{\xi_{q}}{\sqrt{\pi}\lambda}+\frac{1}{2} \Big{\rfloor} , 
    \nonumber\\
    n_{p}  &= \Big{\lfloor} \frac{\xi_{p}}{\sqrt{\pi}/\lambda}+\frac{1}{2} \Big{\rfloor} . 
\end{align}
For the idling locations in the data qubits of the surface-GKP code, we always use $\lambda=1$ but for the ancilla qubits, we use $0.8\le \lambda \le 1.2$. With these two integers $n_{q}$ and $n_{p}$, we apply
\begin{align}
    &\begin{cases}
    I & \textrm{if }n_{q}  \in 2\mathbb{Z} \\
    X & \textrm{if }n_{q} \in 2\mathbb{Z}+1 
    \end{cases} , 
\end{align}
and then apply
\begin{align}
    &\begin{cases}
    I & \textrm{if }n_{p}  \in 2\mathbb{Z} \\
    Z & \textrm{if }n_{p} \in 2\mathbb{Z}+1 
    \end{cases}  . 
\end{align}

We then need to compute the probabilities (or conditional probabilities) $p_{I},p_{X},p_{Y},p_{Z}$ of the idling Pauli errors, which are used to compute the edge weights in the matching graphs of the surface code. In the case where we do not use the extra analog information, we compute 
\begin{align}
    q_{X} &=   \sum_{n \in 2\mathbb{Z} + 1} \int_{( n-\frac{1}{2} )\sqrt{\pi}/\lambda}^{( n+\frac{1}{2} )\sqrt{\pi}/\lambda}  \frac{d\xi}{\sqrt{2\pi ( 2\sigma_{\mathrm{gkp}}^{2} )}}  \exp\Big{[} -\frac{\xi^{2}}{ 2 ( 2\sigma_{\mathrm{gkp}}^{2} ) } \Big{]} , 
    \nonumber\\
    q_{Z} &=   \sum_{n \in 2\mathbb{Z} + 1} \int_{( n-\frac{1}{2} )\sqrt{\pi}\lambda}^{( n+\frac{1}{2} )\sqrt{\pi}\lambda}  \frac{d\xi}{\sqrt{2\pi ( 2\sigma_{\mathrm{gkp}}^{2} )}}  \exp\Big{[} -\frac{\xi^{2}}{ 2 ( 2\sigma_{\mathrm{gkp}}^{2} ) } \Big{]} ,
\end{align}
and set $p_{I},p_{X},p_{Y},p_{Z}$ to be
\begin{align}
    \begin{bmatrix}
    p_{I} & p_{X} \\
    p_{Z} & p_{Y} 
    \end{bmatrix} = \begin{bmatrix}
    (1-q_{Z}) \\
    q_{Z}
    \end{bmatrix} \begin{bmatrix}
    (1-q_{X}) & q_{X}
    \end{bmatrix}. \label{eq:idling q to p}
\end{align}

When the extra analog information is used, we first find 
\begin{align}
    \bar{\xi}_{q} &= \xi_{q} - n_{q}\sqrt{\pi}\lambda, 
    \nonumber\\
    \bar{\xi}_{p} &= \xi_{p} - n_{p}\sqrt{\pi}/ \lambda, 
\end{align}
and then compute $q_{X}$ and $q_{Z}$ as follows: 
\begin{align}
    q_{X} &= \frac{ b   }{  a + b } , \quad q_{Z} = \frac{ d   }{  c+ d} , 
\end{align}
where $a,b,c,d$ are given by 
\begin{align}
    a &= \sum_{n \in 2\mathbb{Z} }  \exp\Big{[} -\frac{( \bar{\xi}_{q} + n\sqrt{\pi}\lambda )^{2}}{ 2 ( 2\sigma_{\mathrm{gkp}}^{2} ) } \Big{]}  , 
    \nonumber\\
    b &= \sum_{n \in 2\mathbb{Z}+1 }   \exp\Big{[} -\frac{( \bar{\xi}_{q} + n\sqrt{\pi}\lambda )^{2}}{ 2 ( 2\sigma_{\mathrm{gkp}}^{2} ) } \Big{]}  , 
    \nonumber\\
    c &= \sum_{n \in 2\mathbb{Z} }    \exp\Big{[} -\frac{( \bar{\xi}_{p} + n\sqrt{\pi}/\lambda )^{2}}{ 2 ( 2\sigma_{\mathrm{gkp}}^{2} ) } \Big{]}  , 
    \nonumber\\
    d &= \sum_{n \in 2\mathbb{Z}+1 }   \exp\Big{[} -\frac{( \bar{\xi}_{p} + n\sqrt{\pi}/\lambda )^{2}}{ 2 ( 2\sigma_{\mathrm{gkp}}^{2} ) } \Big{]}   . 
\end{align}
Then, the conditional Pauli error probabilities $p_{I},p_{X},p_{Y},p_{Z}$ are computed in the same way as in \cref{eq:idling q to p}. Again, in practice, one can replace $\sum_{n \in 2\mathbb{Z} }$ by $\sum_{n \in \lbrace 0 \rbrace }$ and $\sum_{n \in 2\mathbb{Z}+1 }$ by $\sum_{n \in \lbrace -1,1 \rbrace }$ in the parameter range $9\mathrm{dB} \le \sigma_{\mathrm{gkp}}^{(\mathrm{dB})} \le 13\mathrm{dB}$.

\subsubsection{CNOT}

The detailed setup of the error-corrected CNOT gate between two GKP qubits is discussed in great detail in \cref{subsection:Maximum likelihood decoding for GKP}. Here, we focus on describing how to simulate the CNOT gate. Note that the first qubit is the control qubit encoded in the rectangular-lattice GKP code and the second qubit is the target qubit encoded in the square-lattice GKP qubit.

First, we sample eight random shifts independently from the same Gaussian distribution $\mathcal{N}(0, \sigma_{\mathrm{gkp}}^{2})$, i.e., $\xi_{1},\cdots,\xi_{8} \sim_{\mathrm{iid}} \mathcal{N}(0, \sigma_{\mathrm{gkp}}^{2}) $. Then, set $\xi_{q}^{(1)},\xi_{q}^{(2)},\xi_{p}^{(1)},\xi_{p}^{(2)}$ as follows (see \cref{eq:CNOT shift propagation qq main text}): 
\begin{align}
    \xi_{q}^{(1)} &\leftarrow \xi_{1} + \xi_{2}, 
    \nonumber\\
    \xi_{q}^{(2)} &\leftarrow \frac{1 }{  \lambda } \xi_{1} + \xi_{3} + \xi_{4} , 
    \nonumber\\
    \xi_{p}^{(1)} &\leftarrow - \frac{1 }{ \lambda } \xi_{5} + \xi_{6} + \xi_{7}, 
    \nonumber\\
    \xi_{p}^{(2)} &\leftarrow \xi_{5} + \xi_{8} . 
\end{align}
Then, given $(\xi_{q}^{(1)} , \xi_{q}^{(2)})$, compute two integers $(n_{q}^{(1)} , n_{q}^{(2)})$ by using \cref{alg:CNOTnq1nq2} (by setting the inputs to be $(\xi_{q}^{(1)} , \xi_{q}^{(2)})$, instead of $(\sqrt{2}q_{m}^{(1)} , \sqrt{2}q_{m}^{(2)})$). Similarly, given $(\xi_{p}^{(1)} , \xi_{p}^{(2)})$, compute two other integers $(n_{p}^{(1)} , n_{p}^{(2)})$ by using \cref{alg:CNOTnp1np2} with $(\xi_{p}^{(1)} , \xi_{p}^{(2)})$ as an input instead of $(\sqrt{2}p_{m}^{(1)} , \sqrt{2}p_{m}^{(2)})$. 

Given four integers $n_{q}^{(1)} , n_{q}^{(2)}, n_{p}^{(1)} , n_{p}^{(2)}$, we apply  
\begin{align}
    &\begin{cases}
    II & \textrm{if }n_{q}^{(1)}  \in 2\mathbb{Z} \textrm{ and } n_{q}^{(2)} \in 2\mathbb{Z}  \\
    XI & \textrm{if }n_{q}^{(1)}  \in 2\mathbb{Z}+1 \textrm{ and } n_{q}^{(2)} \in 2\mathbb{Z} \\
    IX & \textrm{if }n_{q}^{(1)}  \in 2\mathbb{Z} \textrm{ and } n_{q}^{(2)} \in 2\mathbb{Z}+1 \\
    XX & \textrm{if }n_{q}^{(1)}  \in 2\mathbb{Z}+1 \textrm{ and } n_{q}^{(2)} \in 2\mathbb{Z}+1
    \end{cases} , 
\end{align}
and then apply
\begin{align}
    &\begin{cases}
    II & \textrm{if }n_{p}^{(1)}  \in 2\mathbb{Z} \textrm{ and } n_{p}^{(2)} \in 2\mathbb{Z}  \\
    ZI & \textrm{if }n_{p}^{(1)}  \in 2\mathbb{Z}+1 \textrm{ and } n_{p}^{(2)} \in 2\mathbb{Z} \\
    IZ & \textrm{if }n_{p}^{(1)}  \in 2\mathbb{Z} \textrm{ and } n_{p}^{(2)} \in 2\mathbb{Z}+1 \\
    ZZ & \textrm{if }n_{p}^{(1)}  \in 2\mathbb{Z}+1 \textrm{ and } n_{p}^{(2)} \in 2\mathbb{Z}+1
\end{cases} . 
\end{align}
Then, we need to compute $16$ Pauli error probabilities (or conditional probabilities) $p_{II},p_{XI},p_{YI},p_{ZI}$,$\cdots$, $p_{IZ},p_{XZ},p_{YZ},p_{ZZ}$ which are used to determine the edge weights in the surface code matching graphs. If we do not use the extra analog information, we simply compute 
\begin{align}
    q_{XI} &= \sum_{(n_{1},n_{2}) \in (2\mathbb{Z} + 1, 2\mathbb{Z}) } \int_{ \mathcal{R}_{qq}(n_{1},n_{2}) } dx_{1}dx_{2} F_{qq}(x_{1},x_{2}) , 
    \nonumber\\
    q_{IX} &= \sum_{(n_{1},n_{2}) \in (2\mathbb{Z} , 2\mathbb{Z} + 1) } \int_{ \mathcal{R}_{qq}(n_{1},n_{2}) } dx_{1}dx_{2} F_{qq}(x_{1},x_{2}) ,
    \nonumber\\
    q_{XX} &= \sum_{(n_{1},n_{2}) \in (2\mathbb{Z}+1 , 2\mathbb{Z} + 1) } \int_{ \mathcal{R}_{qq}(n_{1},n_{2}) } dx_{1}dx_{2} F_{qq}(x_{1},x_{2}), 
\end{align}
where the integration domain $\mathcal{R}_{qq}(n_{1},n_{2})$ is defined as
\begin{align}
    &\mathcal{R}_{qq}(n_{1},n_{2}) \equiv \mathrm{Vor}_{\mathrm{CNOT}}^{[qq]}(n_{1} , n_{2}) = \Big{\lbrace } (q_{1} , q_{2}) :  
    \nonumber\\
    &\Big{|} q_{2} - n_{2}\sqrt{\pi} - \Big{(}2\lambda + \frac{1}{\lambda}\Big{)} (q_{1} -n_{1}\sqrt{\pi}\lambda ) \Big{|} <   \frac{(2\lambda^{2} + 1)}{2}\sqrt{\pi} , 
    \nonumber\\
    & \Big{|} q_{2} - n_{2}\sqrt{\pi} - \frac{1}{2\lambda}(q_{1} -n_{1}\sqrt{\pi}\lambda) \Big{|} <   \frac{\sqrt{\pi}}{2}, 
    \nonumber\\
    & \Big{|} q_{2} - n_{2}\sqrt{\pi} + 2\lambda (q_{1} -n_{1}\sqrt{\pi}\lambda) \Big{|} <   \frac{(2\lambda^{2}+1)}{2}\sqrt{\pi}  \Big{ \rbrace } , 
\end{align}
and the probability density function $F_{qq}(x_{1},x_{2})$ is given by 
\begin{align}
    &F_{qq}(x_{1},x_{2}) \equiv P_{\mathrm{CNOT}}^{[qq]}( x_{1} ,x_{2} ) 
    \nonumber\\
    &= \frac{1}{2\pi\sqrt{ (4 + \frac{1}{\lambda^{2}} ) \sigma_{\mathrm{gkp}}^{4} }}   \exp\Big{[} -\frac{ (2 + \frac{1}{\lambda^{2}}) x_{1}^{2} + 2x_{2}^{2} -\frac{2}{\lambda} x_{1}x_{2} }{2( 4 + \frac{1}{\lambda^{2}}  )\sigma_{\mathrm{gkp}}^{2} } \Big{]} . 
\end{align}
Similarly for the momentum shift errors, we compute 
\begin{align}
    q_{ZI} &= \sum_{(n_{1},n_{2}) \in (2\mathbb{Z} + 1, 2\mathbb{Z}) } \int_{ \mathcal{R}_{pp}(n_{1},n_{2}) } dx_{1}dx_{2} F_{pp}(x_{1},x_{2}) , 
    \nonumber\\
    q_{IZ} &= \sum_{(n_{1},n_{2}) \in (2\mathbb{Z} , 2\mathbb{Z} + 1) } \int_{ \mathcal{R}_{pp}(n_{1},n_{2}) } dx_{1}dx_{2} F_{pp}(x_{1},x_{2}) ,
    \nonumber\\
    q_{ZZ} &= \sum_{(n_{1},n_{2}) \in (2\mathbb{Z}+1 , 2\mathbb{Z} + 1) } \int_{ \mathcal{R}_{pp}(n_{1},n_{2}) } dx_{1}dx_{2} F_{pp}(x_{1},x_{2}), 
\end{align}
where the integration domain $\mathcal{R}_{pp}(n_{1},n_{2})$ is defined as  
\begin{align}
    &\mathcal{R}_{pp}(n_{1},n_{2}) \equiv \mathrm{Vor}_{\mathrm{CNOT}}^{[pp]}(n_{1} , n_{2}) \equiv \Big{\lbrace } (p_{1} , p_{2}) :  
    \nonumber\\
    &\Big{|} p_{2} - n_{2}\sqrt{\pi} +2\lambda (p_{1} -n_{1}\sqrt{\pi}/\lambda ) \Big{|} <   \sqrt{\pi}, 
    \nonumber\\
    & \Big{|} p_{2} - n_{2}\sqrt{\pi} + \frac{\lambda}{2\lambda^{2}+1}(p_{1} -n_{1}\sqrt{\pi}/\lambda) \Big{|} <   \frac{\sqrt{\pi}}{2}, 
    \nonumber\\
    & \Big{|} p_{2} - n_{2}\sqrt{\pi} -\frac{1}{2\lambda}(p_{1} -n_{1}\sqrt{\pi}/\lambda) \Big{|} <   \frac{(2\lambda^{2}+1)}{4\lambda^{2}}\sqrt{\pi}  \Big{ \rbrace }, 
\end{align}
and the probability density function $F_{pp}(x_{1},x_{2})$ is given by 
\begin{align}
    &F_{PP}(x_{1},x_{2}) \equiv P_{\mathrm{CNOT}}^{[pp]}( x_{1} ,x_{2} ) 
    \nonumber\\
    &=\frac{1}{2\pi\sqrt{ ( 4 + \frac{1}{\lambda^{2}} )\sigma_{\mathrm{gkp}}^{4} }}  \exp\Big{[} -\frac{ 2x_{1}^{2} + ( 2 + \frac{1}{\lambda^{2}}) x_{2}^{2} + \frac{ 2}{ \lambda} x_{1}x_{2} }{2(4+\frac{1}{\lambda^{2}})\sigma_{\mathrm{gkp}}^{2} } \Big{]}.
\end{align}
Then, the $16$ Pauli error probabilities are given by 
\begin{align}
    &\begin{bmatrix}
    p_{II} & p_{XI} & p_{IX} & p_{XX} \\ 
    p_{ZI} & p_{YI} & p_{ZX} & p_{YX} \\ 
    p_{IZ} & p_{XZ} & p_{IY} & p_{XY} \\ 
    p_{ZZ} & p_{YZ} & p_{ZY} & p_{YY} 
    \end{bmatrix} 
    \nonumber\\
    &= \begin{bmatrix}
    (1-q_{ZI}-q_{IZ} - q_{ZZ}) \\ 
    q_{ZI} \\
    q_{IZ} \\
    q_{ZZ}
    \end{bmatrix} 
    \nonumber\\
    &\times   \begin{bmatrix}
    (1-q_{XI}-q_{IX} - q_{XX}) & q_{XI} & q_{IX} & q_{XX}
    \end{bmatrix} .  \label{eq:CNOT 16 Pauli error probabilities}
\end{align}

On the other hand, if we take into account the extra analog information, we define 
\begin{align}
    \bar{\xi}_{q}^{(1)} &= \xi_{q}^{(1)} - n_{q}^{(1)}\sqrt{\pi}\lambda, \quad \bar{\xi}_{q}^{(2)} = \xi_{q}^{(2)} - n_{q}^{(2)}\sqrt{\pi} , 
    \nonumber\\
    \bar{\xi}_{p}^{(1)} &= \xi_{p}^{(1)} - n_{p}^{(1)}\sqrt{\pi}/ \lambda, \quad  \bar{\xi}_{p}^{(2)} = \xi_{p}^{(2)} - n_{p}^{(2)}\sqrt{\pi} , 
\end{align}
and then compute 
\begin{align}
    q_{XI} &= \frac{ b  }{ a+b+c+d } ,
    \nonumber\\
    q_{IX} &=  \frac{ c  }{ a+b+c+d } ,
    \nonumber\\
    q_{XX} &=  \frac{ d  }{ a+b+c+d } , 
\end{align}
where $a,b,c,d$ are given by 
\begin{align}
    a &= \sum_{(n_{1},n_{2}) \in ( 2\mathbb{Z} , 2\mathbb{Z}) } F_{qq}( \bar{\xi}_{q}^{(1)} + n_{1}\sqrt{\pi}\lambda, \bar{\xi}_{q}^{(2)} + n_{2}\sqrt{\pi} ) , 
    \nonumber\\
    b &= \sum_{(n_{1},n_{2}) \in ( 2\mathbb{Z}+1 , 2\mathbb{Z}) } F_{qq}( \bar{\xi}_{q}^{(1)} + n_{1}\sqrt{\pi}\lambda, \bar{\xi}_{q}^{(2)} + n_{2}\sqrt{\pi} ) , 
    \nonumber\\
    c &= \sum_{(n_{1},n_{2}) \in ( 2\mathbb{Z} , 2\mathbb{Z}+1) } F_{qq}( \bar{\xi}_{q}^{(1)} + n_{1}\sqrt{\pi}\lambda, \bar{\xi}_{q}^{(2)} + n_{2}\sqrt{\pi} ), 
    \nonumber\\
    d &= \sum_{(n_{1},n_{2}) \in ( 2\mathbb{Z}+1 , 2\mathbb{Z}+1 ) } F_{qq}( \bar{\xi}_{q}^{(1)} + n_{1}\sqrt{\pi}\lambda, \bar{\xi}_{q}^{(2)} + n_{2}\sqrt{\pi} ) . 
\end{align}
Similarly, for the momentum shift errors, we compute 
\begin{align}
    q_{ZI} &= \frac{ f  }{ e+f+g+h } ,
    \nonumber\\
    q_{IZ} &=  \frac{ g  }{ e+f+g+h } ,
    \nonumber\\
    q_{ZZ} &=  \frac{ h  }{ e+f+g+h } , 
\end{align}
where $e,f,g,h$ are given by 
\begin{align}
    e &= \sum_{(n_{1},n_{2}) \in ( 2\mathbb{Z} , 2\mathbb{Z}) } F_{pp}( \bar{\xi}_{p}^{(1)} + n_{1}\sqrt{\pi}/\lambda, \bar{\xi}_{p}^{(2)} + n_{2}\sqrt{\pi} ) , 
    \nonumber\\
    f &= \sum_{(n_{1},n_{2}) \in ( 2\mathbb{Z}+1 , 2\mathbb{Z}) } F_{pp}( \bar{\xi}_{p}^{(1)} + n_{1}\sqrt{\pi}/\lambda, \bar{\xi}_{p}^{(2)} + n_{2}\sqrt{\pi} ) , 
    \nonumber\\
    g &= \sum_{(n_{1},n_{2}) \in ( 2\mathbb{Z} , 2\mathbb{Z}+1) } F_{pp}( \bar{\xi}_{p}^{(1)} + n_{1}\sqrt{\pi}/\lambda, \bar{\xi}_{p}^{(2)} + n_{2}\sqrt{\pi} ), 
    \nonumber\\
    h &= \sum_{(n_{1},n_{2}) \in ( 2\mathbb{Z}+1 , 2\mathbb{Z}+1 ) } F_{pp}( \bar{\xi}_{p}^{(1)} + n_{1}\sqrt{\pi}/\lambda, \bar{\xi}_{p}^{(2)} + n_{2}\sqrt{\pi} ) . 
\end{align}
Then, the $16$ conditional Pauli error probabilities are computed by using \cref{eq:CNOT 16 Pauli error probabilities}. In practice, $( 2\mathbb{Z} , 2\mathbb{Z})$, $( 2\mathbb{Z}+1 , 2\mathbb{Z})$, $( 2\mathbb{Z} , 2\mathbb{Z}+1)$, and $( 2\mathbb{Z}+1 , 2\mathbb{Z}+1)$ can be replaced by $\lbrace (0,0) \rbrace$, $\lbrace (1,0) , (-1,0) \rbrace$, $\lbrace (0,1), (0,-1) \rbrace$, and $\lbrace (1,1),(1,-1),(-1,1),(-1,-1) \rbrace$, respectively. 

\subsubsection{CZ} 

We consider an error-corrected CZ gate between two GKP qubits (see \cref{fig:GKP error corrected two qubit gates} (d)). We assume that the first qubit is encoded in the rectangular-lattice GKP code and the second qubit is encoded in the square-lattice GKP qubit. Thus, unlike the usual CZ gate, there is no permutation symmetry unless the first qubit is also encoded in the square-lattice GKP code. 

To simulate the CZ gate, we first sample eight random shifts independently from the Gaussian distribution $\mathcal{N}(0, \sigma_{\mathrm{gkp}}^{2} )$, i.e., $\xi_{1},\cdots,\xi_{8} \sim_{\mathrm{iid}} \mathcal{N}( 0, \sigma_{\mathrm{gkp}}^{2} ) $ and set 
\begin{align}
    \xi_{q}^{(1)} &\leftarrow \xi_{1} + \xi_{2}, 
    \nonumber\\
    \xi_{q}^{(2)} &\leftarrow \xi_{5} + \xi_{6}, 
    \nonumber\\
    \xi_{p}^{(1)} &\leftarrow \frac{1}{\lambda}\xi_{5} + \xi_{7} + \xi_{8}  , 
    \nonumber\\
    \xi_{p}^{(2)} &\leftarrow \frac{1}{\lambda}\xi_{1} + \xi_{3} + \xi_{4} . 
\end{align}
Note that $\xi_{q}^{(1)}$ is correlated with $\xi_{p}^{(2)}$ and $\xi_{p}^{(1)}$ is correlated with $\xi_{q}^{(2)}$. First, given $(\xi_{q}^{(1)} , \xi_{p}^{(2)} )$, we need to find two integers $(n_{q}^{(1)},n_{p}^{(2)})$ such that $(\xi_{q}^{(1)} , \xi_{p}^{(2)} )$ is in the Voronoi cell $\mathcal{R}_{qp}(n_{q}^{(1)} , n_{p}^{(2)} )$, where $\mathcal{R}_{qp}(n_{1} , n_{2} )$ is defined as 
\begin{align}
    &\mathcal{R}_{qp}(n_{1},n_{2}) \equiv \mathrm{Vor}_{\mathrm{CZ}}^{[qp]}(n_{1} , n_{2}) \equiv \Big{\lbrace } (q_{1} , p_{1}) :  
    \nonumber\\
    &\Big{|} p_{2} - n_{2}\sqrt{\pi} - \Big{(}2\lambda + \frac{1}{\lambda}\Big{)} (q_{1} -n_{1}\sqrt{\pi}\lambda ) \Big{|} <   \frac{(2\lambda^{2} + 1)}{2}\sqrt{\pi} , 
    \nonumber\\
    & \Big{|} p_{2} - n_{2}\sqrt{\pi} - \frac{1}{2\lambda}(q_{1} -n_{1}\sqrt{\pi}\lambda) \Big{|} <   \frac{\sqrt{\pi}}{2}, 
    \nonumber\\
    & \Big{|} p_{2} - n_{2}\sqrt{\pi} + 2\lambda (q_{1} -n_{1}\sqrt{\pi}\lambda) \Big{|} <   \frac{(2\lambda^{2}+1)}{2}\sqrt{\pi}  \Big{ \rbrace } . 
\end{align}
This is done in the exactly the same way as in \cref{alg:CNOTnq1nq2}. 

Similarly, given $(\xi_{p}^{(1)} , \xi_{q}^{(2)} )$, we need to find two other integers $(n_{p}^{(1)},n_{q}^{(2)})$ such that $(\xi_{p}^{(1)} , \xi_{q}^{(2)} )$ is in the Voronoi cell $\mathcal{R}_{pq}(n_{p}^{(1)} , n_{q}^{(2)} )$, where $\mathcal{R}_{pq}(n_{1} , n_{2} )$ is defined as
\begin{align}
    &\mathcal{R}_{pq}(n_{1},n_{2}) \equiv \mathrm{Vor}_{\mathrm{CZ}}^{[pq]}(n_{1} , n_{2}) \equiv \Big{\lbrace } (p_{1} , q_{2}) :  
    \nonumber\\
    &\Big{|} q_{2} - n_{2}\sqrt{\pi} - 2\lambda (p_{1} -n_{1}\sqrt{\pi}/\lambda ) \Big{|} <  \sqrt{\pi} , 
    \nonumber\\
    & \Big{|} q_{2} - n_{2}\sqrt{\pi} - \frac{\lambda}{2\lambda^{2}+1}(p_{1} -n_{1}\sqrt{\pi}/\lambda) \Big{|} <   \frac{\sqrt{\pi}}{2}, 
    \nonumber\\
    & \Big{|} q_{2} - n_{2}\sqrt{\pi} + \frac{1}{2\lambda} (p_{1} -n_{1}\sqrt{\pi}/\lambda) \Big{|} <   \frac{(2\lambda^{2}+1)}{4\lambda^{2}}\sqrt{\pi}  \Big{ \rbrace } .
\end{align}
This can be done by using a slightly modified version of \cref{alg:CNOTnp1np2} with proper sign changes. 

Similarly as in the case of the CNOT gate, given four integers $n_{q}^{(1)} , n_{q}^{(2)}, n_{p}^{(1)} , n_{p}^{(2)}$, we apply  
\begin{align}
    &\begin{cases}
    II & \textrm{if }n_{q}^{(1)}  \in 2\mathbb{Z} \textrm{ and } n_{q}^{(2)} \in 2\mathbb{Z}  \\
    XI & \textrm{if }n_{q}^{(1)}  \in 2\mathbb{Z}+1 \textrm{ and } n_{q}^{(2)} \in 2\mathbb{Z} \\
    IX & \textrm{if }n_{q}^{(1)}  \in 2\mathbb{Z} \textrm{ and } n_{q}^{(2)} \in 2\mathbb{Z}+1 \\
    XX & \textrm{if }n_{q}^{(1)}  \in 2\mathbb{Z}+1 \textrm{ and } n_{q}^{(2)} \in 2\mathbb{Z}+1
    \end{cases} , 
\end{align}
and then apply
\begin{align}
    &\begin{cases}
    II & \textrm{if }n_{p}^{(1)}  \in 2\mathbb{Z} \textrm{ and } n_{p}^{(2)} \in 2\mathbb{Z}  \\
    ZI & \textrm{if }n_{p}^{(1)}  \in 2\mathbb{Z}+1 \textrm{ and } n_{p}^{(2)} \in 2\mathbb{Z} \\
    IZ & \textrm{if }n_{p}^{(1)}  \in 2\mathbb{Z} \textrm{ and } n_{p}^{(2)} \in 2\mathbb{Z}+1 \\
    ZZ & \textrm{if }n_{p}^{(1)}  \in 2\mathbb{Z}+1 \textrm{ and } n_{p}^{(2)} \in 2\mathbb{Z}+1
\end{cases} . 
\end{align}
Then, in the case where we neglect the extra analog information, we compute 
\begin{align}
    q_{XI} &= \sum_{(n_{1},n_{2}) \in (2\mathbb{Z} + 1, 2\mathbb{Z}) } \int_{ \mathcal{R}_{qp}(n_{1},n_{2}) } dx_{1}dx_{2} F_{qp}(x_{1},x_{2}) , 
    \nonumber\\
    q_{IZ} &= \sum_{(n_{1},n_{2}) \in (2\mathbb{Z} , 2\mathbb{Z} + 1) } \int_{ \mathcal{R}_{qp}(n_{1},n_{2}) } dx_{1}dx_{2} F_{qp}(x_{1},x_{2}) ,
    \nonumber\\
    q_{XZ} &= \sum_{(n_{1},n_{2}) \in (2\mathbb{Z}+1 , 2\mathbb{Z} + 1) } \int_{ \mathcal{R}_{qp}(n_{1},n_{2}) } dx_{1}dx_{2} F_{qp}(x_{1},x_{2}), 
\end{align}
and 
\begin{align}
    q_{ZI} &= \sum_{(n_{1},n_{2}) \in (2\mathbb{Z} + 1, 2\mathbb{Z}) } \int_{ \mathcal{R}_{pq}(n_{1},n_{2}) } dx_{1}dx_{2} F_{pq}(x_{1},x_{2}) , 
    \nonumber\\
    q_{IX} &= \sum_{(n_{1},n_{2}) \in (2\mathbb{Z} , 2\mathbb{Z} + 1) } \int_{ \mathcal{R}_{pq}(n_{1},n_{2}) } dx_{1}dx_{2} F_{pq}(x_{1},x_{2}) ,
    \nonumber\\
    q_{ZX} &= \sum_{(n_{1},n_{2}) \in (2\mathbb{Z}+1 , 2\mathbb{Z} + 1) } \int_{ \mathcal{R}_{pq}(n_{1},n_{2}) } dx_{1}dx_{2} F_{pq}(x_{1},x_{2}), 
\end{align}
to get the $16$ Pauli error probabilities 
\begin{align}
    &\begin{bmatrix}
    p_{II} & p_{ZI} & p_{IX} & p_{ZX} \\ 
    p_{XI} & p_{YI} & p_{XX} & p_{YX} \\ 
    p_{IZ} & p_{ZZ} & p_{IY} & p_{ZY} \\ 
    p_{XZ} & p_{YZ} & p_{XY} & p_{YY}
    \end{bmatrix} 
    \nonumber\\
    &= \begin{bmatrix}
    (1-q_{XI}-q_{IZ} - q_{XZ}) \\ 
    q_{XI} \\
    q_{IZ} \\
    q_{XZ}
    \end{bmatrix} 
    \nonumber\\
    &\times   \begin{bmatrix}
    (1-q_{ZI}-q_{IX} - q_{ZX}) & q_{ZI} & q_{IX} & q_{ZX}
    \end{bmatrix} . \label{eq:CZ 16 Pauli error probabilities}
\end{align}
Here, the probability density functions $F_{qp}(x_{1},x_{2})$ and $F_{pq}(x_{1},x_{2})$ are defined as
\begin{align}
    &F_{qp}(x_{1},x_{2}) \equiv  P_{\mathrm{CZ}}^{[qp]}( x_{1} ,x_{2} ) 
    \nonumber\\
    &= \frac{1}{2\pi\sqrt{ (4 + \frac{1}{\lambda^{2}} ) \sigma_{\mathrm{gkp}}^{4} }}   \exp\Big{[} -\frac{ (2 + \frac{1}{\lambda^{2}}) x_{1}^{2} + 2x_{2}^{2} -\frac{2}{\lambda} x_{1}x_{2} }{2( 4 + \frac{1}{\lambda^{2}}  )\sigma_{\mathrm{gkp}}^{2} } \Big{]} , 
\end{align}
and 
\begin{align}
    &F_{pq}(x_{1},x_{2}) \equiv  P_{\mathrm{CZ}}^{[pq]}( x_{1} ,x_{2} ) 
    \nonumber\\
    &=\frac{1}{2\pi\sqrt{ ( 4 + \frac{1}{\lambda^{2}} )\sigma_{\mathrm{gkp}}^{4} }}  \exp\Big{[} -\frac{ 2x_{1}^{2} + ( 2 + \frac{1}{\lambda^{2}}) x_{2}^{2} - \frac{ 2}{ \lambda} x_{1}x_{2} }{2(4+\frac{1}{\lambda^{2}})\sigma_{\mathrm{gkp}}^{2} } \Big{]} . 
\end{align}

If we use the extra analog information, we define  
\begin{align}
    \bar{\xi}_{q}^{(1)} &= \xi_{q}^{(1)} - n_{q}^{(1)}\sqrt{\pi}\lambda, \quad \bar{\xi}_{q}^{(2)} = \xi_{q}^{(2)} - n_{q}^{(2)}\sqrt{\pi} , 
    \nonumber\\
    \bar{\xi}_{p}^{(1)} &= \xi_{p}^{(1)} - n_{p}^{(1)}\sqrt{\pi}/ \lambda, \quad  \bar{\xi}_{p}^{(2)} = \xi_{p}^{(2)} - n_{p}^{(2)}\sqrt{\pi} , 
\end{align}
and then compute 
\begin{align}
    q_{XI} &= \frac{ b  }{ a+b+c+d } ,
    \nonumber\\
    q_{IZ} &=  \frac{ c  }{ a+b+c+d } ,
    \nonumber\\
    q_{XZ} &=  \frac{ d  }{ a+b+c+d } , 
\end{align}
where $a,b,c,d$ are given by 
\begin{align}
    a &= \sum_{(n_{1},n_{2}) \in ( 2\mathbb{Z} , 2\mathbb{Z}) } F_{qp}( \bar{\xi}_{q}^{(1)} + n_{1}\sqrt{\pi}\lambda, \bar{\xi}_{p}^{(2)} + n_{2}\sqrt{\pi} ) , 
    \nonumber\\
    b &= \sum_{(n_{1},n_{2}) \in ( 2\mathbb{Z}+1 , 2\mathbb{Z}) } F_{qp}( \bar{\xi}_{q}^{(1)} + n_{1}\sqrt{\pi}\lambda, \bar{\xi}_{p}^{(2)} + n_{2}\sqrt{\pi} ) , 
    \nonumber\\
    c &= \sum_{(n_{1},n_{2}) \in ( 2\mathbb{Z} , 2\mathbb{Z}+1) } F_{qp}( \bar{\xi}_{q}^{(1)} + n_{1}\sqrt{\pi}\lambda, \bar{\xi}_{p}^{(2)} + n_{2}\sqrt{\pi} ), 
    \nonumber\\
    d &= \sum_{(n_{1},n_{2}) \in ( 2\mathbb{Z}+1 , 2\mathbb{Z}+1 ) } F_{qp}( \bar{\xi}_{q}^{(1)} + n_{1}\sqrt{\pi}\lambda, \bar{\xi}_{p}^{(2)} + n_{2}\sqrt{\pi} ) . 
\end{align}
Similarly, we also compute 
\begin{align}
    q_{ZI} &= \frac{ f  }{ e+f+g+h } ,
    \nonumber\\
    q_{IX} &=  \frac{ g  }{ e+f+g+h } ,
    \nonumber\\
    q_{ZX} &=  \frac{ h  }{ e+f+g+h } , 
\end{align}
where $e,f,g,h$ are given by 
\begin{align}
    e &= \sum_{(n_{1},n_{2}) \in ( 2\mathbb{Z} , 2\mathbb{Z}) } F_{pq}( \bar{\xi}_{p}^{(1)} + n_{1}\sqrt{\pi}/\lambda, \bar{\xi}_{q}^{(2)} + n_{2}\sqrt{\pi} ) , 
    \nonumber\\
    f &= \sum_{(n_{1},n_{2}) \in ( 2\mathbb{Z}+1 , 2\mathbb{Z}) } F_{pq}( \bar{\xi}_{p}^{(1)} + n_{1}\sqrt{\pi}/\lambda, \bar{\xi}_{q}^{(2)} + n_{2}\sqrt{\pi} ) , 
    \nonumber\\
    g &= \sum_{(n_{1},n_{2}) \in ( 2\mathbb{Z} , 2\mathbb{Z}+1) } F_{pq}( \bar{\xi}_{p}^{(1)} + n_{1}\sqrt{\pi}/\lambda, \bar{\xi}_{q}^{(2)} + n_{2}\sqrt{\pi} ), 
    \nonumber\\
    h &= \sum_{(n_{1},n_{2}) \in ( 2\mathbb{Z}+1 , 2\mathbb{Z}+1 ) } F_{pq}( \bar{\xi}_{p}^{(1)} + n_{1}\sqrt{\pi}/\lambda, \bar{\xi}_{q}^{(2)} + n_{2}\sqrt{\pi} ) . 
\end{align}
Then, the $16$ conditional Pauli error probabilities are computed by using \cref{eq:CZ 16 Pauli error probabilities}. In practice, $( 2\mathbb{Z} , 2\mathbb{Z})$, $( 2\mathbb{Z}+1 , 2\mathbb{Z})$, $( 2\mathbb{Z} , 2\mathbb{Z}+1)$, and $( 2\mathbb{Z}+1 , 2\mathbb{Z}+1)$ can be replaced by $\lbrace (0,0) \rbrace$, $\lbrace (1,0) , (-1,0) \rbrace$, $\lbrace (0,1), (0,-1) \rbrace$, and $\lbrace (1,1),(1,-1),(-1,1),(-1,-1) \rbrace$, respectively.

\subsection{Edge weights in the matching graphs}
\label{subappedix:Edge weights in the matching graphs}

\begin{figure*}
	\centering
	\subfloat[\label{fig:XMatchingGraph2D}]{%
		\includegraphics[width=0.48\textwidth]{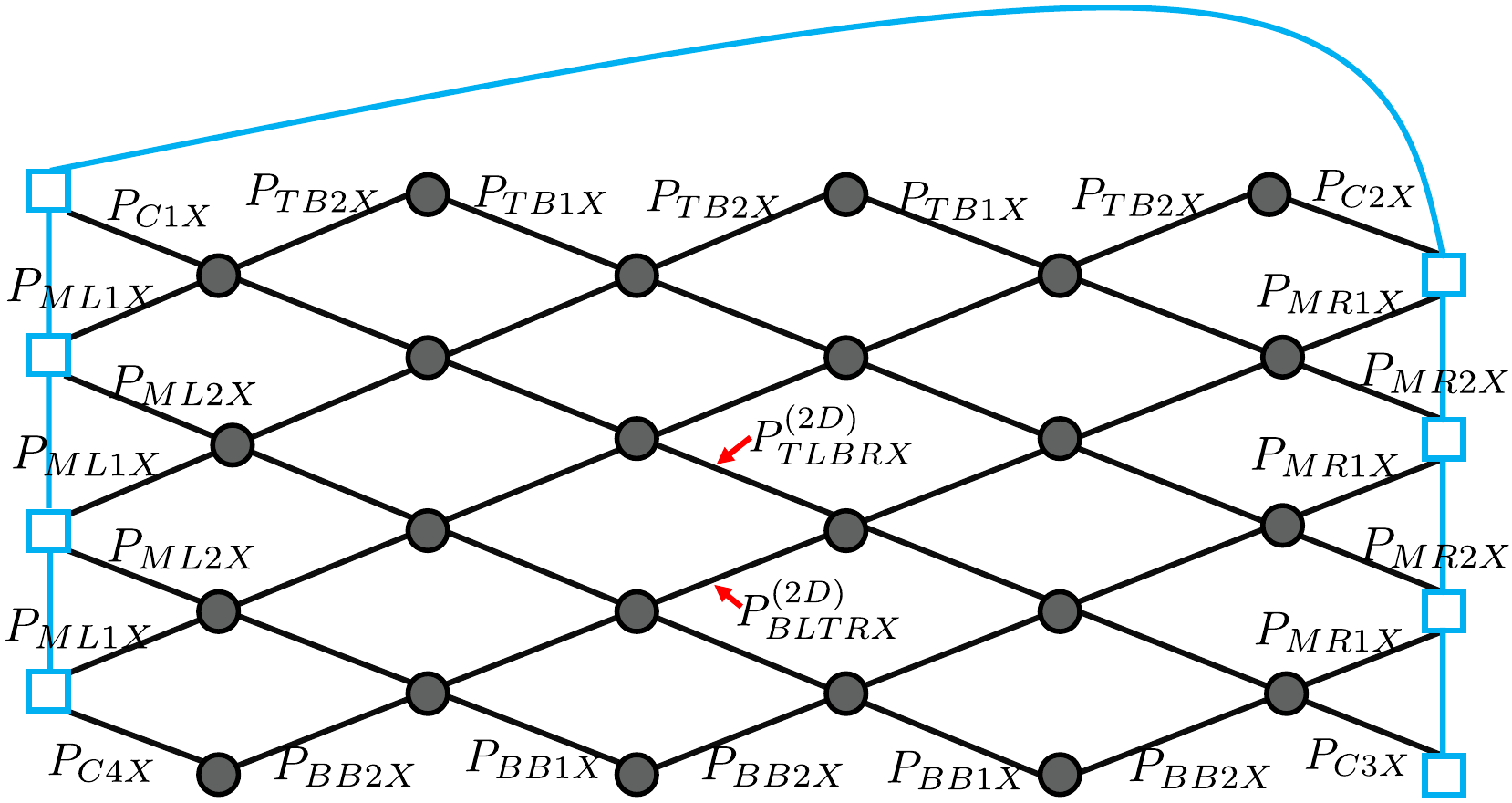}
	}
	\subfloat[\label{fig:XMatchingGraph3D}]{%
		\includegraphics[width=0.48\textwidth]{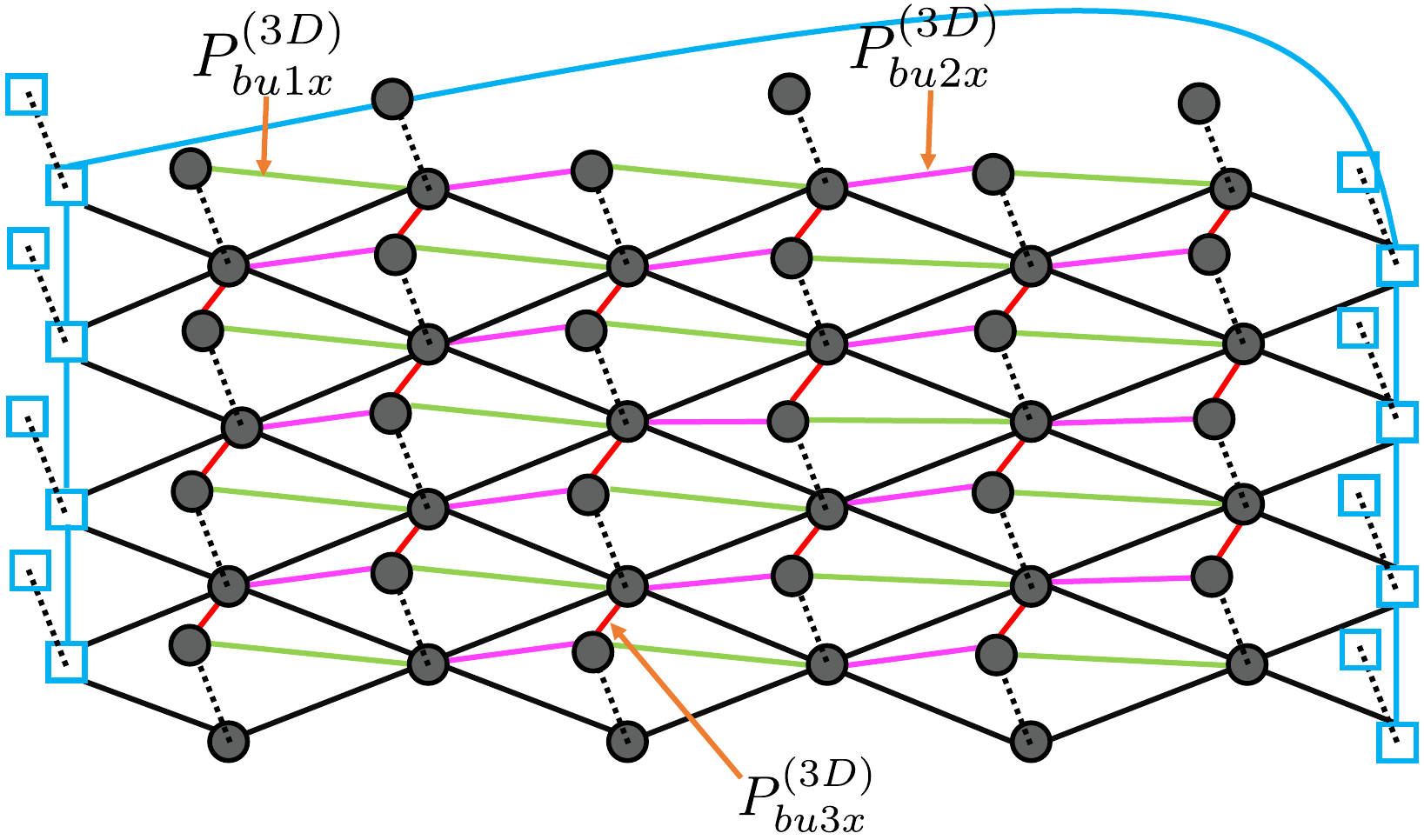}
	}
	\vfill
	\subfloat[\label{fig:ZMatchingGraph2D}]{%
		\includegraphics[width=0.48\textwidth]{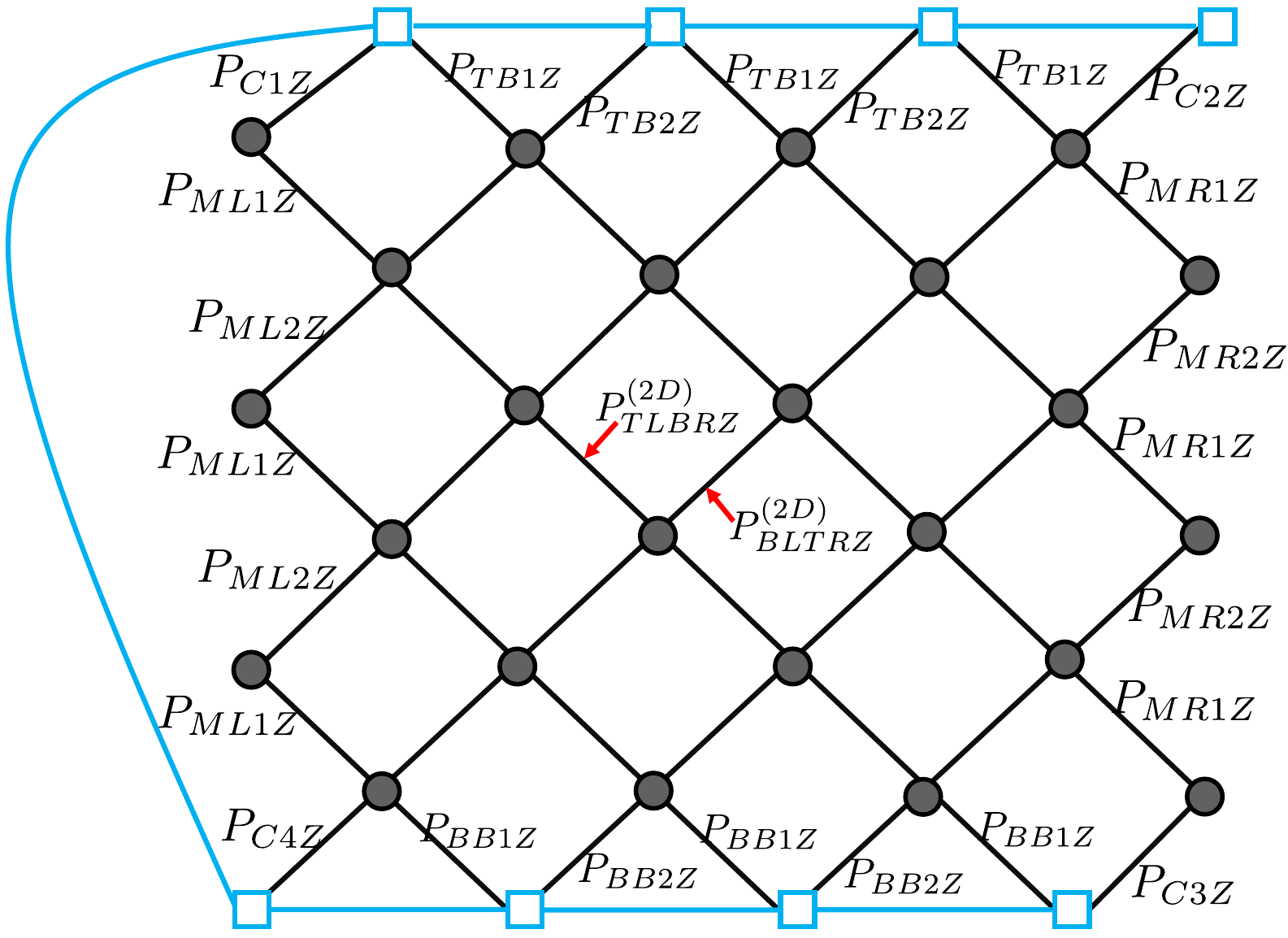}
	}
	\subfloat[\label{fig:ZMatchingGraph3D}]{%
		\includegraphics[width=0.48\textwidth]{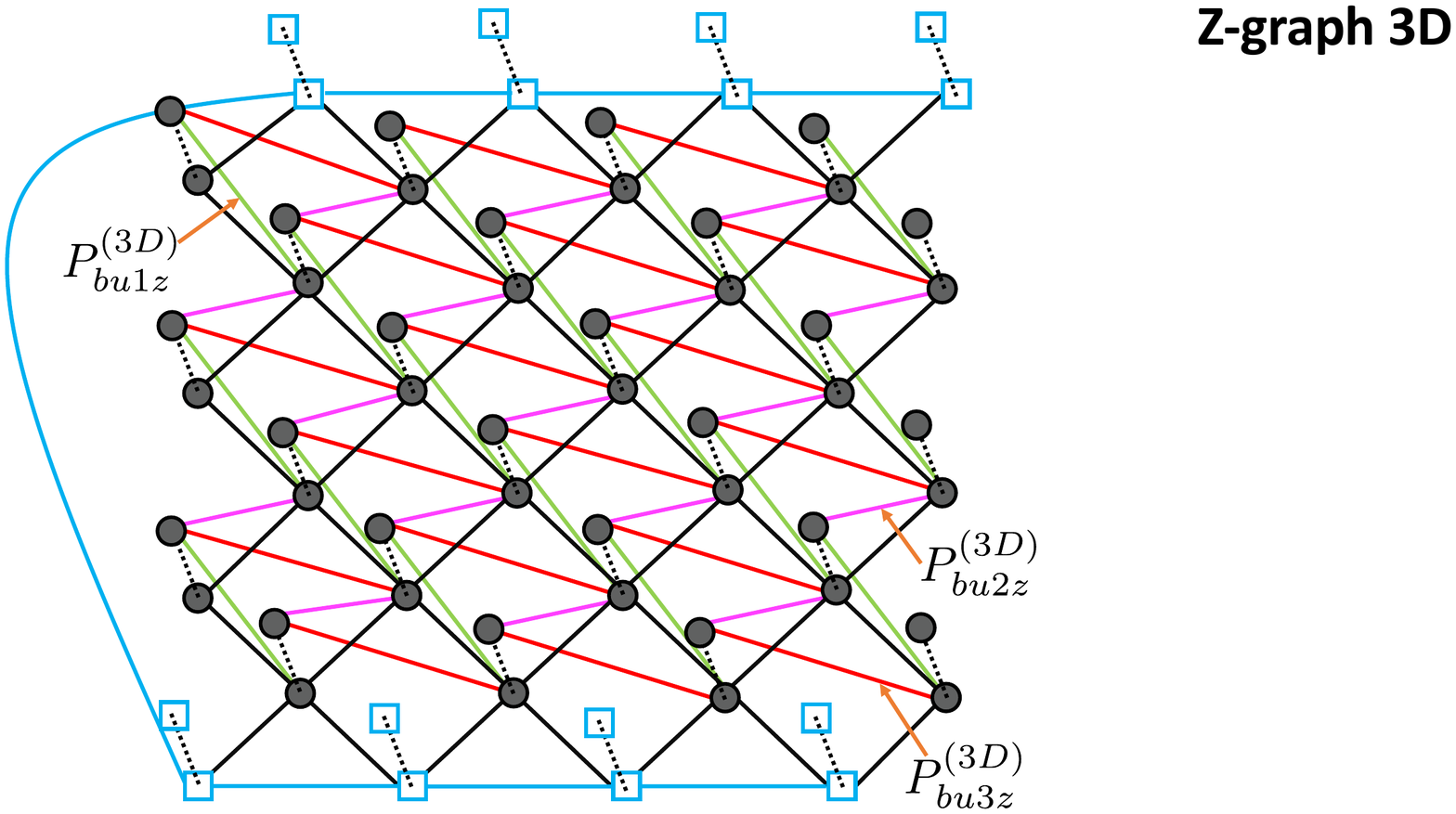}
	}

	\caption{\label{fig:MatchingGraphs} Matching graphs corresponding to the surface code lattice in \cref{fig:SurfaceCodeLattice}. (a) Two-dimensional matching graph with labeled boundary and bulk-type edge for $X$-type stabilizer measurement outcomes of the $d_x=d_z=7$ surface code. (b) Same matching graph as in (a) but with vertical (dashed) and space-time correlated edges (with each type identified by a different color) connecting to an adjacent two-dimensional layer. (c) Two-dimensional matching graph with labeled boundary and bulk-type edge for $Z$-type stabilizer measurement outcomes of the $d_x=d_z=7$ surface code. (d) Same matching graph as in (c) but with vertical and space-time correlated edges connecting to an adjacent two-dimensional layer. In all graphs, blue edges are boundary edges with zero weight, and blue squares correspond to virtual vertices connecting the boundary edges. Grey vertices correspond to measurement outcomes of the surface code stabilizers.  }
\end{figure*}

In this section we describe the methods used to perform the simulations of the surface-GKP code for various code distances. As a first step, in \cref{fig:MatchingGraphs}, we illustrate the matching graphs used in the implementation of the MWPM algorithm with weighted edges. Two-dimensional bulk and boundary edges are distinguished by their associated labels since such edges will have different weights. Vertical edges used for detecting measurement errors (see Ref.~\cite{FMMC12}) are dashed. Lastly, consider the space-time correlated edges in \cref{fig:XMatchingGraph3D,fig:ZMatchingGraph3D} used for detecting errors arising from CNOT/CZ failures which result in errors with a spatial and temporal component (see Refs.~\cite{FowlerSpaceTime2011,ChamberlandPRX,CKYZ20,Chamberland2020_building}). Such edges are distinguished by their labels and associated colors. For instance, space-time correlated edges colored in green in \cref{fig:XMatchingGraph3D} are labelled $P^{(3D)}_{bu1x}$. 

We now describe how to compute the weights for edges in the matching graphs of \cref{fig:MatchingGraphs}. As shown in \cref{subappedix:Noise model}, if the extra analog information from GKP error corrections is used, the conditional probabilities for the failure rates of each circuit component is computed dynamically for every gate, idle, state-preparation and measurement location at each time step. As such, all circuit components can have different conditional probabilities at every location and in each syndrome measurement round. Hence edge weights for the matching graphs cannot be pre-computed and must be updated dynamically. 

Given the above, let $t$ be the number of syndrome measurement rounds for a surface with minimum weight logical $X$ operator $d_x$ and minimum weight logical $Z$ operator $d_z$.  We define a $ 1 \times t(3n + 62m_1 + 62m_2)$ matrix $P_{e}$ where $n = d_xd_z$, $m_1 = (d_x+1)(d_z-1)/2$ and $m_2 = (d_z+1)(d_x-1)/2$. Here, $t$ is the number of syndrome measurement rounds, $n$ corresponds to the number of data qubits, $m_1$ corresponds to the number of $X$-type stabilizers and $m_2$ corresponds to the number of $Z$-type stabilizers.

The $P_{e}$ matrix can be decomposed into $t$ blocks and each block can be decomposed into three sub-matrices $N^{(j)} = [ M^{(j)}_1 | M^{(j)}_2 | M^{(j)}_3 ]$ where $j \in \{1,2, \cdots, t\}$. The matrix $M^{(j)}_1$ is $1 \times 3n$, $M^{(j)}_2$ is $1 \times 62m_1$ and $M^{(j)}_3$ is $1 \times 62m_2$. We now describe the constituents of each sub-matrix for a given block. 

The $3n$ columns of the sub-matrix $M^{(j)}_1$ can be decomposed into $n$ blocks, where the $k$'th block stores the conditional probabilities for $X$, $Y$ and $Z$ Pauli errors arising from idling fault locations on the $k$'th data qubit and in the $j$'th syndrome measurement round (ancilla qubits have no idling fault locations). Note that data qubits on the boundaries of the lattice can have more than one idling fault locations in a given syndrome measurement round (for instance, if three two-qubit gates interact with a data qubit with an idling location in between two gates). However, because we do not consider photon loss in this work, all idling fault locations can be set at the beginning of a syndrome measurement round and thus incorporates the measurement and reset wait time. As such, we can write
\begin{align}
    M^{(j)}_1 = [P^{(X)}_{q_1,j}, P^{(Y)}_{q_1,j}, P^{(Z)}_{q_1,j}, \cdots , P^{(X)}_{q_n,j}, P^{(Y)}_{q_n,j}, P^{(Z)}_{q_n,j}],
    \label{eq:M1Def}
\end{align}
where for instance $P^{(X)}_{q_k,j}$ is the conditional probability of a Pauli $X$ error on the $k$'th data qubit during the $j$'th syndrome measurement round arising from an idling fault location. 

The $62m_1$ columns of $M^{(j)}_2$ can be decomposed into $m_1$ blocks, where the $k$'th block stores all conditional probabilities arising from fault locations on the $k$'th ancilla for $X$-type stabilizer measurements during the $j$'th syndrome measurement round. We note that by default, we consider a fault location for a two-qubit gate be on the control-qubit of such a gate, and we set the control qubit af a CZ gate to be on the ancilla qubit. On a given ancilla qubit, there are 62 fault locations: one for the $|+\rangle$ state preparation, one for the $X$-basis measurement and each of the four two-qubit gates can fail introducing a two-qubit Pauli error $P_1 \otimes P_2$ (of which there are 15 non-trivial possibilities). Since the ordering is important for what follows, we define
\begin{align}
    \tilde{P}^{(t_s;j)}_{a^{(x)}_m} &= [P^{(t_s;j)}_{IX}, P^{(t_s;j)}_{IZ}, P^{(t_s;j)}_{IY}, 
    \nonumber\\
    &\qquad P^{(t_s;j)}_{XI}, P^{(t_s;j)}_{XX}, P^{(t_s;j)}_{XZ}, P^{(t_s;j)}_{XY}, 
    \nonumber\\
    &\qquad P^{(t_s;j)}_{ZI}, P^{(t_s;j)}_{ZX}, P^{(t_s;j)}_{ZZ}, P^{(t_s;j)}_{ZY}, 
    \nonumber\\
    &\qquad P^{(t_s;j)}_{YI}, P^{(t_s;j)}_{YX}, P^{(t_s;j)}_{YZ}, P^{(t_s;j)}_{YY}],
    \label{eq:PaxmCond}
\end{align}
where $t_s \in \{ 1,2,3,4\}$, $a^{(x)}_m$ corresponds to the $m$'th ancilla for $X$-type stabilizers and $P^{(t_s;j)}_{P_1P_2}$ corresponds to conditional probability for the error $P_1 \otimes P_2$ occurring on a CNOT gate in time-step $t_s$ during the $j$'th syndrome measurement round. Hence $\tilde{P}^{(t_s;j)}_{a^{(x)}_m}$ stores the conditional probabilities of all 15 two-qubit Pauli errors for the CNOT gate applied in time step $t_s$ during the $j$'th syndrome measurement round and which interacts with the ancilla $a^{(x)}_m$. We note that for ancillas used in measuring a weight-two $X$-type stabilizer, if no CNOT gates are applied in time step $t_s$, we set all terms in \cref{eq:PaxmCond} to zero. 
Using \cref{eq:PaxmCond}, we can write 
\begin{align}
    M^{(j)}_2 &= [P^{(S_z)}_{a^{(x)}_1}, \tilde{P}^{(t_1;j)}_{a^{(x)}_1}, \tilde{P}^{(t_2;j)}_{a^{(x)}_1}, \tilde{P}^{(t_3;j)}_{a^{(x)}_1}, \tilde{P}^{(t_4;j)}_{a^{(x)}_1}, P^{(M_z)}_{a^{(x)}_1}, \cdots, \nonumber \\ &P^{(S_z)}_{a^{(x)}_{m_1}},\tilde{P}^{(t_1;j)}_{a^{(x)}_{m_1}}, \tilde{P}^{(t_2;j)}_{a^{(x)}_{m_1}}, \tilde{P}^{(t_3;j)}_{a^{(x)}_{m_1}}, \tilde{P}^{(t_4;j)}_{a^{(x)}_{m_1}}, P^{(M_z)}_{a^{(x)}_{m_1}}],
    \label{eq:Mj2Def}
\end{align}
where $P^{(S_z)}_{a^{(x)}_l}$ and $P^{(M_z)}_{a^{(x)}_l}$ are the conditional probabilities for state-preparation and measurement errors on the $l$'th ancilla qubit for $X$-type stabilizer measurements. We also note that we use the convention where ancillas for $X$-type stabilizers are labeled top-bottom and left to right in the surface code lattice shown in \cref{fig:SurfaceCodeLattice}.

The $62m_2$ columns of $M^{(j)}_3$ into $m_2$ blocks where the $k$'th block stores all conditional probabilities arising from fault locations on the $k$'th ancilla for $Z$-type stabilizer measurements during the $j$'th syndrome measurement round. We define the vector $\tilde{P}^{(t_s;j)}_{a^{(z)}_m}$ in the same way as in \cref{eq:PaxmCond} but where the terms $P^{(t_s;j)}_{P_1P_2}$ correspond to the conditional probability of a $P_1 \otimes P_2$ error accuring during a CZ gate in time step $t_s$ during the $j$'th syndrome measurement round. As such, we can write
\begin{align}
    M^{(j)}_3 &= [P^{(S_z)}_{a^{(z)}_1}, \tilde{P}^{(t_1;j)}_{a^{(z)}_1}, \tilde{P}^{(t_2;j)}_{a^{(z)}_1}, \tilde{P}^{(t_3;j)}_{a^{(z)}_1}, \tilde{P}^{(t_4;j)}_{a^{(z)}_1}, P^{(M_z)}_{a^{(z)}_1}, \cdots, \nonumber \\ & P^{(S_z)}_{a^{(z)}_{m_2}}, \tilde{P}^{(t_1;j)}_{a^{(z)}_{m_2}}, \tilde{P}^{(t_2;j)}_{a^{(z)}_{m_2}}, \tilde{P}^{(t_3;j)}_{a^{(z)}_{m_2}}, \tilde{P}^{(t_4;j)}_{a^{(z)}_{m_2}}, P^{(M_z)}_{a^{(z)}_{m_2}}],
    \label{eq:Mj3Def}
\end{align}
where $P^{(S_z)}_{a^{(z)}_l}$ and $P^{(M_z)}_{a^{(z)}_l}$ are the conditional probabilities for state-preparation and measurement errors on the $l$'th ancilla qubit for $Z$-type stabilizer measurements.

From the above, we can then write the $P_e$ matrix as
\begin{align}
    P_e = [N^{(1)}, N^{(2)}, \cdots, N^{(t)}],
\end{align}
where $N^{(j)}$ is computed using \cref{eq:M1Def,eq:Mj2Def,eq:Mj3Def}. After performing all $t$ syndrome measurement rounds and generating the $P_e$ matrix, edge weights are computed by extracting the appropriate terms in the $P_e$ matrix. 

We now illustrate the methods to compute the edge weight for bulk edges labeled $P^{(2D)}_{BLTRX}$ in \cref{fig:XMatchingGraph2D} using the $P_e$ matrix formalism (edge weights for other edges can be computed in an analogous fashion). In what follows, we will refer to such an edge as $e^{(2D)}_{BLTRX}$.

\begin{figure*}
    \centering
    \includegraphics[width = 0.7\textwidth]{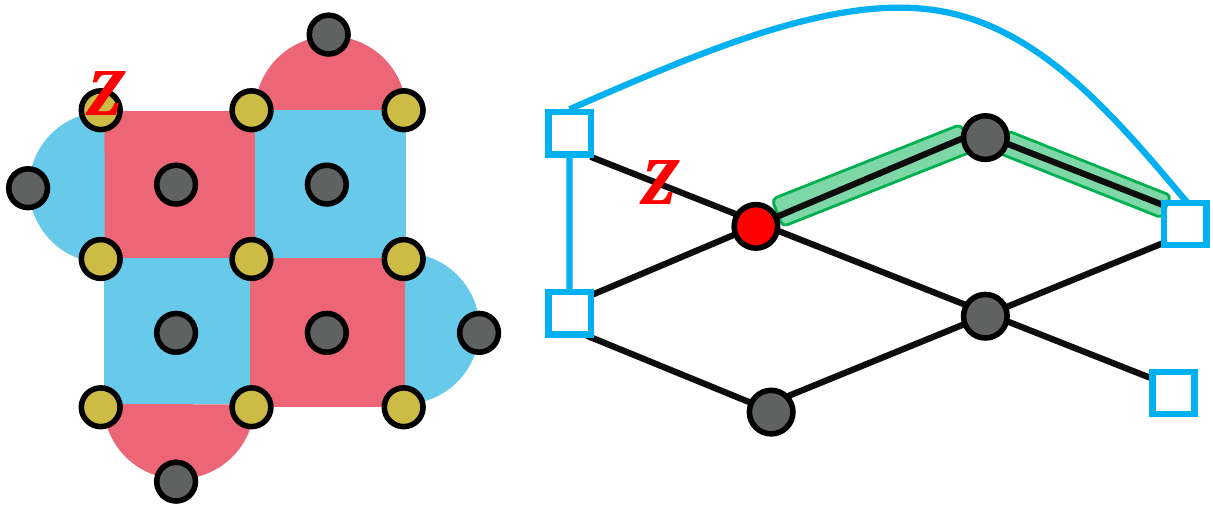}
    \caption{On the left, we illustrate an example of a single failure resulting in a $Z$ data qubit error (shown in red) for a $d_x = d_z = 3$ surface code. On the right, we show the decoding graph for $Z$-type errors, the resulting highlighted vertex and the actual path chosen during MWPM is highlighted in green. Such a path is chosen due to the lower edge weights (along the green path) computed from the conditional probabilities of the GKP error correction protocol. Such events are very rare and have a negligible effect on the total logical failure rates, especially for code distances $d \ge 5$.}
    \label{fig:ExampleSingleFault}
\end{figure*}

Let $e_j \in \{1,2,\cdots, n\}$ correspond to the data qubit represented by t$e^{(2D)}_{BLTRX}$. Upon looking at the circuit in \cref{fig:SurfaceCodeLattice}, it can be seen that to leading order, the only fault locations which can result in $e^{(2D)}_{BLTRX}$ being highlighted when implementing MWPM are failures on CNOT gates in the second and fifth time step of an $X$-type stabilizer whose target qubit is $e_j$. In addition, a $Z$ or $Y$ data qubit error during an idling location on $e_j$ can also result in $e^{(2D)}_{BLTRX}$ to be highlighted. For the CNOT gate in the second time step, only errors from the set $\{ Z\otimes Z, Z\otimes Y, Y \otimes Z, Y \otimes Y\}$ can contribute to $e^{(2D)}_{BLTRX}$ being highlighted. Similarly, for the CNOT in time step 5, only errors from the set $\{ I \otimes Z, X \otimes Z, I \otimes Y, X \otimes Y\}$ can contribute to $e^{(2D)}_{BLTRX}$ being highlighted. Next we define the following functions:
\begin{align}
    m^{(t_j)}_{(1,k,j)} &= (3n + 62m_1 + 62m_2)(j-1) \nonumber \\ 
    &+ 3n + 62(k-1) + 1 + 15(j-1),
    \label{eq:m1k}
\end{align}
\begin{align}
    m^{(t_j)}_{(2,k,j)} &= (3n + 62m_1 + 62m_2)(j-1) \nonumber \\ 
    &+ 3n + 62m_1 + 62(k-1) + 1 + 15(j-1).
    \label{eq:m2k}
\end{align}
and
\begin{align}
    n_j(d) = (3n + 62m_1 + 62m_2)(j-1) + 3(d-1).
    \label{eq:nk}
\end{align}
Furthermore, let $\gamma_s$ be the label for the ancilla qubit of an $X$-type stabilizer where the CNOT gate in the second time step has $e_j$ as the target qubit. Define $\tilde{\gamma}_s = \gamma_s + (d_x+1)/2$ so that $\tilde{\gamma}_s$ labels the ancilla qubit of an $X$-type stabilizer where the CNOT gate in the fifth time step has $e_j$ as the target qubit. From \cref{eq:m1k}, the ordering convention used in \cref{eq:PaxmCond} and the construction of the $Pe$ matrix, the total probability of $e^{(2D)}_{BLTRX}$ being highlighted from a failure of the CNOT gate in the second time step during the $j$'th syndrome measurement round is 
\begin{align}
    &P^{(1,m^{(t_1)}_{(1,\gamma_s,j)})}_{ZZ;CX} 
    \nonumber\\
    &= Pe[m^{(t_1)}_{(1,\gamma_s,j)} + 10] + Pe[m^{(t_1)}_{(1,\gamma_s,j)} + 11] 
    \nonumber\\ 
    &+ Pe[m^{(t_1)}_{(1,\gamma_s,j)} + 14] + Pe[m^{(t_1)}_{(1,\gamma_s,j)} + 15],
    \label{eq:PZZCX1}
\end{align}
where $Pe[q]$ is the element in the $q$'th row of the $Pe$ matrix. Similarly, the total probability of $e^{(2D)}_{BLTRX}$ being highlighted from a failure of the CNOT gate in the fifth time step during the $j$'th syndrome measurement round (assuming $j > 1$) is
\begin{align}
    &P^{(1,m^{(t_4)}_{(1,\tilde{\gamma_s},j-1)})}_{IZ;CX} 
    \nonumber\\
    &= Pe[m^{(t_4)}_{(1,\tilde{\gamma_s},j-1)} + 2] + Pe[m^{(t_4)}_{(1,\tilde{\gamma_s},j-1)} + 6] 
    \nonumber\\
    &+ Pe[m^{(t_4)}_{(1,\tilde{\gamma_s},j-1)} + 3] + 
    Pe[m^{(t_4)}_{(1,\tilde{\gamma_s},j-1)} + 7].
    \label{eq:PIZCX1}
\end{align}
Note that in \cref{eq:PIZCX1}, the index for the syndrome measurement round is $j-1$ and not $j$. This is because failures from the CNOT gate in the fifth time step add Pauli errors which are only detected in the next round. As such, this example highlights the importance of storing the conditional probabilities for the syndrome history and not just the conditional probabilities for the current syndrome measurement round.

The total probability of $e^{(2D)}_{BLTRX}$ being highlighted from an idle failure on the data qubit $e_j$ during the $j$'th syndrome measurement round is
\begin{align}
    P^{n_j(e_j)}_d = Pe[n_j(e_j) + 2] + Pe[n_j(e_j) + 3].
    \label{eq:Pnk2}
\end{align}

Using \cref{eq:PZZCX1,eq:PIZCX1,eq:Pnk2}, the total probability to leading order for $e^{(2D)}_{BLTRX}$ being highlighted during the $j$'th syndrome measurement round if $j > 1$ is
\begin{align}
    &P^{(2D;e_j)}_{BLTRX} 
    \nonumber\\
    &= P^{(1,m^{(t_1)}_{(1,\gamma_s,j)})}_{ZZ;CX}(1-P^{(1,m^{(t_4)}_{(1,\tilde{\gamma_s},j-1)})}_{IZ;CX})(1-P^{n_j(e_j)}_d)
    \nonumber\\
    &+ P^{(1,m^{(t_4)}_{(1,\tilde{\gamma_s},j-1)})}_{IZ;CX}(1-P^{(1,m^{(t_1)}_{(1,\gamma_s,j)})}_{ZZ;CX})(1-P^{n_j(e_j)}_d)
    \nonumber\\
    &+ P^{n_j(e_j)}_d(1-P^{(1,m^{(t_1)}_{(1,\gamma_s,j)})}_{ZZ;CX})(1-P^{(1,m^{(t_4)}_{(1,\tilde{\gamma_s},j-1)})}_{IZ;CX}).
    \label{eq:Pbltrxe}
\end{align}
If $j = 1$, we omit the term $P^{(1,m^{(t_4)}_{(1,\tilde{\gamma_s},j-1)})}_{IZ;CX}$ in \cref{eq:Pbltrxe}. The edge weight for the edge $e^{(2D)}_{BLTRX}$ is then 
\begin{align}
    W_{e^{(2D)}_{BLTRX}} = -\log{(P^{(2D;e_j)}_{BLTRX})}.
\end{align}

We conclude this section by noting that the function in \cref{eq:m2k} is used when fault locations arising from $Z$-type syndrome measurements must be taken into account when computing leading order edge weights. 

\subsection{Fault-tolerance properties of the surface-GKP code when using the analog information}
\label{subappedix:FaultToleranceSurfaceGKP}

In this section we show how a single fault in a distance-three surface-GKP code can result in a logical failure when taking into account the analog information for dynamically computing the edge weights of the matching graphs. To be clear, here we make an important distinction between faults and shift errors. A fault will always be taken to be an event such that after implementing the GKP error correction or performing a homodyne measurement, a non-trivial Pauli error is added to GKP qubits or a flip in the Pauli measurement outcome occurs. 

In \cref{fig:ExampleSingleFault}, we illustrate a case where a single fault results in a $Z$ data qubit error on the first data qubit. Such a data qubit error results in the red highlighted vertex of the matching graph to the right of the figure. Typically, the minimum weight path would be to choose a single edge connecting the red highlighted vertex to a blue boundary vertex on the left-hand side of the matching graph. However, there are very rare cases where, after obtaining all conditional probabilities from the GKP error correction, the two edges highlighted in green in \cref{fig:ExampleSingleFault} will have a combined total weight which is less than either edge connecting the red highlighted vertex to a left boundary vertex. Further, such a scenario occurs even though no data qubit errors are present on the green highlighted edges. This is possible when, for instance, the two data qubits in these green highlighted edges have large but correctable shift errors. In this case, these shift errors are correctly inferred and countered (hence not causing a Pauli error). Nevertheless, they come with large conditional error probabilities (i.e., false alarm) because they are too close to the decision boundary that they could have been an uncorrectable shift which does result in a Pauli error. As such, when performing MWPM, the path highlighted in green in \cref{fig:ExampleSingleFault} will be chosen resulting in a logical fault.

The above example shows that when using the analog information to dynamically compute the edge weights of the matching graphs, edges which have no errors may nevertheless be favored by the MWPM algorithm (due to the false alarm as discussed above), thus making the MWPM decoder choose a wrong path when pairing highlighted vertices.  
However it is important to keep in mind that the probability of such events occurring is very small (for the example above at $\sigma_{\mathrm{gkp}}^{(\mathrm{dB})} = 11$dB, we found such probabilities to be approximately $10^{-6}$ whereas the logical error rate is $8.8\times 10^{-4}$). Further, for larger code distances, higher order events need to occur such that fewer than $(d-1)/2$ faults resulting in non-trivial data qubit errors still lead to a logical failure. The probabilities of these higher order events are smaller than the computed logical failure rates and thus have a negligible contribution to the total logical failure rates in \cref{fig:GKPl1AlldAnalog}. To conclude, incorporating the extra analog information provides an overall benefit despite the possible, but very rare, false alarms (compare \cref{fig:GKPl1AlldNoAnalog} with \cref{fig:GKPl1AlldAnalog}).  


\bibliography{GKP}






\end{document}